\documentclass[a4paper,fleqn,usenatbib]{mnras}

\usepackage{newtxtext,newtxmath}

\usepackage[T1]{fontenc}
\usepackage{ae,aecompl}

\usepackage{graphicx}	
\usepackage{amsmath}	
\usepackage{breqn}
\usepackage{threeparttable,adjustbox,array,multirow,makecell}
\usepackage{xcolor}

\newenvironment{altfont}{\fontfamily{lmtt}\selectfont}{\par}
\DeclareTextFontCommand{\textaltfont}{\altfont}
\defcitealias{mackey10KpcStellar2016}{M16}
\defcitealias{choiSMASHingLMCTidally2018}{C18}
\defcitealias{vandermarelThirdEpochMagellanicCloud2014}{vdM14}
\defcitealias{C20}{Paper I}
\title[MagES: LMC northern arm]{The Magellanic Edges Survey II. Formation of the LMC's northern arm}

\author[L. R. Cullinane et al.]{L. R. Cullinane$^{1}$\thanks{E-mail: lara.cullinane@anu.edu.au (LRC)},
	A. D. Mackey$^{1}$,
	G. S. Da Costa$^{1}$,
	D. Erkal$^{2}$,
	S. E. Koposov$^{3,4}$,\newauthor
	V. Belokurov$^{4}$
	\\
	$^{1}$Research School of Astronomy and Astrophysics, Australian National University, Canberra, ACT 2611, Australia\\
	$^{2}$Department of Physics, University of Surrey, Guildford GU2 7XH, UK \\
	$^{3}$Institute for Astronomy, University of Edinburgh, Royal Observatory, Blackford Hill, Edinburgh EH9 3HJ, UK\\
	$^{4}$Institute of Astronomy, University of Cambridge, Madingley Road, Cambridge CB3 0HA, UK\\
}

\date{Accepted XXX. Received YYY; in original form ZZZ}

\pubyear{2021}

\begin{document}
	\label{firstpage}
	\pagerange{\pageref{firstpage}--\pageref{lastpage}}
	\maketitle
	
	\begin{abstract}
	The highly-substructured outskirts of the Magellanic Clouds provide ideal locations for studying the complex interaction history between both Clouds and the Milky Way (MW). In this paper, we investigate the origin of a >20$^\circ$ long arm-like feature in the northern outskirts of the Large Magellanic Cloud (LMC) using data from the Magellanic Edges Survey (MagES) and Gaia EDR3. We find that the arm has a similar geometry and metallicity to the nearby outer LMC disk, indicating that it is comprised of perturbed disk material. Whilst the azimuthal velocity and velocity dispersions along the arm are consistent with those in the outer LMC, the in-plane radial velocity and out-of-plane vertical velocity are significantly perturbed from equilibrium disk kinematics. We compare these observations to a new suite of dynamical models of the Magellanic/MW system, which describe the LMC as a collection of tracer particles within a rigid potential, and the SMC as a rigid Hernquist potential. Our models indicate the tidal force of the MW during the LMC’s infall is likely responsible for the observed increasing out-of-plane velocity along the arm. Our models also suggest close LMC/SMC interactions within the past Gyr, particularly the SMC’s pericentric passage \textasciitilde150~Myr ago and a possible SMC crossing of the LMC disk plane \textasciitilde400~Myr ago, likely do not perturb stars that today comprise the arm. Historical interactions with the SMC prior to \textasciitilde1~Gyr ago may be required to explain some of the observed kinematic properties of the arm, in particular its strongly negative in-plane radial velocity.
	\end{abstract}
	
	



\section{Introduction}\label{sec:intro}
The Large and Small Magellanic Clouds (LMC/SMC), as the closest pair of interacting dwarf satellites of the Milky Way \citep[at distances of \textasciitilde50 and \textasciitilde60~kpc respectively:][]{pietrzynskiDistanceLargeMagellanic2019,graczykDistanceDeterminationSmall2020a}, are ideally situated for detailed study of the influence of tidal interactions on galaxy evolution. The SMC has long been known to be heavily distorted, with a line of sight depth of up to 20~kpc \citep[e.g.][]{hatzidimitriouStellarPopulationsLargescale1989,ripepiVMCSurveyXXV2017} which varies as a function of position angle. It possesses an asymmetric, irregular morphology exhibiting striking differences between the locations of young and old stars \citep[e.g.][]{elyoussoufiVMCSurveyXXXIV2019,mackeySubstructuresTidalDistortions2018}, and kinematic evidence for tidal expansion \citep[e.g.][]{deleoRevealingTidalScars2020a,zivickDecipheringKinematicStructure2021}. The LMC, although more kinematically ordered than the SMC, also displays substantial deviations from a simple rotating disk structure. It has multiple warps \citep{olsenWarpLargeMagellanic2002,choiSMASHingLMCTidally2018}, sharp truncations in the outer disk \citep{mackeySubstructuresTidalDistortions2018}, ring-like overdensities \citep{kunkelDynamicsLargeMagellanic1997,choiSMASHingLMCMapping2018}, and an off-centre stellar bar \citep[e.g.][]{vandermarelMagellanicCloudStructure2001}. Each of these features encodes valuable information about the extensive interaction history of the Clouds. 

Precise measurements of the masses and orbits of the LMC and SMC, and their internal kinematics, are key to understanding how interactions between both Clouds, and the Milky Way, form the disturbed features observed. While the Clouds are strongly suspected to have experienced a close passage \textasciitilde150~Myr ago \citep{zivickProperMotionField2018b}, and are likely just past pericentre on their first infall into the Milky Way potential \citep{kallivayalilThirdEpochMagellanicCloud2013}, particulars of their interactions beyond this remain relatively unconstrained. Recent studies of the star-formation history of the Clouds provide evidence of potential past interactions, with spikes in the global star formation rate of both Clouds \textasciitilde1-2~Gyr ago \citep[e.g.][]{rubeleVMCSurveyXXXI2018a,ruiz-laraLargeMagellanicCloud2020a}. However, these studies have lower time-resolution than dynamical studies, and alone provide limited constraints on, for example, the impact parameter or the relative location and orientation of the Clouds during close interactions.

One useful method to explore past dynamical interactions is to study stars in the outskirts of the Clouds. These stars are most strongly susceptible to external perturbations, and the resulting structural and kinematic signatures are more persistent compared to the central regions, where dynamical timescales are much shorter. Recent studies of the Clouds using deep photometric data \citep[e.g.][]{mackey10KpcStellar2016,mackeySubstructuresTidalDistortions2018,pieresStellarOverdensityAssociated2017a} and multi-dimensional phase-space information from Gaia \citep[e.g.][]{belokurovCloudsArms2019,gaiacollaborationGaiaEarlyData2021a} have revealed a wealth of substructure in the periphery of the Magellanic system. Many of these features are thought to be due to dynamical perturbation and, as a result, are ideal targets for studying the history of interactions between the LMC and SMC, and between the Clouds and the Milky Way. 

Of particular interest is a large arm-like feature to the north of the LMC discovered in first year data from the Dark Energy Survey (DES) by \citet[][henceforth referred to as M16]{mackey10KpcStellar2016}. The feature begins \textasciitilde13$^\circ$ due north of the LMC centre where it appears to join the northern outskirts of the LMC disk, and has an on-sky width of \textasciitilde2$^\circ$. Initial photometric analysis, limited by the extent of the DES footprint, traced the substructure for \textasciitilde12.5$^\circ$ eastward. Utilising astrometric proper motion and parallax information provided by Gaia DR2, \citet{belokurovCloudsArms2019} were also able to recover the feature, tracing it for at least an additional \textasciitilde10$^\circ$ beyond the initial discovery. 

Several papers have attempted to elucidate the origin of the feature using dynamical models, with varying conclusions. \citetalias{mackey10KpcStellar2016} present an $N$-body model of the LMC undergoing infall over \textasciitilde2~Gyr into a 3-component MW potential as described in \citet{gomezItMovesDangers2015}. That simulation produces a qualitatively similar stream of debris in the northern outskirts of the LMC disk, due solely to the tidal influence of the Milky Way (i.e., without requiring the presence of the SMC). In contrast, \citet{beslaLowSurfaceBrightness2016} present $N$-body models of an LMC and SMC interacting in isolation for 6~Gyr, before undergoing infall into a MW halo potential for 1~Gyr. Even prior to entering the MW potential, qualitatively similar asymmetrical spiral structures, formed in the LMC disk after repeated SMC passages, are seen in the LMC's northern outskirts; these persist during infall to the MW potential. \citet{belokurovCloudsArms2019} also show a number of simpler models of tracer particles within high-mass and low-mass LMC potentials, undergoing infall for 1~Gyr into the 3-component MW potential described in \citet{bovyGalpyPythonLIBRARY2015}. Models both with and without the presence of an SMC potential form qualitatively similar features in the northern outskirts of the LMC, with the best qualitative match occurring due to the combined influence of both the SMC and MW. With multiple scenarios each reproducing qualitatively similar structures to that observed, the origin of the feature remains uncertain. 

However, these studies have been fundamentally limited by a lack of kinematic data along the arm. This restricts analysis to only qualitatively reproducing the feature’s shape which -- as demonstrated above -- results in ambiguity regarding its origin. Indeed, \citetalias{mackey10KpcStellar2016} note that line-of-sight (LOS) velocities would assist in distinguishing between material tidally stripped from the LMC, and overdense features in the extended LMC disk. An investigation into the kinematics of the northern arm is therefore critical. 

In this paper, we present a comprehensive analysis of the LMC's northern arm using data from the Magellanic Edges Survey \citep[MagES:][]{C20}. This spectroscopic survey targets red clump (RC) and red giant branch (RGB) stars in the extreme Magellanic periphery, using the 2dF/AAOmega instrument \citep{lewisAngloAustralianObservatory2dF2002,sharpPerformanceAAOmegaAAT2006} on the 3.9~m Anglo-Australian Telescope (AAT) at Siding Spring Observatory. In conjunction with Gaia astrometry, it is the first large-scale survey to study 3D kinematics in the outskirts of the Clouds. MagES fields are specifically selected to cover low-surface-brightness substructures in the Magellanic periphery – including the northern arm. With seven fields located across the length of the feature, providing 3D kinematics for hundreds of individual stars, detailed study of the arm's dynamical properties becomes possible. 

The paper is arranged as follows. Section~\ref{sec:data} presents an overview of the data, and \S\ref{sec:obsprops} describes the derived kinematic, structural, and abundance properties of the feature. In \S\ref{sec:models} we present new dynamical models of the LMC and SMC undergoing infall into the Milky Way potential, aimed at quantitatively reproducing the kinematics of the northern arm, and discuss the main implications for the origin of this structure. Our conclusions are presented in \S\ref{sec:concs}.

\section{Data}\label{sec:data}

MagES utilises the 2dF multi-object fibre positioner, and the dual-beam AAOmega spectrograph on the AAT. The 2dF positioner allows for the observation of \textasciitilde350 science targets per 2 degree diameter field. As described in \citet[][henceforth referred to as Paper I]{C20}, we configure the blue arm on AAOmega with the 1500V grating, to give coverage of the MgIb triplet with resolution R\textasciitilde$3700$, and the red arm with the 1700D grating to give coverage of the near-infrared CaII triplet with R\textasciitilde$10000$. \citetalias{C20} also outlines in detail the target selection procedures, observation characteristics, and data reduction pipeline for MagES; here we briefly present details of the observations specific to the northern arm.

\begin{figure*}
	\includegraphics[height=12cm]{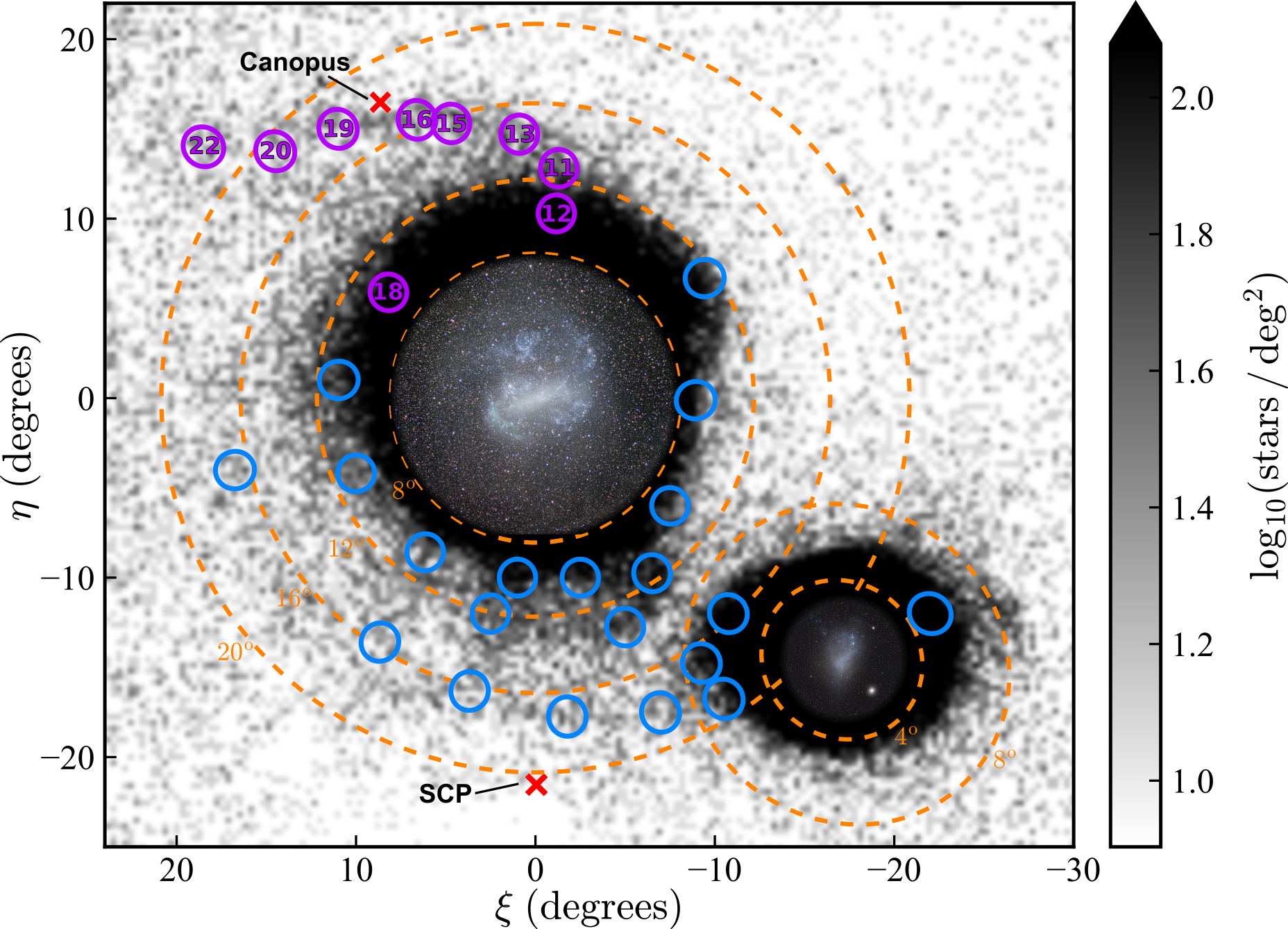}
	\caption{Location of observed MagES fields across the Magellanic periphery. Purple circles indicate fields along the LMC's northern arm analysed in this paper, with blue circles indicating other MagES fields. The background image shows the log density of Magellanic red clump and red giant stars per square degree, selected from Gaia DR2 (the target catalogue from which most MagES stars are drawn) as per \protect\cite{belokurovCloudsArms2019}. On this map, north is up and east is to the left; ($\eta, \xi$) are coordinates in a tangent-plane projection centred on the LMC ($\alpha_0=82.25^{\circ}$, $\delta_0=-69.5^{\circ}$). Orange dashed circles mark angular separations of $8^\circ$, $12^{\circ}$, $16^{\circ}$ and $20^{\circ}$ from the LMC centre and $4^\circ$, $8^\circ$ from the SMC centre. The red x-signs mark the location of Canopus -- the second brightest star in the sky, which limits MagES field placement on the northern arm to avoid spectral contamination from scattered light -- and the south celestial pole.}
	\label{fig:map}
\end{figure*}

Seven MagES fields are located along the arm; field positions are shown in Fig.~\ref{fig:map}. We note that with the exception of field 22 (as well as fields 12 and 18 located in the northern LMC disk) all fields along the arm were observed prior to the release of Gaia DR2, and thus selection for those fields was performed without using parallax and proper motion information. As a consequence, the selection efficiency for true Magellanic members in these fields is relatively low -- these correspond to `D' and `M' fields as defined in \citetalias{C20}. We discuss the implications of this in greater detail below.

Reduction of the spectra using the \textaltfont{2dFDR} pipeline, and derivation of LOS velocities, are described in \citetalias{C20}. Stars with heliocentric velocity estimates are cross-matched against the Gaia EDR3 catalogue\footnote{While \citetalias{C20} describes cross-matching against Gaia DR2, we have updated our procedures to incorporate the latest astrometry from Gaia EDR3 \citep{gaiacollaborationGaiaEarlyData2021b}.}, and further quality cuts based on Gaia parameters \textaltfont{ruwe}<1.4 and $C^*$<4$\sigma_{C^*}$\footnote{$C^*$ and $\sigma_{C^*}$ are defined using Eqs. 6 and 18 of \cite{rielloGaiaEarlyData2021} respectively.} applied. 

The resulting sample of stars includes both true Magellanic stars, and foreground contaminants. We use a statistical framework, described in detail in \citetalias{C20}, to probabilistically associate stars, based on their kinematics, to either the Clouds, or one of several possible Milky Way contaminant populations. These association probabilities are used to weight the fitting of a multi-dimensional Gaussian distribution describing the aggregate Magellanic kinematic properties of each field: the LOS velocity ($V_{\text{LOS}}$) and dispersion ($\sigma_{\text{LOS}}$), plus the two components of proper motion ($\mu_\alpha$, $\mu_\delta$)\footnote{$\mu_\alpha$ refers to proper motion in the $\alpha\cos(\delta)$ direction, as obtained directly from the Gaia EDR3 source catalogue using the column \textaltfont{PMRA}.} and their dispersions ($\sigma_\alpha$, $\sigma_\delta$). We assume there is no covariance between the LOS velocity and either proper motion component, but do account for covariance between the two proper motion components as presented in Gaia EDR3. Fitting is performed using the Markov Chain Monte Carlo ensemble sampler \textsc{emcee} \citep{foreman-mackeyEmceeMCMCHammer2013} in order to maximise the log-likelihood of the Gaussian model given the data; we report the 68 per cent confidence interval as the $1\sigma$ uncertainty in each of the six fitted parameters. As part of this process, we additionally obtain a fitted estimate of the total fraction of likely Magellanic stars per field. 

Table~\ref{tab:fieldbase} provides the inferred kinematic properties for each of the seven fields along the northern arm, as well as the number of stars in the field with an individual probability $P_i\geq$50\% of being associated with the Clouds. This number is typically very similar ($\pm$1-2 stars) to that inferred from the fitted total fraction of Magellanic stars. In each case, the number of likely Magellanic stars is significantly lower than the total number of stars observed in the field. This is primarily due to the relatively inefficient target selection used in all fields except field 22 (and disk fields 12 and 18 as discussed above). These fields were observed prior to the release of Gaia DR2, and thus target selection was based only on colour-magnitude diagram (CMD) position. As there is moderate Milky Way contamination within the selection boxes used to isolate Magellanic red clump stars (see Fig.~2 of \citetalias{C20}), a significant fraction of the targets observed in these fields are not genuinely Magellanic members. Fields observed later in the survey, after the release of Gaia DR2 (`G' fields), use updated target selection procedures that incorporate kinematic priors, and consequently suffer far less from contamination by non-members. This is demonstrated in field 22, which uses the updated selection procedure. Despite being located near the end of the arm -- where the density of Magellanic stars is intrinsically low, and the density of contaminants is high due to the field’s proximity to the Galactic plane -- a comparable number of Magellanic stars are detected as in e.g. field 15, located much closer to the LMC disk in areas where the density of members is significantly higher. Table~\ref{tab:fieldbase} also provides kinematic data for two fields in the northern LMC disk located close to the northern arm, previously discussed in \citetalias{C20} and re-analysed using Gaia EDR3 data in this paper. 

Notable in Table~\ref{tab:fieldbase} is field 20, which contains no stars with a significant probability of being Magellanic. In addition to using the relatively inefficient CMD-only selection procedure, the field centre is \textasciitilde1$^\circ$ offset from the feature track. This offset was not apparent when the field was initially observed in 2017, as at the time it was located at the extreme limit of the known structure. It is only with astrometric cuts as afforded by Gaia that the feature could be traced further, revealing the offset. As a result, no stars in this field are convincingly Magellanic in origin, and we therefore exclude this field from further analysis. 

\begin{table*}
	\centering
	\caption{MagES fields along the northern arm and in the nearby northern LMC disk. Columns give the field number and classification as described in \protect\citetalias{C20}; location of the field centre as RA($\alpha$), DEC($\delta$) in J2000.0; on-sky distance of the field from the centre of the LMC ($R_{\text{LMC}}$), number of likely Magellanic stars per field, and aggregate kinematic parameters (described in \S\ref{sec:data}).}
	\label{tab:fieldbase}
	\begin{adjustbox}{max width=\textwidth}
		\begin{tabular}{lllllllllll}
			\hline
			\multicolumn{1}{>{\centering\arraybackslash}m{1.5cm}}{Field (Class)} & \multicolumn{1}{c}{RA} & \multicolumn{1}{c}{DEC} & \multicolumn{1}{>{\centering\arraybackslash}m{1.3cm}}{$R_{\text{LMC}}$ ($^\circ$)} & \multicolumn{1}{>{\centering\arraybackslash}p{1.4cm}}{$N_{\text{Magellanic}}$ ($P_i\geq50\%$)} & \multicolumn{1}{>{\centering\arraybackslash}p{1.2cm}}{$V_{\text{LOS}}$ \newline(km~s$^{-1}$)} & \multicolumn{1}{>{\centering\arraybackslash}p{1.2cm}}{$\sigma_{\text{LOS}}$\newline (km~s$^{-1}$)} & \multicolumn{1}{>{\centering\arraybackslash}p{1.2cm}}{$\mu_\alpha$\newline (mas~yr$^{-1}$)} & \multicolumn{1}{>{\centering\arraybackslash}p{1.2cm}}{$\sigma_\alpha$\newline (mas~yr$^{-1}$)} & \multicolumn{1}{>{\centering\arraybackslash}p{1.2cm}}{$\mu_\delta$\newline (mas~yr$^{-1}$)} & \multicolumn{1}{>{\centering\arraybackslash}p{1.2cm}}{$\sigma_\delta$\newline (mas~yr$^{-1}$)} \\ \hline
			11 (D) & 05 19 42.63 & -56 53 06.88 & 12.7 & 75 & $ 280.8\pm2.2 $ & $ 17.2\pm2.0 $ & $ 1.72\pm0.03 $ & $ 0.13\pm0.03 $ & $ 0.06\pm0.04 $ & $ 0.24\pm0.03 $  \\
			13 (D) & 05 35 05.69  & -55 06 03.11 & 14.6 & 38 & $ 294.3\pm1.7 $ & $ 8.0\pm1.9 $ & $ 1.58\pm0.04 $ & $ 0.12\pm0.06 $ & $ -0.03\pm0.04 $ & $ 0.16\pm0.06 $ \\
			15 (D) & 06 00 07.40 & -54 17 53.14 & 16.0 & 32 & $ 311.9\pm2.7 $ & $ 12.6\pm2.1 $ & $ 1.50\pm0.04 $ & $ 0.09\pm0.06 $ & $ 0.12\pm0.04 $ & $ 0.06\pm0.05 $ \\
			16 (D) & 06 12 13.07 & -53 52 32.45 & 16.8 & 25 & $ 323.2\pm2.0$ & $ 8.3\pm1.7 $ & $ 1.50\pm0.05 $ & $ 0.11\pm0.07  $& $ 0.24\pm0.04 $ & $ 0.07\pm0.06 $ \\
			19 (M) & 06 40 29.00 & -53 29 04.00 & 18.6 & 13 & $ 351.3\pm4.8 $ & $ 14.5\pm4.5 $ & $ 1.16\pm0.09 $ & $ 0.23\pm0.10 $ & $ 0.57\pm0.07 $ & $0.13\pm0.09 $ \\
			20 (M) & 07 04 01.00 & -53 37 01.00 & 19.9 & 0 & - & - & - & - & - & - \\
			22 (G) & 07 25 34.00 & -52 04 52.00 & 22.8 & 27 & $ 372.9\pm1.6 $ & $ 7.1\pm1.3 $ & $ 1.15\pm0.03 $ & $ 0.06\pm0.04 $ & $ 0.71\pm0.02 $ & $ 0.04\pm0.03 $ \\
			18 (G) & 06 40 00.00 & -62 30 00.00  & 10.7 & 299 & $ 324.5\pm1.2 $ & $ 20.3\pm0.9 $ & $ 1.49\pm0.01 $ &$  0.11\pm0.01 $ & $ 1.00\pm0.01 $ & $ 0.12\pm0.01 $ \\
			12 (G) & 05 20 00.00 & -59 18 00.00 & 10.3 & 284  & $ 287.1\pm1.5 $& $ 24.8\pm1.1 $ & $ 1.78\pm0.01 $ & $ 0.12\pm0.01 $ & $ 0.20\pm0.01 $ & $ 0.19\pm0.01 $ \\ \hline
		\end{tabular}
	\end{adjustbox}
\end{table*}

In addition to kinematic properties, MagES also reports [Fe/H] estimates for sufficiently bright red giant branch stars, derived from the equivalent width of the 8542\AA{} and 8662\AA{} CaII triplet lines (see \citetalias{C20} and \citealt{dacostaCaIiTriplet2016} for details). However, such stars are only included in the target selection for field 22 along the arm (as well as fields 12 and 18, previously described in \citetalias{C20}). For the fainter red clump stars observed in the remaining fields along the arm, the S/N for any individual star is too low to accurately measure the equivalent width of the two lines, particularly as the 8662\AA{} line is within a region of the spectrum relatively heavily contaminated by night sky emission. Therefore, in order to derive metallicity estimates for these fields, spectra for likely ($P_i$$\geq$50\%) Magellanic stars are shifted into the rest frame using their (geocentric) LOS velocities and then stacked to create a single `representative' RC spectrum for the field. This increases the contrast of the two CaII lines relative to the (stochastically over- or under-subtracted) residual night-sky emission, allowing for equivalent width measurements to be performed. As the stacked clump stars only occupy a small magnitude range, stacking spectra is not expected to substantially bias the derived equivalent widths, and the resulting [Fe/H] estimates are expected to tend towards the mean metallicity within a given field. All metallicity estimates are assumed to have systematic uncertainties of 0.2dex (we refer the interested reader to \citetalias{C20} for details). 

\subsection{An arm-like coordinate system}\label{sec:armloc}

For the following analysis, it is convenient to have a coordinate system in which the northern arm, as projected on the sky, is straight -- similar to coordinate systems used to describe stellar streams in the MW halo. However, while coordinate systems for most halo streams can be derived by assuming that the stream follows a great circle on the sky, this is not the case for the northern arm. Consequently, in this section we describe derivation of a custom coordinate system which follows the track of the structure on the sky, with the origin of the feature nearest the LMC disk. In this derivation, we neglect uncertainties on the position of each individual star, as these are negligibly small (\textasciitilde0.15” in each component).

To locate the feature, we use a catalogue of Magellanic red clump and red giant branch stars selected from Gaia DR2 according to \citet{belokurovCloudsArms2019}, which incorporates astrometric, photometric, and quality cuts. This catalogue provides a relatively clean selection of Magellanic stars with contiguous coverage across the entire length of the arm. We calculate an orthographic projection of the stars into Cartesian X,Y coordinates, as per Eq.~2 of \cite{gaiacollaborationGaiaDataRelease2018b}, repeated here as Eq.~\ref{eq:xy}. Here, ($\alpha_0,\delta_0= 79.88^\circ,-69.59^\circ$): the LMC centre-of-mass (COM) position reported by \citet[][henceforth referred to as vdM14]{vandermarelThirdEpochMagellanicCloud2014} for their ‘PMs+Old vLOS Sample’. This is a kinematic centre, derived from a simultaneous fit of HST field-aggregate proper motions, combined with LOS velocities for an ‘old’\footnote{Comprised of carbon stars, AGB and RGB stars that are predominantly older than 1–2~Gyr and therefore similar in age to the red clump population.} stellar sample. This is as similar as possible to the data used in the present work.

\begin{equation} \label{eq:xy}
	\begin{split}
		X &= -\cos(\delta)\sin(\alpha-\alpha_0) \\
		Y &= \sin(\delta)\cos(\delta_0) - \cos(\delta)\sin(\delta_0)\cos(\alpha - \alpha_0)
	\end{split}
\end{equation}

\begin{figure*}
	\setlength\tabcolsep{1.5pt}
	\begin{tabular}{cc}
		\includegraphics[width=0.52\textwidth]{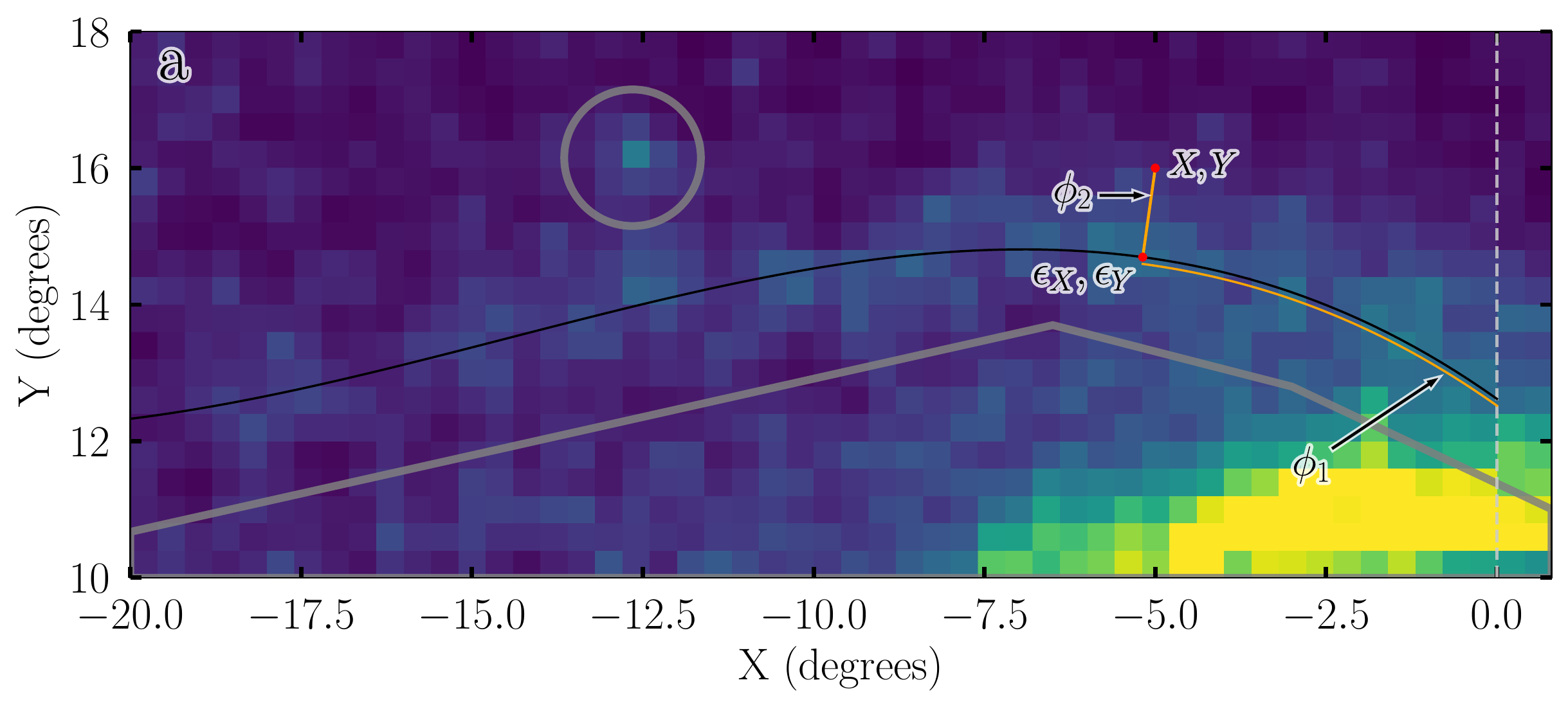} & \vspace{-1pt}\includegraphics[width=0.5\textwidth]{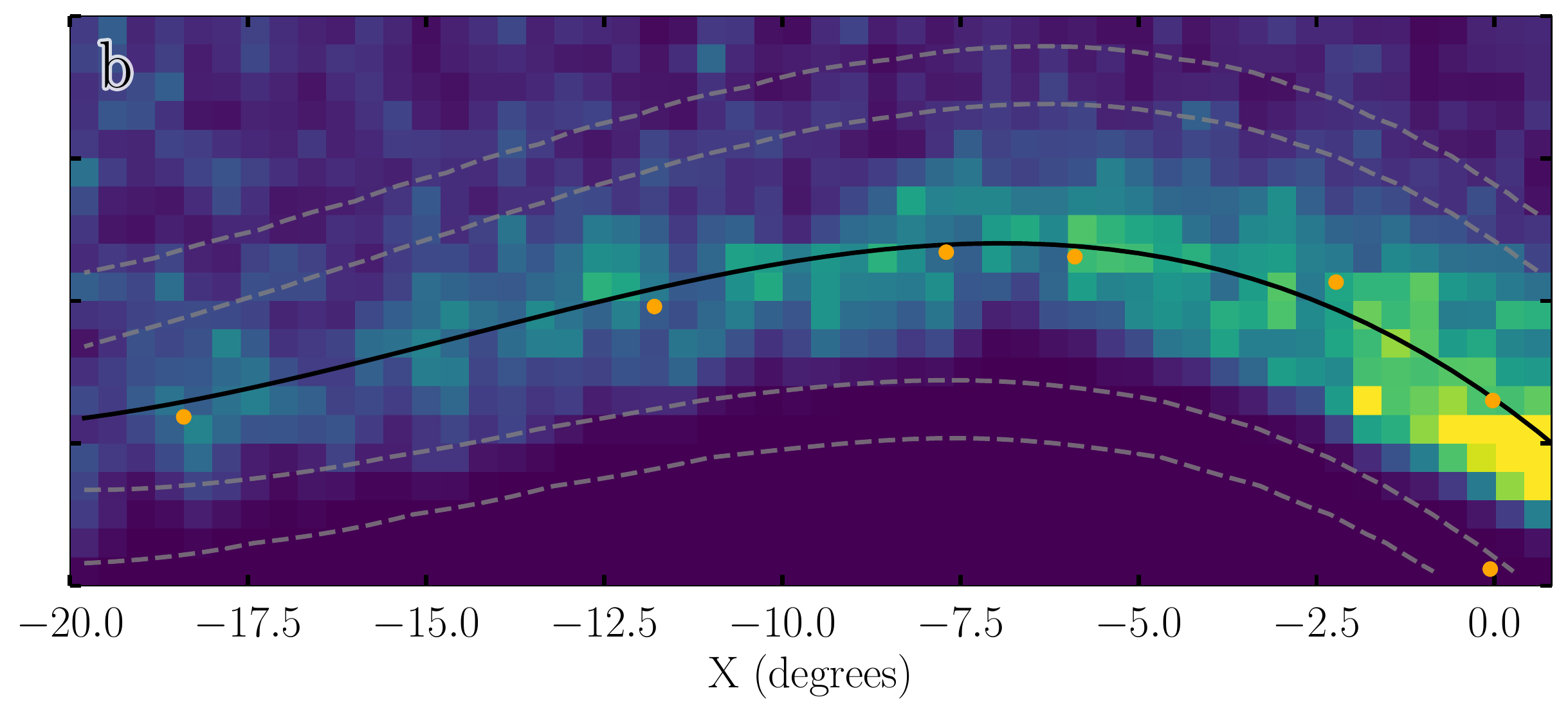} \\ \multicolumn{2}{c}{\includegraphics[height=5cm]{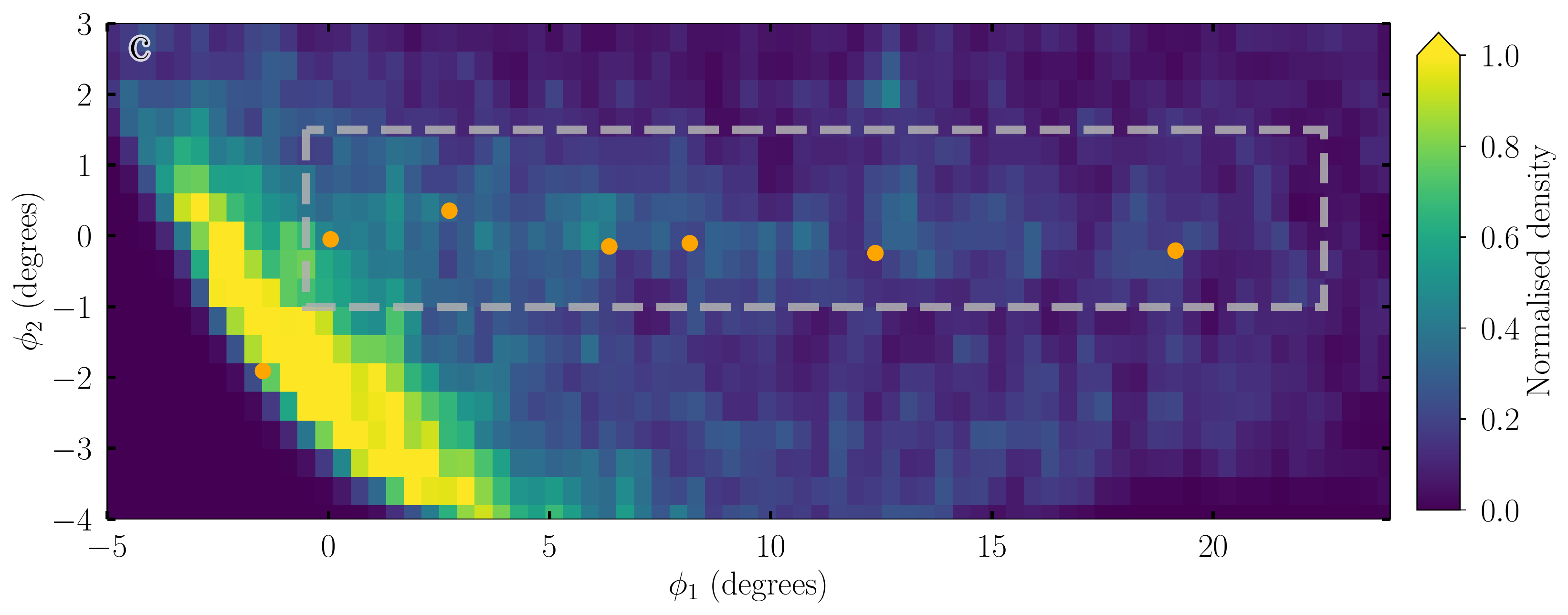}} 			
	\end{tabular}
	
	\caption{Normalised density of red clump and red giant branch stars, selected as per \protect\cite{belokurovCloudsArms2019}, in the region surrounding the northern arm, used to derive a feature coordinate system. Panels \textit{a} and \textit{b} have data binned in $0.4^\circ\times0.4^\circ$ squares within an orthographic projection in Cartesian coordinates. Panel \textit{a} shows the full selection of stars, with solid grey lines indicating the regions surrounding the LMC disk and the Carina dwarf galaxy masked prior to fitting the feature track. Also shown is an example demonstrating the calculation of the two feature coordinates ($\phi_1$ and $\phi_2$) for any point ($X,Y$) in the feature; ($\epsilon_X,\epsilon_Y$) is the nearest point on the feature track. Panel \textit{b} shows the post-masking data used in the fitting routine. The best-fitting polynomial (as described in Eq.~\ref{eq:track}) is shown in solid black, with fitted 1$\sigma$ and $2\sigma$ contours represented by dashed grey lines. Centres of MagES fields along the arm are overplotted in orange; the fitted feature track passes close to the centre of each field. Panel \textit{c} shows the full data selection after transformation into the feature coordinates. In this coordinate system, the feature track is a straight line at $\phi_2=0^\circ$. The selection box used to describe the feature location is shown in dashed grey.}
	\label{fig:xysel}
\end{figure*}

To describe the feature, we select stars in the region $-20$<X<0.8 and 10<Y<18, as seen in Fig.~\ref{fig:xysel}. In addition to containing stars associated with the northern arm feature, this region also includes a significant number of stars associated with the outer LMC disk, the high density of which necessitates masking before fitting the northern arm itself. All stars within the solid selection box in panel \textit{a} of Fig.~\ref{fig:xysel} are masked. Additionally, we mask stars within a two-degree diameter circle centred on the Carina dwarf galaxy ($\alpha_C,\delta_C=100.40^\circ, -50.97^\circ$), just north of the feature; many stars associated with the Carina dwarf pass the selection criteria described in \citet{belokurovCloudsArms2019}. 

Remaining stars are binned into $0.4^\circ\times0.4^\circ$ bins to smooth their distribution; smaller bins contain too few stars near the low-density end of the feature, while larger bins do not sufficiently resolve the feature. We describe the resulting binned distribution $Z$ by a Gaussian profile in $Y$, as in Eq.~\ref{eq:gaussfit}, where the peak height ($A_Y$), centre ($Y_{\text{T}}$), and width ($\sigma_Y$) are each allowed to vary as an $n$th-order polynomial as a function of $X$. 

\begin{equation} \label{eq:gaussfit}
	\begin{split}
		Z(X,Y) &= A_Y(X) \exp{\left[\frac{-\left(Y - Y_{\text{T}}\left(X\right)\right)^2}{2 \left(\sigma_Y\left(X\right)\right)^2}\right]} \\
		A_Y(X) &= a_nX +a_{n-1}X^{n-1}+...+a_0 \\
		\sigma_Y(X) &= b_nX +b_{n-1}X^{n-1}+...+b_0 \\
		Y_{\text{T}}(X) &= c_nX +c_{n-1}X^{n-1}+...+c_0 
	\end{split}
\end{equation}

We perform a least-squares fit to the polynomial coefficients, for all combinations of $n$th-order polynomials up to a maximum of 2\textsuperscript{nd} order in $A_Y$, 2\textsuperscript{nd} order in $\sigma_Y$, and 5\textsuperscript{th} order in $Y_{\text{T}}$; polynomials of higher orders overfit the data, resulting in unrealistic contours particularly near the ends of the feature. The set of coefficients with the lowest sum-of-square residuals are taken as the final track parameters for the arm: Eq.~\ref{eq:track} gives the resulting best-fit equations describing the on-sky feature track. 

\begin{equation} \label{eq:track}
	\begin{split}
		A_Y(X) &= 2.642\times10^{-2}X + 0.7229\\
		\sigma_Y(X) &= 8.198\times10^{-3}X + 1.168 \\
		Y_{\text{T}}(X) &= -3.386\times10^{-6}X^4 - 1.669\times10^{-3}X^3 \\
		&- 6.825\times10^{-2}X^2- 0.7099X + 12.62
	\end{split}
\end{equation}

We note coefficients describing the variation in width and peak height are, for the process of deriving the feature track, nuisance parameters; it is the polynomial describing the centre position that is of interest. However, we do find the peak height ($A_Y$), indicative of the stellar density, decreases by \textasciitilde70\%, and the feature width ($\sigma_Y$) decreases by \textasciitilde14\% along the length of the structure. The track and $1\sigma$ width contours from the fit are shown in panel \textit{b} of Fig.~\ref{fig:xysel}, with the MagES field centres marked in orange. Whilst the polynomial fit is not at all constrained by the locations of MagES fields -- which were deliberately selected to be centred on the feature -- it nonetheless passes very closely to each field centre. We define the origin, where the arm appears to meet the LMC disk, to sit at $X=0^\circ$.

We use the best-fit track to define a coordinate system for the arm, with components denoted $\phi_1$ and $\phi_2$, in which the central track is a straight line at $\phi_2=0$. For each star, we determine the nearest point on the track given by Eq.~\ref{eq:track} (which we refer to as $\epsilon$). The coordinate $\phi_1$ is defined as the distance (or line integral) along Eq.~\ref{eq:track} from $X=0$ to $X=\epsilon_X$. We calculate the direction normal to Eq.~\ref{eq:track} at $\epsilon$, and define $\phi_2$ as the distance along the normal vector from $\epsilon$ to the star’s location\footnote{We note this procedure does not result in a 1:1 mapping of X,Y to $\phi_1$,$\phi_2$ across the entire X,Y domain, as due to the shape of the feature normal vectors to Eq.~\ref{eq:track} for negative values of $\phi_2$ eventually intersect. However, these intersections only occur at relatively large negative values of $\phi_2$, within the LMC disk where $\phi_1$,$\phi_2$ coordinates are not meaningful. In the vicinity of the northern arm X,Y locations are mapped to unique $\phi_1$,$\phi_2$ coordinates.}. The outcome of this process is a set of $\phi_1$, $\phi_2$ coordinates for each star; the resulting density plot for stars along the northern arm is shown in panel \textit{c} of Fig.~\ref{fig:xysel}. For convenience, we also transform the MagES field centres into the feature coordinate system. Table~\ref{tab:fieldloc} presents the location of MagES fields along the arm in both cartesian and feature coordinates. When selecting member stars later in our analysis, we define a box of width 2.5$^\circ$ in $\phi_2$, between $-0.5^\circ\leq\phi_1\leq25^\circ$ as in panel \textit{c} of Fig.~\ref{fig:xysel}, which describes the location of the feature. 

\begin{table}
	\centering
	\caption{Orthographic cartesian coordinates centred on the LMC COM, and feature coordinates along and across the northern arm (calculated as in \S\ref{sec:armloc}) for MagES fields.}
	\label{tab:fieldloc}
	\begin{tabular}{cllll}
		\hline
		Field & \multicolumn{1}{c}{X (deg)} & \multicolumn{1}{c}{Y (deg)} & \multicolumn{1}{c}{$\phi_1$ (deg)} & \multicolumn{1}{c}{$\phi_2$ (deg)} \\ \hline
		11 & -0.03  & 12.60  & 0.05  & -0.05  \\
		13 & -2.23  & 14.26  & 2.74  & 0.36  \\
		15 & -5.89  & 14.62  & 6.35  & -0.15  \\
		16 & -7.70  & 14.68  & 8.17  & -0.10  \\
		19 & -11.80  & 13.92  & 12.36 & -0.24 \\
		22 & -18.40  & 12.37  & 19.15 & -0.21 \\ \hline
	\end{tabular}
\end{table}

\section{Observed Properties of the Northern Arm}\label{sec:obsprops}

\subsection{Metallicity}\label{sec:met}

\begin{figure}
	\includegraphics[width=\columnwidth]{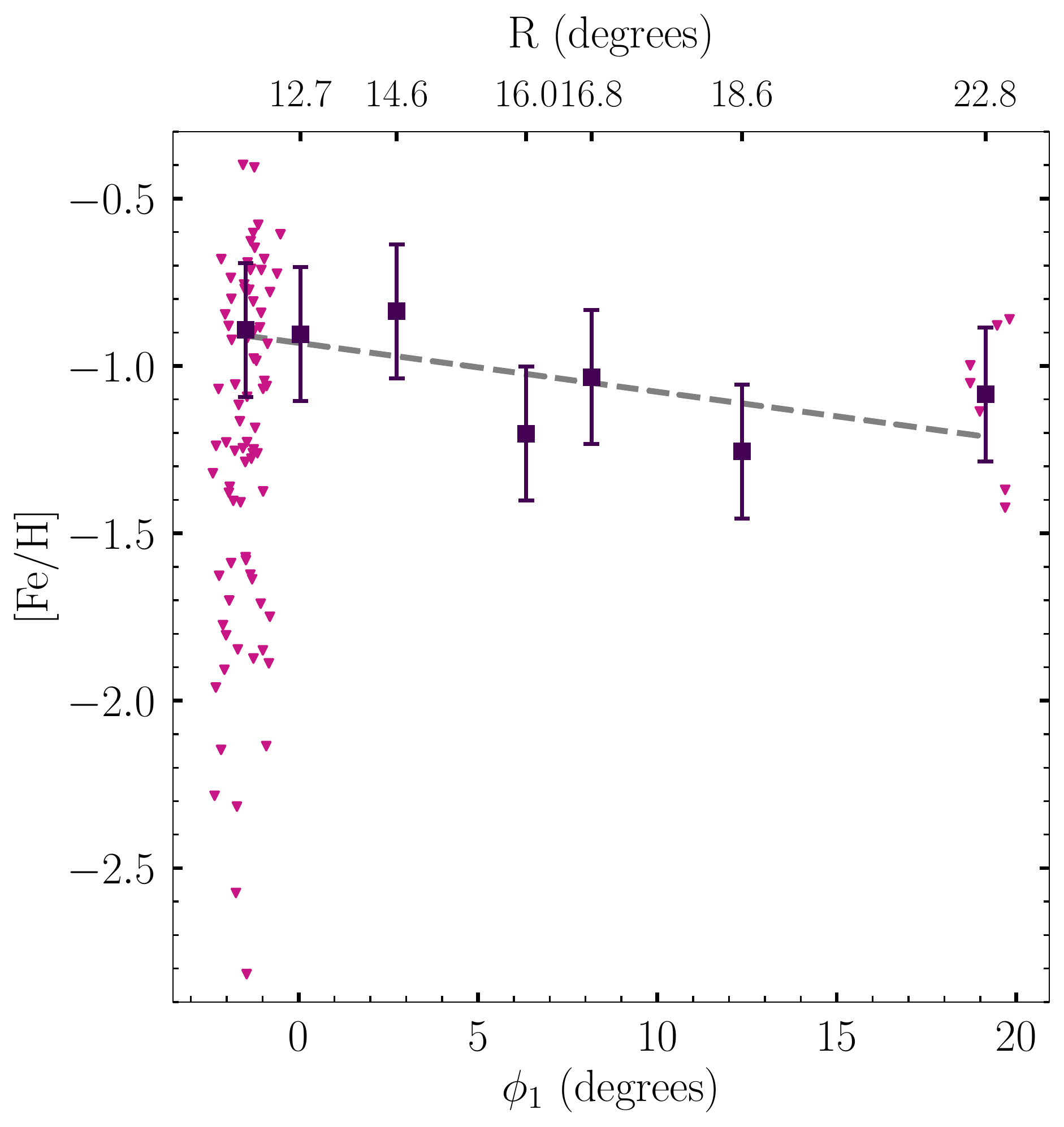}
	\caption{[Fe/H] measurements for stars in the northern arm and nearby outer LMC disk, as a function of LMC galactocentric radius ($R$: top) and $\phi_1$ distance along the feature (bottom). Red triangles indicate MagES measurements for individual stars, while squares indicate metallicities derived from stacked spectra, which tend to the mean metallicity of the field. The dashed grey line shows the best-fitting metallicity gradient along the feature.}
	\label{fig:met}
\end{figure}

[Fe/H] measurements for MagES fields along the arm, as a function of both LMC galactocentric radius ($R$) and $\phi_1$ distance along the feature, are presented in Fig.~\ref{fig:met}. We find very weak (<2$\sigma$) evidence for a negative metallicity gradient along the feature when performing a least-squares fit to the stacked field measurements, which are expected to trend to the field mean. The gradients we derive ($-0.015\pm0.007$ dex per degree in $R$, and $-0.025\pm0.014$ dex per degree in $\phi_1$) both imply a drop from [Fe/H]\textasciitilde$-0.9$ at the base of the feature, to [Fe/H]\textasciitilde$-1.2$ at the most distant measured point ($R$\textasciitilde$23^\circ$, $\phi_1$\textasciitilde20$^\circ$). Whilst only fields 22 and 12 have multiple metallicity measurements, we do note in these fields a relatively large scatter in [Fe/H], with metallicity measurements covering an \textasciitilde0.5~dex range even in the outermost field 22. We discuss the implications of the potential decrease in mean [Fe/H] along the feature on estimates of its structure using RC photometry in \S\ref{sec:phot}. 

Whilst literature [Fe/H] measurements in the outskirts of the LMC are sparse, they are generally consistent with our results. \citet{gradyMagellanicMayhemMetallicities2021} reports photometric metallicities along the feature utilising Gaia DR2 photometry of RC/RGB stars, finding similar [Fe/H] values along the feature to our spectroscopic measurements. Any potential gradient along the feature, however, is masked by a large dispersion (defined in that paper as the difference between the 10\textsuperscript{th} and 90\textsuperscript{th} percentile of the distribution) of up to \textasciitilde0.6~dex within each square degree pixel. Both \citet{majewskiDiscoveryExtendedHalolike2008a} and \citet{carreraMETALLICITIESAGEMETALLICITYRELATIONSHIPS2011} find a decrease in the mean metallicity of RGB stars beyond a LMC galactocentric radius of \textasciitilde$7^\circ$, with mean [Fe/H]\textasciitilde$-1$ at distances of $\geq$10$^\circ$, with a scatter of \textasciitilde1dex at these large distances. We also note that \citet{munozExploringHaloSubstructure2006} measure a mean [Fe/H]$=-0.67$ and a dispersion of $0.62$dex -- somewhat higher than our measurements -- for a group of 15 stars in the vicinity of the Carina dwarf (located near the armlike feature at $\phi_1$\textasciitilde$12.5^\circ$) with heliocentric velocities indicating a potential LMC association. However, cross-matching these stars against Gaia EDR3 returns at least three stars with proper motions strongly inconsistent with being associated with the LMC, suggesting their reported mean metallicity could be too high (assuming the non-Magellanic stars are metal-rich Galactic contaminants). Unfortunately, \citet{munozExploringHaloSubstructure2006} do not report individual [Fe/H] measurements so we cannot calculate a corrected value.

Our [Fe/H] measurements indicate the feature is likely comprised of disturbed LMC disk material. The median metallicity near the base of the feature is consistent with measurements in the nearby outer LMC disk fields \citepalias{C20}, and given the negative metallicity gradient at smaller LMC radii \citep{carreraMETALLICITIESAGEMETALLICITYRELATIONSHIPS2011,majewskiDiscoveryExtendedHalolike2008a}, a mild negative metallicity gradient could be expected under the assumption that the feature is an overdensity in the extreme LMC disk outskirts, such that stars currently located at large distances along the feature had their origin at larger radii than stars currently located at smaller galactocentric radii. We explore formation mechanisms for the feature using models to test this idea in \S\ref{sec:models}. 

We can, however, rule out the feature being the disrupted remains of an accreted dwarf satellite of the LMC as discussed in \citetalias{mackey10KpcStellar2016}. Considering the mass-metallicity relation for dwarf galaxies \citep[as presented in][]{hidalgoMassmetallicityRelationDwarf2017,kirbyUNIVERSALSTELLARMASSSTELLAR2013}, a stellar mass of $\geq$10$^{7.6}$M$_\odot$ is required for a mean [Fe/H]$\gtrsim$-1.2, corresponding to an integrated luminosity $M_V\lesssim-11.5$ \citep{mcconnachieOBSERVEDPROPERTIESDWARF2012}. In contrast, \citetalias{mackey10KpcStellar2016} find the integrated luminosity of the feature is only $M_V$\textasciitilde$-7.4$. Even accounting for the increased spatial extent of the feature traced using more recent data, and uncertainties in the mass-metallicity relation, this is still $\gtrsim$30 times fainter than the luminosity of the required satellite. 

\subsection{Structure}\label{sec:phot}
In order to place constraints on the geometry of the feature, we carefully analyse Gaia EDR3 photometry of stars along its length. Although Gaia parallaxes lack the precision to provide useful distances for the Clouds \citep{gaiacollaborationGaiaEarlyData2021a}, the apparent magnitude of the red clump can instead be used as a standardizable candle to provide information about the relative geometry of the feature. However, the apparent magnitude of the red clump is not purely distance dependent: population effects including the age and metallicity of clump stars affect their intrinsic luminosity  \citep[see][for a review]{girardiRedClumpStars2016}, and interstellar extinction along the line-of-sight also affects the measured clump magnitude. To determine the relative geometry of the feature therefore requires dereddened photometry, as well as assumptions about its constituent stellar populations.

Following the procedure described in \cite{gaiacollaborationGaiaEarlyData2021c}, we deredden our photometry utilising the \citet{schlegelMapsDustInfrared1998} dust maps, corrected as described in \citet{schlaflyMEASURINGREDDENINGSLOAN2011}, in conjunction with the mean extinction coefficients for the Gaia passbands described in \citet{casagrandeUseGaiaMagnitudes2018}. No correction is made for reddening internal to the Clouds as this is not expected to be significant in the low-density peripheral regions targeted by MagES \citep[c.f.][henceforth referred to as C18]{choiSMASHingLMCTidally2018}. We correct the Gaia G-band photometry for the 6-parameter solution as described in \cite{rielloGaiaEarlyData2021} prior to applying the dereddening procedure.


In order to effectively utilise the clump magnitude as a distance estimator along the feature, we initially assume that the stellar population comprising the clump does not vary, and is identical to that in the nearby LMC outer disk. This implies any magnitude differences observed along the feature are due entirely to distance effects. However, as discussed in \S\ref{sec:met}, there is weak evidence for a mild negative metallicity gradient along the feature. In the Gaia G passband (which substantially overlaps the optical V-band investigated in \citealt{girardiPopulationEffectsRed2001}), this is expected to result in an increase in clump luminosity along the feature, as well as a shift to bluer colours. We discuss the scale of this potential population effect and its implications on our results in detail below. 

\begin{figure}
	\includegraphics[width=0.455\columnwidth]{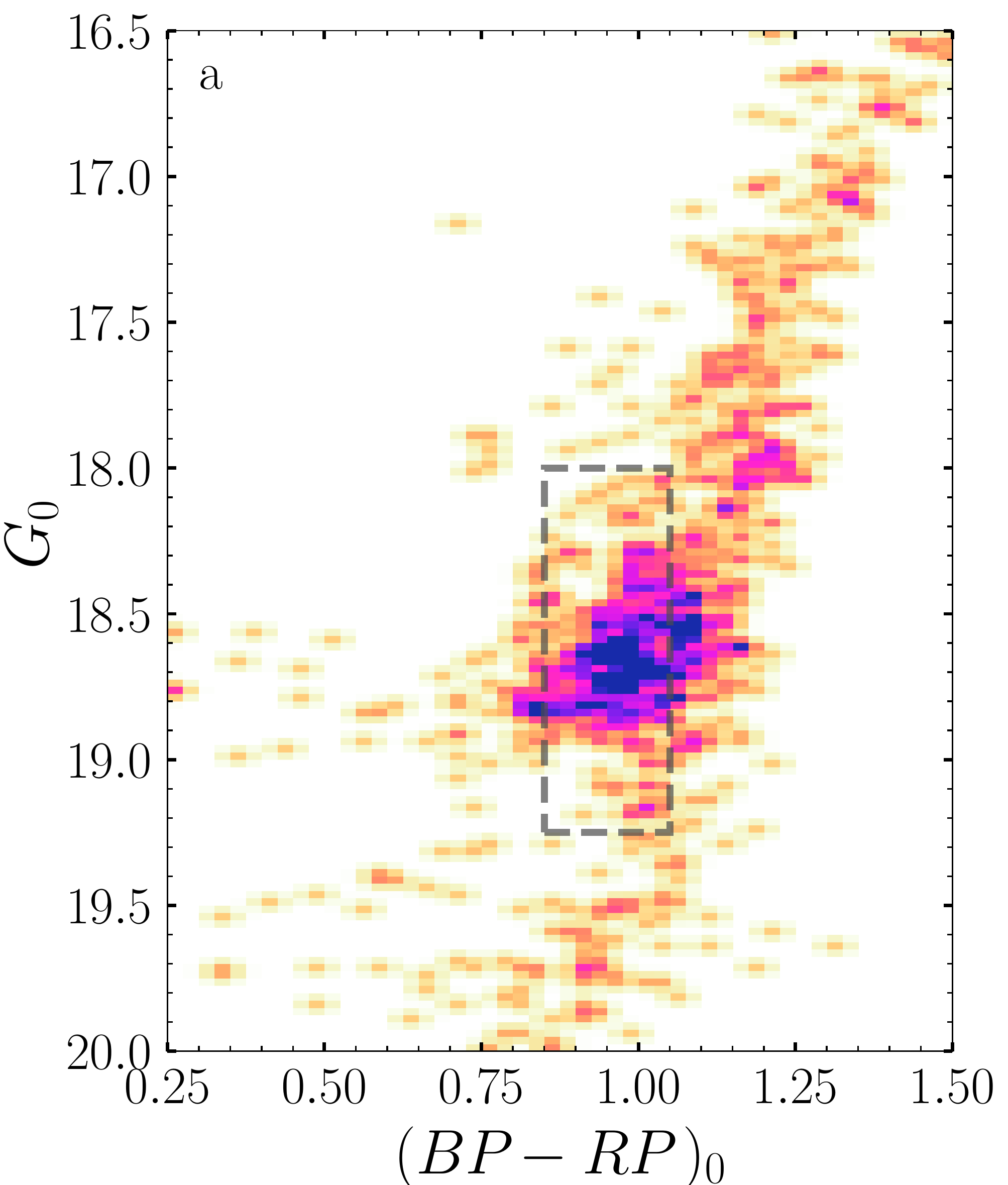}
	\includegraphics[width=0.499\columnwidth]{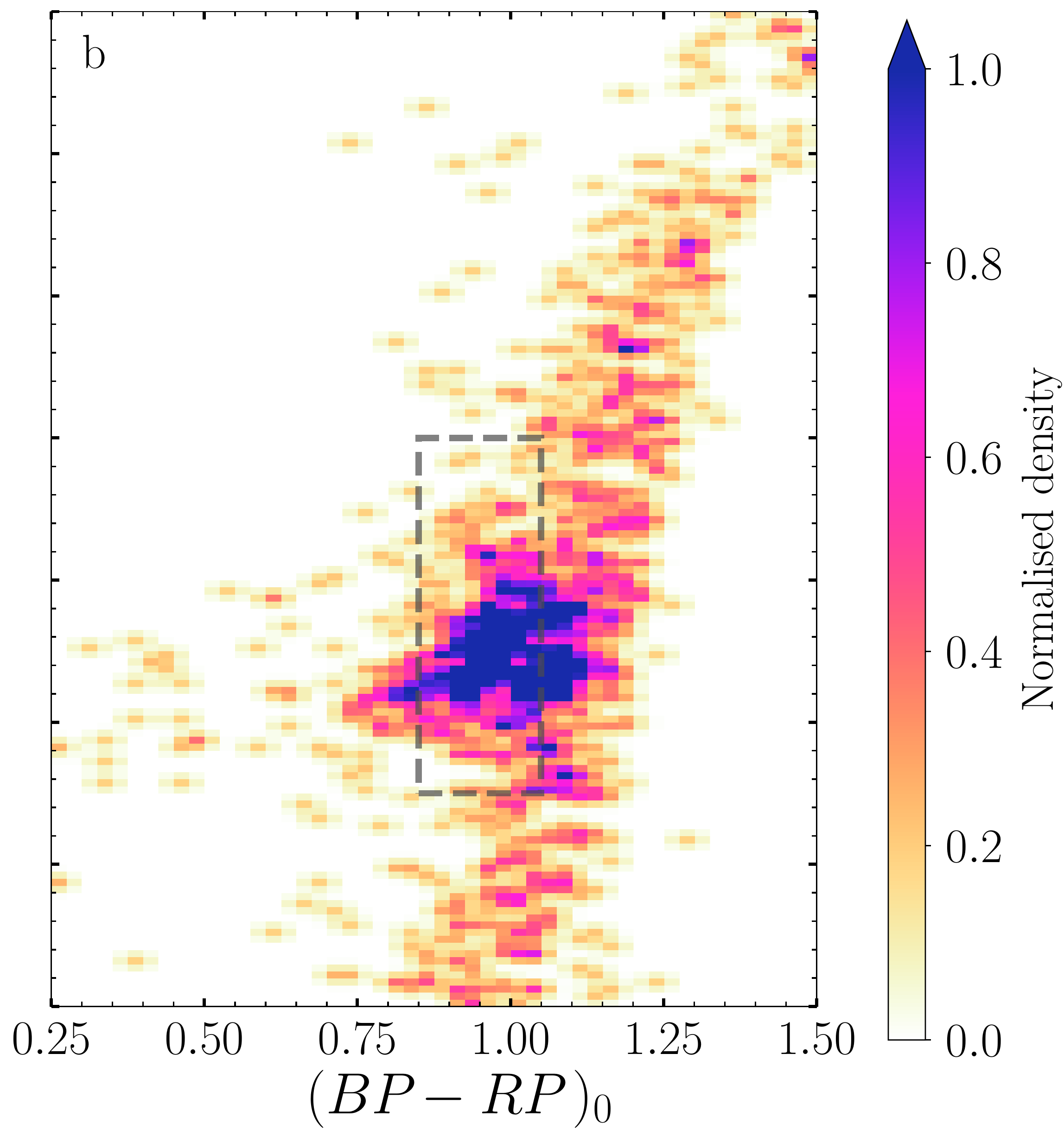}
	\caption{Colour-magnitude selection boxes used to isolate red clump stars, overlaid on Hess diagrams of stars within 2$^\circ$ diameter fields centred on MagES fields 18 (panel \textit{a}) and 12 (panel \textit{b}), located in the northern LMC disk. Only stars passing proper motion, parallax, and quality cuts as described in the text are included. The RC selection box (dashed grey line) is designed to minimise contamination from non-RC populations (including RGB, horizontal branch, and blue loop stars) whilst allowing for colour and magnitude shifts along the arm.}
	\label{fig:diskcmd}
\end{figure}

To determine an appropriate CMD selection box for RC stars, we utilise dereddened Gaia EDR3 photometry within the northern LMC disk, where the clump is well-populated. We select stars within a 1$^\circ$ radius of two MagES disk fields (fields 12 and 18: see Fig.~\ref{fig:map}), with parallax<0.2 and proper motions within a box of full width five times the dispersion of the field median motions reported in Table~\ref{tab:fieldbase} (i.e. $\pm$2.5$\sigma_{\alpha/\delta}$), and passing the quality cuts \textaltfont{ruwe}<1.4 and $C^*$<4$\sigma_{C^*}$. Fig.~\ref{fig:diskcmd} shows the resulting Hess diagrams for the two fields. We define a selection box of 0.85<$(G_{\text{BP}}-G_{\text{RP}})_0$<1.05, and 18.0<$G_0$<19.25, to select red clump stars. The selection is designed to minimise contamination from the RGB and potential RGB bump (which, at a similar magnitude to the RC, could bias estimation of the clump magnitude), whilst being sufficiently wide in magnitude range to accommodate potential distance variations along the northern arm. The resulting median clump magnitude and colour, and associated dispersion calculated as the standard deviation, are provided for the two fields in Table~\ref{tab:diskphot}. Note the observed \textasciitilde0.1~mag difference in median $G_0$ for these fields is expected due to the inclined disk geometry of the LMC. We test small (\textasciitilde0.25~mag) adjustments to the selection box (including stricter blue and bright magnitude cutoffs to minimise contamination from horizontal branch and blue loop stars respectively) and find these do not significantly affect our results. 

\begin{table}\label{tab:diskphot}
	\caption{Median RC magnitude $\left\langle G_0\right\rangle$ and colour $\left\langle(G_{\text{BP}}-G_{\text{RP}})_0\right\rangle$, and associated dispersion, for two MagES northern disk fields. Standard errors on both the median and dispersion are reported. }
	\begin{adjustbox}{max width=\columnwidth}
		\begin{tabular}{lllll}
			\hline
			Field & $\left\langle G_0\right\rangle$  & $\sigma_{G_0}$ &  $\left\langle(G_{\text{BP}}-G_{\text{RP}})_0\right\rangle$ & $\sigma_{(G_{\text{BP}}-G_{\text{RP}})_0}$ \\ \hline
			18   & $18.66\pm0.02$ & $0.24\pm0.01$ & $0.975\pm0.004$   & $0.052\pm0.002$   \\
			12   & $18.75\pm0.02$ & $0.23\pm0.01$ & $0.972\pm0.004$   & $0.052\pm0.002$\\  \hline
		\end{tabular}
	\end{adjustbox}
\end{table}

\begin{figure}
	\includegraphics[width=\columnwidth]{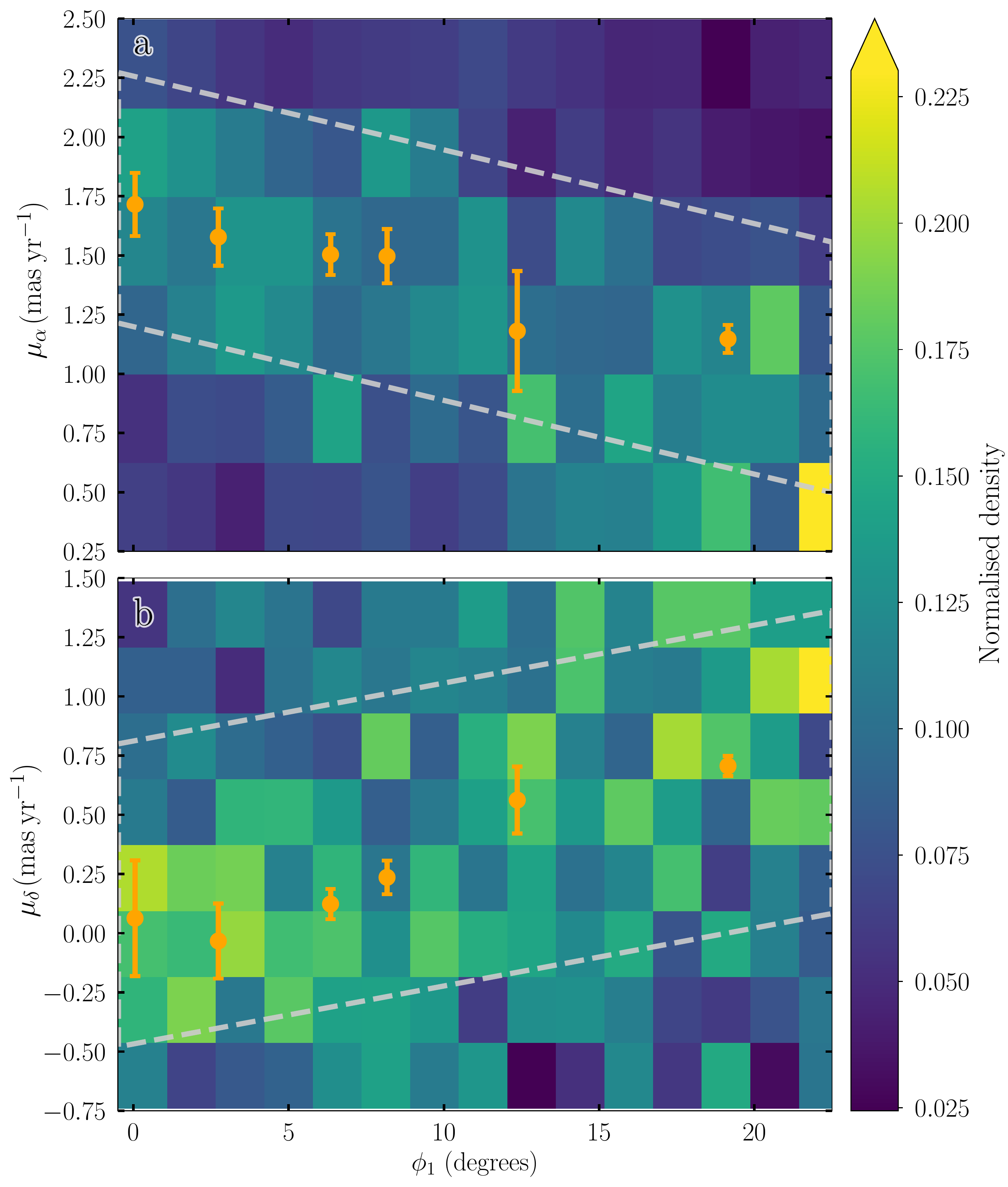}
	\caption{Proper motion selection boxes (dashed grey) used to isolate likely LMC stars along the northern arm. Orange points indicate MagES field aggregate motions, with error bars representing the field aggregate $1\sigma$ dispersion. These are overlaid on 2D histograms of $\mu_\alpha$ (panel \textit{a}) and $\mu_\delta$ (panel \textit{b}) as a function of $\phi_1$ location along the feature for RC stars in the vicinity of the northern arm. }
	\label{fig:photpmbox}
\end{figure}

We apply the same CMD, parallax, and quality cuts to stars within the feature selection box in $\phi_1/\phi_2$ coordinates presented in \S\ref{sec:armloc}. Unlike in an individual disk field, the mean proper motion varies along the length of the feature, and therefore a simple global proper motion cut is insufficient to minimise contamination. As such, we perform a least-squares fit to each of the two proper motion components measured for the MagES fields along the feature as a function of $\phi_1$, weighted by the proper motion uncertainty. We define each proper motion selection to be a box centred on the resulting fit, with a full width 5 times the mean proper motion dispersion of all MagES fields along the arm (i.e. $\pm$2.5$\left\langle\sigma_{\alpha/\delta}\right\rangle$). The resulting selections are presented in Fig.~\ref{fig:photpmbox}, overlaid on 2D histograms of the underlying proper motion distribution (limited to stars with proper motions 0<$\mu_\alpha$<3 and $-2$<$\mu_\delta$<2); the fitted relations follow the underlying overdensities in proper motion space associated with the feature. Our final selection includes only stars which pass the selection in both proper motion components. 

\begin{figure}
	\centering	\includegraphics[width=0.8\columnwidth]{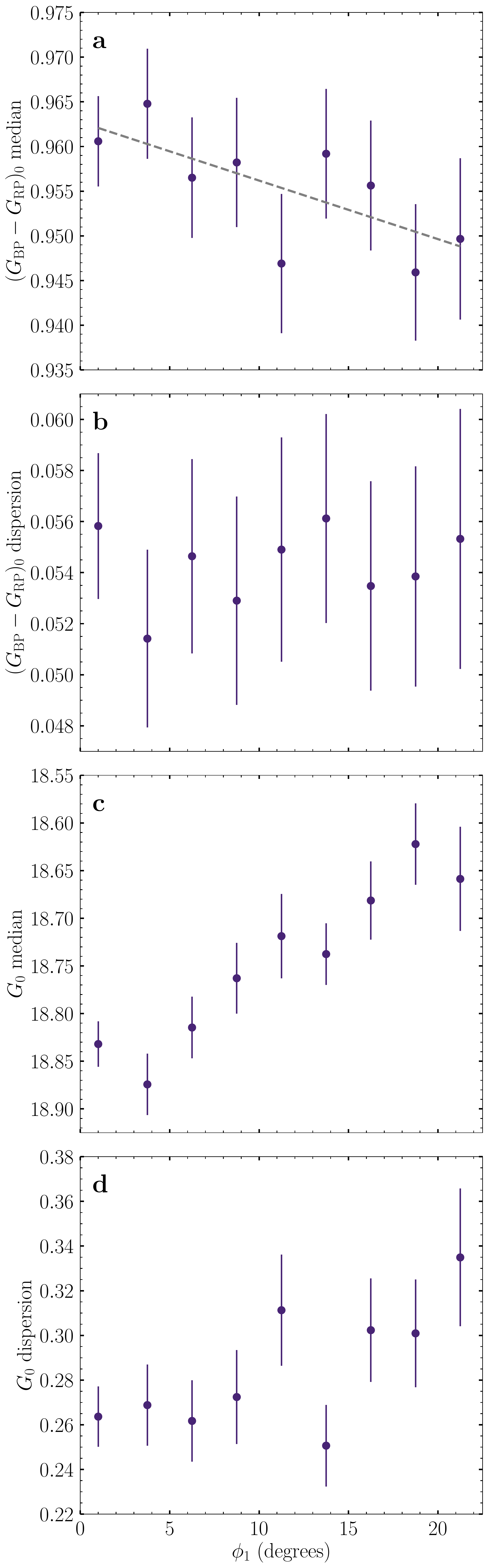}
	\caption{Photometric properties of red clump stars calculated within $2.5^\circ$ bins in $\phi_1$ distance along the feature. Panels show (\textit{a}) median $(G_{\text{BP}}-G_{\text{RP}})_0$ colour, (\textit{b}) standard deviation in the $(G_{\text{BP}}-G_{\text{RP}})_0$ distribution, (\textit{c}) median $G_0$ magnitude, and (\textit{d}) standard deviation in the $G_0$ distribution. The dashed grey line in panel \textit{a} shows a linear least-squares fit to the median colour as a function of $\phi_1$.}
	\label{fig:photprops}
\end{figure}

We bin our final selection into segments of $2.5^\circ$ in $\phi_1$, and determine the median $(G_{\text{BP}}-G_{\text{RP}})_0$ colour, $G_0$ magnitude, and associated dispersions (calculated as the standard deviation of the distribution) for each bin. Bins are chosen such that at least 60 stars are present in each bin, in order to provide robust estimates of the colour-magnitude distributions. Fig.~\ref{fig:photprops} shows the resulting photometric trends as a function of the $\phi_1$ distance along the feature; error bars represent the standard error on each parameter. The standard error on all quantities increases along the feature due to the decreasing density of Magellanic stars further from the LMC disk.

An underlying assumption of the following analysis of the structure’s photometric properties is that any underlying contaminant (i.e. non-Magellanic) population of stars within our final selection is uniformly distributed within the CMD selection box along the length of the feature. To test this, we utilise the Besan{\c c}on Model of the Galaxy \citep{robinSyntheticViewStructure2003}\footnote{Accessed as version 1603 through the web service\\ \url{https://model.obs-besancon.fr/}}. We generate an empirical representation of the observed Milky Way contaminant profile within the feature selection box by applying the same CMD, position, parallax, and proper motion selection cuts as our observed sample to the Model. We find the underlying MW population is distributed relatively uniformly within the CMD selection box, and remains so along the length of the feature, indicating this does not bias either the median RC colour or magnitude inferred from our final selection. The number of contaminant stars within the selection does increase along the length of the feature; we discuss this in further detail below. 

As seen in Fig.~\ref{fig:photprops}, there is a mild (\textasciitilde0.01~mag) trend to bluer colours as the $\phi_1$ distance along the feature increases, such that the mean colours at either end of the feature differ by approximately 1$\sigma$. A linear least-squares fit, weighted by the standard error on the median colour, has a slope of $(-6\pm2)\times10^{-4}$ mag/degree. Such a trend is qualitatively consistent with the mild trend to lower metallicities with increasing $\phi_1$ as observed in \S\ref{sec:met}: red clump stars become bluer at lower metallicities \citep{girardiRedClumpStars2016}. We test whether the magnitude of this colour shift is consistent with that expected from the metallicity gradient along the feature using PARSEC isochrones \citep{bressanParsecStellarTracks2012}\footnote{Accessed as version 3.4 of the web form \\ \url{http://stev.oapd.inaf.it/cmd}}, assuming the default parameters for IMF and mass loss, in Gaia EDR3 passbands. For isochrones of an 11~Gyr old population\footnote{the best-fitting isochrone for the outer LMC disk in \cite{mackeySubstructuresTidalDistortions2018}.}, at metallicities of [Fe/H]$=-0.9$ and [Fe/H]$=-1.2$ (the maximum inferred metallicity difference along the feature), we calculate the luminosity-weighted mean magnitude in the $G$, $G_{\text{BP}}$, and $G_{\text{RP}}$ passbands for core He-burning stars\footnote{with label `4' in the isochrone.} as in Eqs. 3 and 4 of \citet{girardiPopulationEffectsRed2001}. 

The calculated (reddening-free) $G_{\text{BP}}-G_{\text{RP}}$ colour difference between the two metallicities is \textasciitilde0.06~mag: significantly larger than the measured \textasciitilde0.01~mag difference in $(G_{\text{BP}}-G_{\text{RP}})_0$ along the feature. This is not unexpected: dispersion in the clump age and metallicity (\textasciitilde0.5~dex as measured in \S\ref{sec:met}), as well as photometric uncertainties, will act to `smear out' the clump colour and reduce the measured colour difference along the feature. We also note the dispersion in $(G_{\text{BP}}-G_{\text{RP}})_0$ remains constant along the length of the feature, implying the underlying scatter within the RC population remains relatively constant along the arm. 

Whilst the most straightforward interpretation of the shift to bluer colours along the arm is due to an underlying metallicity gradient, it is not the only possibility. Stellar age also affects the median RC colour, with young ($\lesssim$2~Gyr) RC stars significantly bluer than older RC stars \citep{girardiPopulationEffectsRed2001}. However, DECam CMDs in the vicinity of the feature reveal a lack of main sequence stars above an ancient \citep[\textasciitilde$10^{11}$~Gyr:][]{mackeySubstructuresTidalDistortions2018} turnoff. We can thus infer age is not the dominant driver of the shift in RC colour. In contrast, we cannot rule out the possibility of systematics in the reddening correction affecting the median $(G_{\text{BP}}-G_{\text{RP}})_0$ colour at the level of 0.01~mag, noting the mean $E(B-V)$ value along the feature remains relatively constant at \textasciitilde0.08 between 0$^\circ$<$\phi_1$<15$^\circ$, but increases to \textasciitilde0.25 at $\phi_1$\textasciitilde22.5$^\circ$. However, we do note the minimal change in $(G_{\text{BP}}-G_{\text{RP}})_0$ colour implies any systematic in the reddening correction must be small. 

We now consider the gradient in $G_0$ observed along the northern arm. The median magnitude increases from $G_0$\textasciitilde18.83 at the base of the feature to $G_0$\textasciitilde18.63 far from the LMC disk. To check the possible effect of changing metallicity on G-band magnitude, we utilise the same isochrones as described above to quantify the maximum potential difference in $G_0$ from metallicity, and find only a \textasciitilde0.06~mag increase in magnitude for more metal-poor populations. This is similar to the standard error on the $G_0$ magnitude per bin. Further, as in the case of the RC colour, the large metallicity scatter along the feature is expected to reduce the severity of the observed magnitude difference due to metallicity along the feature. We therefore conclude that the gradient in $G_0$ observed along the northern arm can be entirely attributed to a change in distance, with the structure becoming closer with increasing $\phi_1$: as would be expected given the inclination ($i$) and orientation ($\Omega$) of the LMC disk.

To investigate how closely the northern arm follows the plane of the LMC disk, we calculate the expected magnitude difference along the feature under the assumption of disk geometry. As multiple estimates of the LMC disk geometry exist, even considering only `old' stellar populations, we investigate two geometries which span the range of recent measurements reported in the literature: that from \citetalias{vandermarelThirdEpochMagellanicCloud2014} ($i=34.0^\circ\pm7^\circ$, $\Omega= 139.1^\circ\pm4.1^\circ$; similar to that reported by \citealt{gaiacollaborationGaiaEarlyData2021a}), and that from \citetalias{choiSMASHingLMCTidally2018} ($i=25.86^\circ\pm1.4^\circ$, $\Omega=149.23^\circ\pm8.35^\circ$). Accounting for changes in both the LMC galactocentric radius and position angle along the feature, the expected magnitude difference along the feature is $0.16\pm0.03$~mag under the \citetalias{choiSMASHingLMCTidally2018} geometry, and $0.21\pm0.08$~mag under the \citetalias{vandermarelThirdEpochMagellanicCloud2014} geometry: corresponding to end-to-end changes in distance of $3.4\pm0.6$~kpc and $4.1\pm1.4$~kpc respectively. Our measured difference of $0.21\pm0.05$~mag is, within uncertainty, consistent with both of these estimates, if somewhat closer that of \citetalias{vandermarelThirdEpochMagellanicCloud2014}. We can therefore infer the feature does, to first order, follow the plane of the LMC disk, though the precision of our measurements limits our ability to isolate a preferred disk geometry at these large radii. 

We also investigate the thickness of the feature using the $G_0$ dispersion of the RC. Within a given bin and passband, the measured dispersion $\sigma_{G_0}$ can be parameterised by Eq.~\ref{eq:disp}, where $\sigma_{\text{geo}}$ is the apparent dispersion due to global distance differences along the length of the feature, $\sigma_{\text{int}}$ is the intrinsic dispersion of the clump due to population effects, $\sigma_{\text{err}}$ is dispersion introduced through photometric uncertainties, $\sigma_{\text{depth}}$ is due to the intrinsic thickness of the feature, and $\sigma_{\text{cont}}$ is the apparent broadening of the RC due to the presence of an underlying uniformly-distributed model contaminant population within the selection box. Note that under this parameterisation, $\sigma_{\text{cont}}$ only accounts for Milky Way contamination, and not contamination from non-RC Magellanic populations, such as RGB stars, discussed further below. Of interest is whether $\sigma_{\text{depth}}$ within the feature is comparable to that in the outer LMC disk. 

\begin{equation}\label{eq:disp}
	\sigma_G^2= \sigma_{\text{geo}}^2 +\sigma_{\text{int}}^2+\sigma_{\text{err}}^2+\sigma_{\text{depth}}^2+\sigma_{\text{cont}}^2
\end{equation}

As we bin the data into 2.5$^\circ$ lengths along the feature, within each bin we expect the global distance gradient $\sigma_{\text{geo}}$ to be small: assuming either \citetalias{choiSMASHingLMCTidally2018} or \citetalias{vandermarelThirdEpochMagellanicCloud2014} geometries results in a maximum magnitude difference of \textasciitilde0.05~mag across each bin due to a global distance gradient. We expect $\sigma_{\text{geo}}$ to be similar, if slightly smaller, within our two LMC disk reference fields (MagES fields 12 and 18) as these fields also have a diameter of \textasciitilde$2^\circ$. We subtract the predicted $\sigma_{\text{geo}}$ effect from the measured $G_0$ dispersion both along the feature and within the disk fields prior to comparison. 

As discussed above, the dispersion in RC colour remains constant along the feature, implying similar population effects along its length, and in \S\ref{sec:met} a metallicity dispersion of \textasciitilde0.5~dex is measured along the feature: consistent with the dispersions measured for the two disk fields in \citetalias{C20}. As such, it is not unreasonable to assume the stellar populations within the feature are similar to those in the outer LMC disk, and we can infer that $\sigma_{\text{int}}$ is constant both along the feature, and within the two reference disk fields. Similarly, as we utilise the same photometric dataset and implement the same quality cuts throughout our analysis, we expect $\sigma_{\text{err}}$ to be approximately constant both along the feature, and within the disk. 

Under these assumptions, any difference in $G_0$ dispersion between the feature and the disk fields is due entirely to a difference in feature thickness, or the effects of contamination. We expect $\sigma_{\text{cont}}$ to be effectively zero within the disk fields, due to the very high density of Magellanic stars compared to the expected MW contamination within the selection (a factor of $\gtrsim$100). In contrast, the level of predicted MW contamination within the selection increases along the length of the feature, increasing by a factor of \textasciitilde3 from the disk fields to the outermost bins along the feature. We hypothesise that this increase in contamination, and associated increase in $\sigma_{\text{cont}}$, is the dominant driver of the increased $G_0$ dispersion measured beyond 10$^\circ$ along the feature. In contrast, for bins within the first 10$^\circ$ of the feature, the predicted number of contaminant stars is significantly less than the total number of observed stars per bin. This implies $\sigma_{\text{cont}}$ is much smaller, if not zero, within these bins. When we compare the measured $G_0$ dispersion within these bins to that for the two disk fields, we find these are equal within uncertainty: implying the thickness of the feature is approximately the same as the thickness of the LMC disk. This is further evidence the feature is made from perturbed LMC disk material. 

To test the hypothesis that contamination is responsible for the observed increase in $G_0$ dispersion along the arm, and to check that contamination is not adversely affecting any of the other measured parameters, we fitted a mixture model that explicitly tries to account for non-RC populations within each bin. The model assumes the density of red clump stars takes the form of a two-dimensional Gaussian on the CMD, while the background density is described by linearly varying terms in both colour and magnitude. The relative fraction of contaminants and members in a given bin is left as a free parameter. We sample the posterior probability distributions for the model parameters using the Markov Chain Monte Carlo ensemble sampler {\sc emcee}. 

Whilst this approach is more comprehensive in modelling the stellar populations within the selection box than our original method, a disadvantage is that it requires a substantial number of stars per bin to robustly converge. As a result, we can only reliably perform the fit within four bins (of length \textasciitilde5$^\circ$, compared to 2.5$^\circ$ in our original method) along the arm. Nonetheless, our results agree closely with the RC $(G_{\text{BP}}-G_{\text{RP}})_0$ colour and $G_0$ magnitude trends determined from the simple medians, as well as the $(G_{\text{BP}}-G_{\text{RP}})_0$ colour dispersion. The fitted RC dispersion in $G_0$ also remains constant along the full length of the arm, with the non-member fraction increasing by a factor of \textasciitilde2 from $\phi_1$=0 to $\phi_1$=20 degrees. This supports our conclusion that it is the contaminating populations that drive the increasing $G_0$ dispersion along the arm in our simple measurements, while the intrinsic thickness does not substantially change.

\subsection{Kinematics}\label{sec:kinematics}
Whilst Table~\ref{tab:fieldloc} reports kinematic information in observable units, our finding that the northern arm sits close to the expected plane of the LMC disk and is likely comprised of disk material means it is more informative to consider its kinematics in the LMC disk frame. As such, the framework presented in \citet{vandermarelMagellanicCloudStructure2001} and \citet{vandermarelNewUnderstandingLarge2002} is used to transform the observed components into velocities in a cylindrical coordinate system. This coordinate system is aligned with the LMC disk, and has its origin at the LMC centre of mass (COM). As in \citetalias{C20}, we choose the COM to be ($\alpha_0=79.88^\circ$, $\delta_0=-69.59^\circ$) as reported by \citetalias{vandermarelThirdEpochMagellanicCloud2014} for their ‘PMs+Old $V_{\text{LOS}}$ Sample’, and the associated systemic motions applicable for this choice of centre: i.e. $\mu_{\delta,0}=0.287\pm0.054$ mas~yr$^{-1}$, $\mu_{\alpha,0}=1.895\pm0.024$ mas~yr$^{-1}$, and $V_{\text{LOS},0} = 261.1\pm2.2$km~s$^{-1}$.  

\begin{figure*}
	\includegraphics[width=\textwidth]{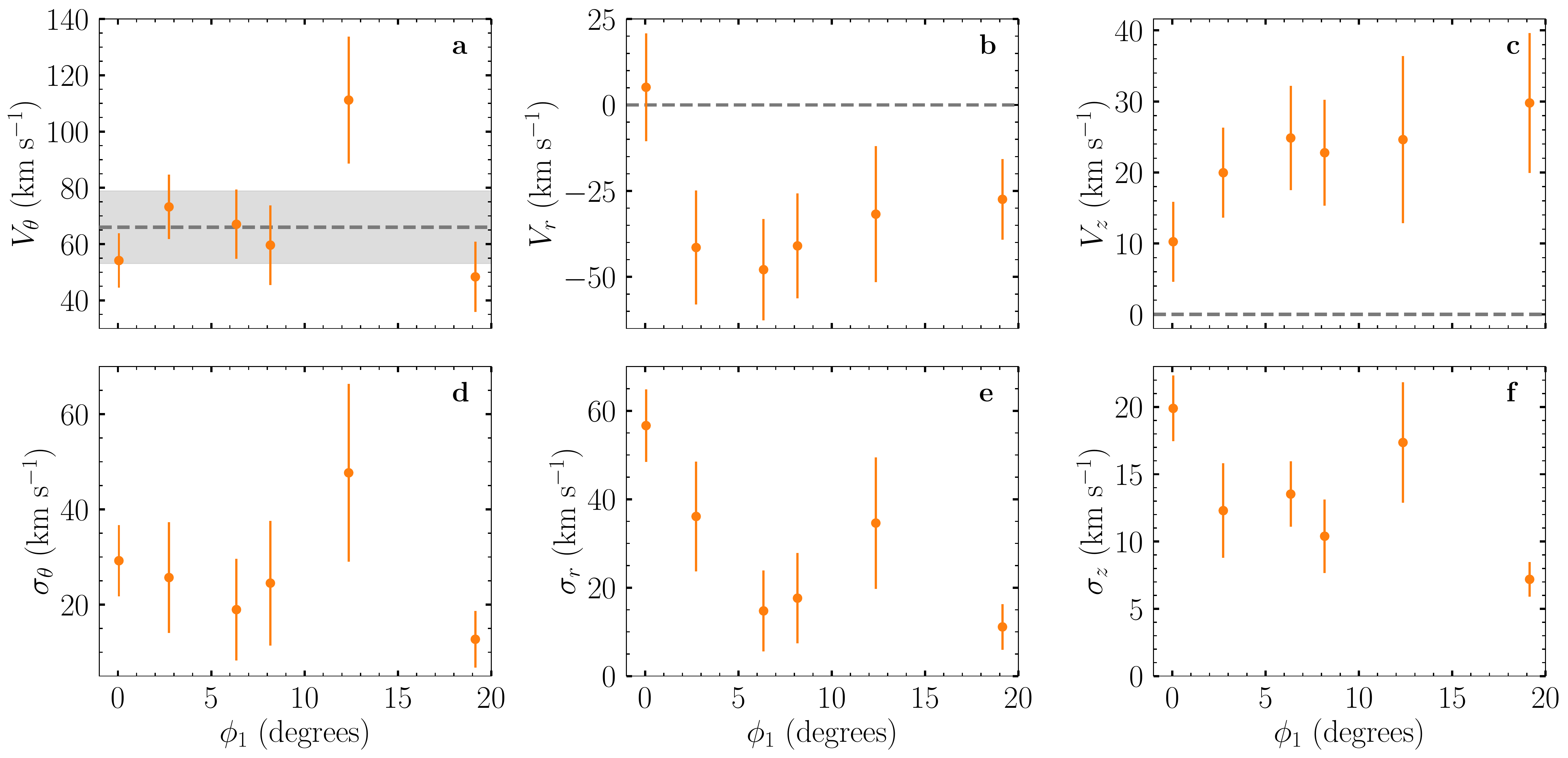}
	\caption{Observed velocities and dispersions for MagES fields along the northern arm as a function of $\phi_1$, calculated assuming \protect\citetalias{choiSMASHingLMCTidally2018} disk geometry. Top panels show, in order, the azimuthal, in-plane radial, and out-of-plane vertical velocities; bottom panels show the velocity dispersions in each component. Positive azimuthal velocities indicate clockwise rotation (i.e. in a direction from North towards West), positive radial velocities indicate movement outward from the LMC COM in the LMC disk plane, and positive vertical velocities indicate movement perpendicular to the disk plane, in a direction primarily towards the observer. Grey dashed lines indicate the expected kinematics for an equilibrium disk, with the rotational velocity (and uncertainties) taken from \protect\citetalias{C20}, recalculated assuming the \protect\citetalias{choiSMASHingLMCTidally2018} disk geometry for consistency.}
	\label{fig:armkin}
\end{figure*}

The orientation of the LMC disk relative to the line-of-sight must also be assumed during this coordinate transform. From \S\ref{sec:phot}, the feature remains roughly within the plane of the LMC disk, though the moderate uncertainties in our measurement preclude distinguishing between varying literature measurements of the disk geometry. As such, for this paper we choose to utilise the \citetalias{choiSMASHingLMCTidally2018} geometry when calculating kinematics in the plane of the LMC disk. This is motivated by preliminary results from Mackey et al (in prep), which indicate the inclination of the LMC decreases at large radii. Using this assumed geometry, we transform the observed kinematic parameters for the feature fields into physical velocities and dispersion in the LMC disk frame. We calculate $V_\theta$, the azimuthal streaming or rotation velocity; $V_r$, the radial velocity in the disk plane; and $V_z$, the vertical velocity perpendicular to the disk plane, as well as dispersions ($\sigma_\theta$, $\sigma_r$, $\sigma_z$) in each of these components. These disk measurements are reported in Table~\ref{tab:armkin}, and Fig.~\ref{fig:armkin} plots each component as a function of $\phi_1$ position along the feature.

\begin{table*}
	\caption{Disk velocities for northern arm feature fields, calculated assuming \citetalias{choiSMASHingLMCTidally2018} disk geometry. In $V_\theta$, positive values indicate clockwise rotation. In $V_r$, positive values indicate movement outward from the LMC COM in the LMC disk plane. In $V_z$, positive values indicate movement perpendicular to the disk plane, in a direction primarily towards the observer: `in front' of the LMC disk. We also give the number of likely Magellanic stars per field, repeated from Table~\ref{tab:fieldloc}.}
	\label{tab:armkin}
	\begin{tabular}{clllllll}
		\hline
		Field & $N_{\text{Magellanic}}$ ($P_i\geq50\%$) &\multicolumn{1}{c}{$V_\theta$ (km~s$^{-1}$)}& \multicolumn{1}{c}{$\sigma_\theta$ (km~s$^{-1}$)} & \multicolumn{1}{c}{$V_r$ (km~s$^{-1}$)} & \multicolumn{1}{c}{$\sigma_r$ (km~s$^{-1}$)}  & \multicolumn{1}{c}{$V_z$ (km~s$^{-1}$)} & \multicolumn{1}{c}{$\sigma_z$ (km~s$^{-1}$)}\\ \hline
		11 & 75 &$54.2 \pm   9.8$  & $29.2\pm7.5$  & $5.7\pm16.7$    & $56.7\pm8.2$  & $9.4\pm6.2$      & $19.9   \pm2.4$ \\ 
		13 & 38 &$73.4 \pm   11.7$ & $25.7\pm   11.7$ & $-39.3   \pm 16.3$ & $36.1   \pm12.4$ & $19.5 \pm   6.0$ & $12.3   \pm3.5$ \\
		15 & 32 &$67.3 \pm   12.9$ & $19.0   \pm10.7$ & $-47.1   \pm15.0$  & $14.7   \pm9.1$  & $24.2\pm6.2$     & $13.5   \pm2.4$ \\
		16 & 25 &$59.2 \pm   14.2$ & $24.5\pm   13.1$ & $-40.4   \pm 15.8$ & $17.6   \pm10.2$ & $21.6 \pm   6.6$ & $10.4   \pm2.7$ \\
		19 & 13 &$108.8   \pm22.9$ & $47.7\pm   18.6$ & $-30.2   \pm20.0$  & $34.6\pm   14.9$ & $22.4   \pm9.9$  & $17.4   \pm4.5$ \\
		22 & 27 &$47.6   \pm14.6$  & $12.7\pm   5.9$  & $-26.3   \pm 9.8$  & $11.1\pm   5.2$  & $28.3 \pm   6.1$ & $7.2   \pm1.3$ \\ \hline
	\end{tabular}
\end{table*}

With the exception of field 19 (discussed below), clear trends are observed in each of the disk velocity components and their dispersions. Within uncertainty, the azimuthal rotation velocity in each field is consistent with that derived from the two MagES disk fields 12 and 18 in \citetalias{C20}, recalculated using Gaia EDR3 astrometry and the assumption of a \citetalias{choiSMASHingLMCTidally2018} disk geometry to maintain consistency with fields along the northern arm. The dominant source of uncertainty in estimates of the disk kinematics -- both in the \citetalias{C20} values, and those measured here -- are uncertainties in the assumed disk geometry. The measured azimuthl velocity is also within the uncertainty of that derived for RC sources in \citet{gaiacollaborationGaiaEarlyData2021a}, which is approximately flat at \textasciitilde70~km~s$^{-1}$ (with perhaps a very mild \textasciitilde5~km~s$^{-1}$ downturn at their outermost radii of \textasciitilde8~kpc). 

In contrast to the relatively ordered and disk-like kinematics observed in the azimuthal velocity, the in-plane radial velocity ($V_r$) and out-of-plane vertical velocity ($V_z$) along the feature are strongly out of equilibrium. Both these values are expected to be near zero in an equilibrium disk, and measurements of these in MagES northern disk fields reported in \citetalias{C20} are consistent with zero within uncertainty. However, the in-plane radial velocity drops sharply from approximately zero (in field 11) to $-40$~km~s$^{-1}$ by the next field \textasciitilde3$^\circ$ away along the feature. This strongly-inward velocity remains roughly constant along the feature before a slight decrease in magnitude to approximately $-30$~km~s$^{-1}$ in the outermost field 22. In addition, the vertical velocity gradually increases along the feature from near zero to a maximum of nearly 30~km~s$^{-1}$ in field 22: a significant out of plane motion. Such dramatic kinematic signatures strongly suggest perturbation by an external gravitational potential: we investigate this possibility in greater detail in \S\ref{sec:models}. 

Since the geometry of the outer LMC disk is uncertain, it is possible the observed kinematic perturbations are simply reflections of incorrectly assuming \citetalias{choiSMASHingLMCTidally2018} parameters for the disk geometry. To test this, we solve for the disk orientation required for the feature kinematics to simply be a projection of purely rotational motion within the LMC disk plane by simultaneously minimizing the sum of squares of $V_r$ and $V_z$. However, we find the derived disk orientations (typically with inclinations approximately $-40^\circ$) are strongly inconsistent with the constraints derived in \S\ref{sec:phot}, indicating genuinely perturbed kinematics. 

In general, the velocity dispersion in each component decreases along the length of the feature. The azimuthal ($\sigma_\theta$) and vertical ($\sigma_z$) velocity dispersions in field 11 nearest the outer LMC disk are similar to those observed in the nearby outer disk in \citetalias{C20}; the dispersions gradually decrease to approximately half their disk values by the outermost field along the structure. As moving along the feature also increases the galactocentric radius of the fields, this is not surprising: the velocity dispersion is expected to decrease with radius for material within an axisymmetric disk potential \citep{gaiacollaborationGaiaEarlyData2021a,vasilievInternalDynamicsLarge2018b,binneyGalacticDynamics2008b}. More interesting is the dispersion of the in-plane radial velocity. The dispersion in the two innermost feature fields (>40~km~s$^{-1}$) are similar to that in the nearby outer LMC disk \citepalias[\textasciitilde45~km~s$^{-1}$:][]{C20}; the innermost feature field is even larger than this value. This is in stark contrast to the canonical disk value of \textasciitilde25~km~s$^{-1}$ measured at large ($\geq8^\circ$) radii in undisturbed fields \citep{C20,gaiacollaborationGaiaEarlyData2021a,vasilievInternalDynamicsLarge2018b}. As discussed in \citet{wanSkyMapperViewLarge2020}, a large in-plane radial velocity dispersion can be indicative of perturbation by an external gravitational force: in their model, an interaction with the SMC can elevate the radial velocity dispersion in the outskirts of the LMC. In fields further along the feature, the radial velocity dispersion drops to values more consistent with the outer disk, continuing to drop along the length of the feature to \textasciitilde10~km~s$^{-1}$ in the outermost feature field. 

In Fig.~\ref{fig:armkin}, field 19 is a notable outlier when compared with the kinematic trends observed in the other fields. This is especially true of its azimuthal velocity (\textasciitilde1.8$\sigma$ higher than the other fields) and three dispersion components (\textasciitilde1$\sigma$ higher than the other fields). In addition, field 19 has generally larger uncertainties for all parameters. Investigation revealed that these characteristics can be traced to a combination of two issues. Firstly, field 19 has substantially fewer Magellanic members than any of the other fields -- a factor of two smaller than adjacent fields 16 and 22. Secondly, it transpires that the peak of one of the contamination populations we model during our membership analysis (see \citetalias{C20}) sits very close (\textasciitilde0.4~mas~yr$^{-1}$) to the kinematic peak for field 19 in the proper motion plane. As a consequence, our algorithm finds it difficult to robustly distinguish between Magellanic members and non-members. 

In fields where a large number of Magellanic stars are present, this is not an issue: the contamination populations are generally very broad compared to the narrow Magellanic kinematic peaks, allowing reliable association of stars with the appropriate population. However, the low number of genuinely Magellanic stars in field 19  broadens the observed Magellanic peak in proper motion space, resulting in misclassification of some genuinely Magellanic stars as belonging to the contaminant population (and potentially vice-versa, though the large difference in LOS velocity between Magellanic and non-Magellanic stars typically mitigates misclassification in this direction). This biases the derived Magellanic $\mu_\alpha$; indeed, most of the deviance observed in $V_\theta$ can be directly mapped to the $\mu_\alpha$ component of proper motion. Because of these issues, we do not attribute any physical significance to the fact that field 19 appears to be a kinematic outlier, and downweight its importance when comparing our measurements with numerical models in the next section.

\section{Modelling and Analysis}\label{sec:models}
In order to interpret the observed properties of the northern arm, we have created a suite of dynamical models of the LMC+SMC+Milky Way system with which to compare our observations. These comprise an existing $N$-body model of the MW and LMC only, presented in \citetalias{mackey10KpcStellar2016}, and five new model ensembles with varying LMC, SMC, and MW masses. Within each of the five new model ensembles, we sample from literature uncertainties in the 6D phase space properties of the LMC and SMC centres in order to investigate the allowed distribution of orbits -- and hence past interactions -- of the Clouds. We utilise these models to test the relative importance of tidal forces from the MW and SMC in generating structures akin to the northern arm. 

\subsection{General methodology} 
While we analyse several different models, calculated using two distinct numerical methods, we utilise a common procedure for making mock observations of these models. The simulations are evolved in Cartesian coordinates which are centered on the present-day location of the Milky Way. Mock observations are made on the final snapshot from the location of the Sun, which is assumed to be at a distance of 8.178~kpc \citep{gravitycollaborationGeometricDistanceMeasurement2019} from the Galactic center and moving with a velocity of $(11.1,242.5,7.3)$ km~s$^{-1}$ (motivated by the results of \citealt {schonrichLocalKinematicsLocal2010} and \citealt{bovyMILKYWAYCIRCULARVELOCITY2012}). These mock observations are made for the same observables as the real data, i.e. $\alpha$, $\delta$, $D$, $\mu_\alpha$, $\mu_\delta$, $V_{\text{LOS}}$. We subsequently convert these observables into the same (X,Y) coordinate system as the observed data using Eq.~\ref{eq:xy}. Note that in this transformation, we set $\alpha_0$, $\delta_0$ to be the defined LMC centre for each individual model, rather than the observed LMC centre, as the defined centre by design varies between model iterations.

To determine the model kinematics within each field for comparison with our observations, we select all particles within a one-degree radius of the central (X,Y) coordinates of each field reported in Table~\ref{tab:fieldloc} -- the same size as a MagES field observed with 2dF. We calculate the resulting median and dispersion of each kinematic component ($V_{\text{LOS}}$, $\mu_\alpha$, and $\mu_\delta$), which are suitable for direct comparison with the equivalent MagES observations. We further convert the model kinematics for each field into the reference frame of the assumed LMC disk plane using the same process as for the observed data, described in \S\ref{sec:kinematics}, to facilitate comparison with the equivalent observations. However, we make one key change to this process, as unlike the observed stars, the true distance to each model particle is known. We therefore utilise the true particle distances to calculate the out-of-plane distance ($z$) relative to the assumed \citetalias{choiSMASHingLMCTidally2018} LMC disk plane for each particle, rather than making the assumption that all particles are in the LMC disk plane ($z=0$) as required for the observations. We use the calculated out-of-plane distances to assess the accuracy of this earlier assumption (see below). 

We note there are two possible approaches for comparing model fields to MagES fields. The first, which we have adopted, is to select model locations at the same (X,Y) coordinates as each MagES field. However, these positions are not always precisely co-located with any northern overdensity that may appear in a given model. An alternative is to fit a unique feature track following any northern overdensity for each model realisation, using the same method as described in \S\ref{sec:armloc}, and compare model fields centred at the same $\phi_1/\phi_2$ coordinates as each MagES field as reported in Table~\ref{tab:fieldloc}. While this ensures model fields are co-located with any northern overdensity generated in the models, differences in the shape of the feature between model iterations can result in model fields located at significantly different LMC galactocentric radii and position angles.

We adopt the first approach described above as this is equivalent to selecting particles at the same projected LMC galactocentric radius and position angle as the MagES fields, ensuring particles feel comparable gravitational forces from the LMC+SMC+MW as the observed stars. In comparison, under the second approach outlined above, the different galactocentric radii of the model fields means the gravitational forces felt by particles at each field location can differ, potentially significantly, between model iterations. The derived kinematics are thus not strictly comparable, even between individual model realisations. Nonetheless, we have tested the second approach by fitting a feature track to each model realisation, selecting all particles within a one-degree radius of the $\phi_1/\phi_2$ coordinates of each MagES field, and calculating the resulting field kinematics. Comparison of the two approaches reveals the choice of field location does not significantly affect the derived model kinematics, nor the resulting conclusions regarding the origin of the feature. 

\subsection{$N$-body model}\label{sec:nbody}
We first compare our data to an existing $N$-body simulation of an LMC flyby of the MW presented in \citetalias{mackey10KpcStellar2016}. The LMC is modelled as a two-component galaxy (stellar disk and NFW halo), with a total mass of $1.4\times10^{11}$M$_\odot$ and stellar disk mass of $4\times10^{9}$M$_\odot$. The disk and halo are comprised of $10^6$ particles each, with a softening length of 75 and 500~pc respectively. The disk has a scale radius of 1.5~kpc, and a scale height of 0.3~kpc; the total LMC mass within 8.7~kpc is $1.8\times10^{10}$M$_\odot$ and has a circular velocity of \textasciitilde90~km~s$^{-1}$, consistent with \citetalias{vandermarelThirdEpochMagellanicCloud2014}. The Milky Way is modelled as a three-component system with a bulge, disc, and dark matter halo as described in \cite{gomezItMovesDangers2015}. The model was integrated for 2~Gyr, with initial positions and velocities of the Milky Way and LMC chosen using backward integration from the current position as in \cite{gomezItMovesDangers2015}, and initial LMC disk orientation chosen to match that reported in \citetalias{vandermarelThirdEpochMagellanicCloud2014}. The resulting present-day LMC position and systemic velocities were within $2\sigma$ of the Galactocentric Cartesian values reported in \cite{kallivayalilThirdEpochMagellanicCloud2013}.

Whilst the $N$-body model was run prior to the availability of more recent (i.e. post-\textit{Gaia}) structural and kinematic measurements in the outer LMC disk, we still perform a comparison to the $N$-body model as it surpasses our newer models in several aspects. In particular, it captures the self-gravity of the LMC disk and the deformation of the LMC dark matter halo during infall to the MW potential, both potentially significant in forming the northern arm, and follows the evolution of the LMC for twice as long as the newer models, allowing for a better understanding of the arm's formation timescale. 

For the present analysis, we have shifted the final LMC position and systemic velocities to new coordinates ($\alpha_0=80.86^\circ$, $\delta_0=-69.89^\circ$, $D_0=49.74$~kpc, $V_{\text{LOS},0}=262.7$ km~s$^{-1}$, $\mu_{\alpha,0}=1.995$ mas~yr$^{-1}$, $\mu_{\delta,0}=0.265$ mas~yr$^{-1}$) in order to more closely match recent estimates of the LMC’s systemic properties, and facilitate comparison with our new model suites which include realisations with these same central properties. We stress that as the true endpoint values of the simulation are different to these shifted values, the orbital history of the LMC in this $N$-body model is slightly different to that of later models which have the shifted values as their true endpoints. Some small differences are therefore expected when comparing predictions from this $N$-body model to later model suites.

In order to verify the applicability of the $N$-body model, we briefly discuss the model kinematics for the two MagES fields located in the northern LMC disk discussed in \citetalias{C20}: any systematic differences between the observed and modelled kinematics in this comparison will likely also occur for fields along the arm. Table~\ref{tab:diskfield} presents the three velocity components in the plane of the LMC disk ($V_\theta$, $V_r$, and $V_z$), as well as dispersions in each of these components, for MagES fields 18 and 12 and the corresponding $N$-body model predictions. We note the observed values are slightly different to those presented in \citetalias{C20} as we re-calculate these using Gaia EDR3 astrometry and assumption of a \citetalias{choiSMASHingLMCTidally2018} disk geometry to maintain consistency with fields along the northern arm.

\begin{table*}
	\caption{Disk velocities for MagES fields 18 and 12, located in the northern LMC disk, derived assuming \citetalias{choiSMASHingLMCTidally2018} disk geometry. Measurements are presented for both observed data, and for the \citetalias{mackey10KpcStellar2016} $N$-body model. As model particles have precisely known positions and kinematics, no uncertainties are reported for the model fields. }
	\label{tab:diskfield}
	\begin{adjustbox}{max width=\textwidth}
		\begin{tabular}{cllllllllllll} 
			\hline
			Field &
			\multicolumn{2}{c}{$V_\theta$ (km~s$^{-1}$)} &
			\multicolumn{2}{c}{$\sigma_\theta$ (km~s$^{-1}$)} &
			\multicolumn{2}{c}{$V_r$ (km~s$^{-1}$)} &
			\multicolumn{2}{c}{$\sigma_r$ (km~s$^{-1}$)} &
			\multicolumn{2}{c}{$V_z$ (km~s$^{-1}$)} &
			\multicolumn{2}{c}{$\sigma_z$ (km~s$^{-1}$)} \\ \hline
			& Measured      & $N$-body & Measured    & $N$-body & Measured      & $N$-body & Measured    & $N$-body & Measured     & $N$-body & Measured    & $N$-body \\ \hline
			18 & $66.0 \pm 12.9$ & 106.8  & $25.6 \pm2.1$ & 9.0    & $19.7 \pm 8.3$  & 20.5   & $25.6 \pm2.0$ & 9.6    & $5.4 \pm 6.8$  & 3.0    & $20.8 \pm1.1$ & 6.1    \\
			12 & $42.2 \pm 7.3$  & 84.9   & $27.5 \pm2.5$ & 6.1    & $25.3 \pm 14.4$ & $-1.8$   & $43.2 \pm3.8$ & 23.3   & $-3.7 \pm 5.4$ & 1.7    & $25.4 \pm1.3$ & 6.4    \\ \hline
		\end{tabular}
	\end{adjustbox}
\end{table*} 

From Table~\ref{tab:diskfield}, one clear difference between the $N$-body model and the observations is the velocity dispersions: each of the three components are significantly lower in the model than the observations. This directly contributes to the overestimation of the azimuthal velocity ($V_\theta$) by the model. The model does not explicitly set the azimuthal velocity; instead, the circular velocity ($V_{\text{circ}}$) is fixed at a singular radius by enforcing an enclosed mass of $1.8\times10^{10}$M$_\odot$ at a radius of 8.7~kpc \citep{vandermarelThirdEpochMagellanicCloud2014}. However, the circular velocity is higher than the azimuthal velocity due to asymmetric drift, with the difference roughly a factor of the disk velocity dispersion at the large distances of these fields. As such, the too-low velocity dispersions in the model directly contribute to its too-high azimuthal velocity. We consequently expect these same discrepancies in fields along the northern arm. In contrast, the radial and vertical velocities are generally similar in both the model and observations. 

We also assess the geometry of the model, calculating the median out-of-plane distance ($z$)\footnote{Following convention we consider positive $z$ to indicate `above' the disk and negative $z$ to indicate `below' the disk. More informative is to note that in the \citet{vandermarelNewUnderstandingLarge2002} framework, $z$ increases in the direction of the observer such that `above' corresponds to `in front of' the disk plane while `below' corresponds to `behind' the disk plane relative to the observer.} at the location of the two fields. If the model geometry matches the assumed disk geometry, the median distance above the disk plane should be zero. We find the median out-of-plane distances are smaller under the assumption of \citetalias{choiSMASHingLMCTidally2018} disk geometry ($\leq$0.5~kpc for both fields, with field 18 above the disk plane and field 12 below the disk plane) than the assumption of \citetalias{vandermarelThirdEpochMagellanicCloud2014} disk geometry (\textasciitilde0.8-1.6~kpc below the disk plane for fields 18 and 12 respectively). The smaller out-of-plane distances calculated using \citetalias{choiSMASHingLMCTidally2018} geometry imply this is closer to the model inclination, and supports our choice in \S\ref{sec:kinematics} to assume this geometry when transforming observed MagES kinematics into the LMC disk plane. 

\begin{figure*}
	\includegraphics[width=\textwidth]{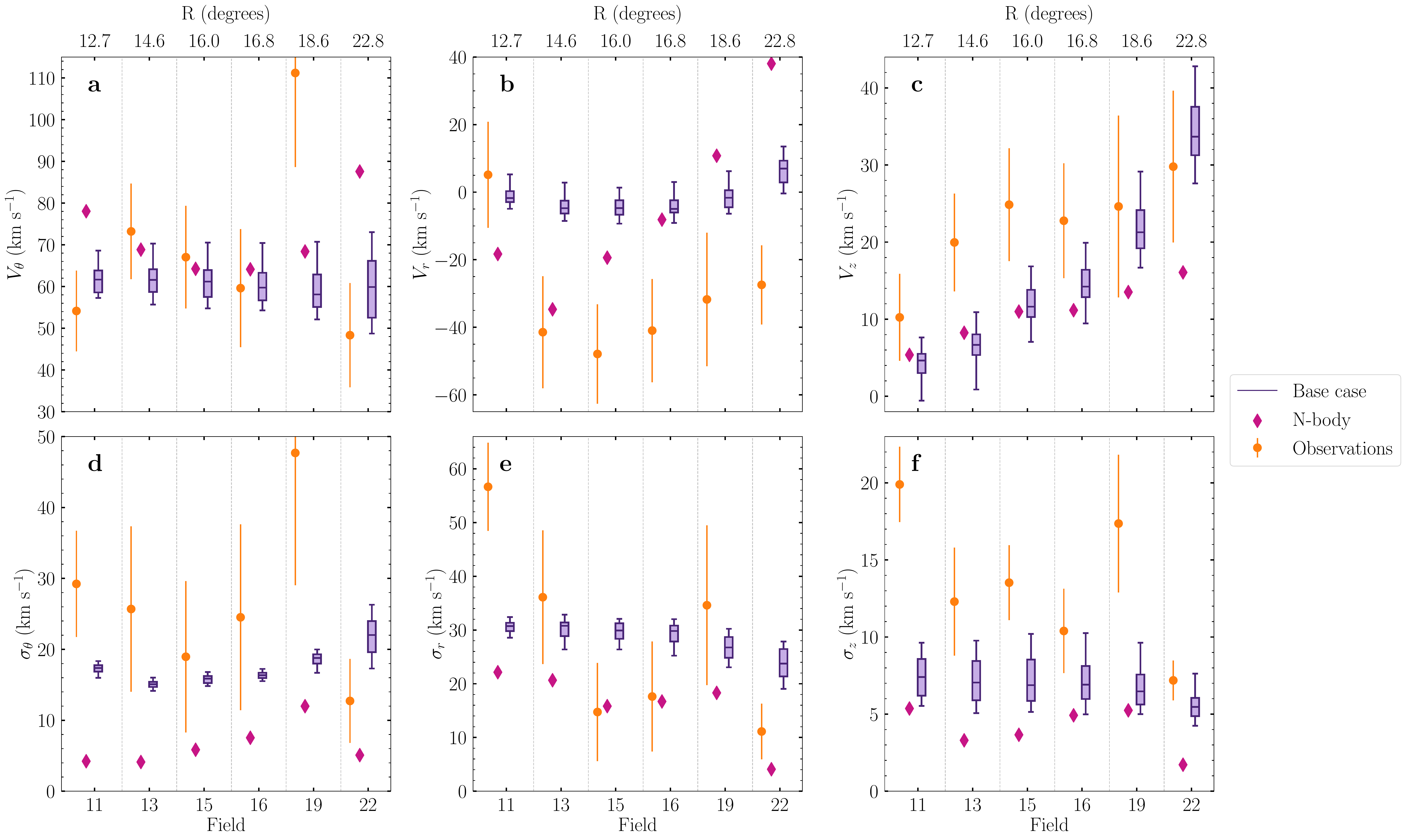}
	\caption{Modelled velocities and dispersions for MagES fields along the northern arm, calculated assuming a \protect\citetalias{choiSMASHingLMCTidally2018} disk geometry. Top panels show, in order, the azimuthal ($V_\theta$), radial ($V_r$), and vertical ($V_z$) velocity components, with bottom panels showing the corresponding velocity dispersion in each component. Orange points show the observations and associated $1\sigma$ uncertainties, and magenta diamonds show results from the $N$-body model. Purple box‐and‐whisker plots show the distribution of the new base-case model suite across 100 realisations: the shaded box shows the 25\textsuperscript{th}‐75\textsuperscript{th} percentiles of the distribution, with whiskers representing the 5\textsuperscript{th} and 95\textsuperscript{th} percentiles, and the central shaded line the 50\textsuperscript{th} percentile of the distribution. For clarity, fields are artificially spaced equally along the x-axis, with model points slightly offset. The top axis lists the LMC galactocentric radius $R$, in degrees, for each field.}
	\label{fig:nbody}
\end{figure*}

Having established caveats associated with the kinematics of the model, and substantiated the assumption of \citetalias{choiSMASHingLMCTidally2018} disk geometry in calculating these, we now compare the model kinematics to MagES fields along the northern arm. Fig.~\ref{fig:nbody} shows the three disk velocity components, as well as dispersions in each of these, for both the $N$-body model (represented by magenta points) and observations within each field. The figure also shows results for the base-case suite of newer models, discussed in more detail below. To improve figure clarity, particularly when comparing several model suites to observations, we plot each field spaced equally along the x-axis, with model points slightly offset from observations. We list the LMC galactocentric radius for each field on the top axis for reference. Whilst overall kinematic trends as a function of position along the feature are similar in both the $N$-body model and the observations, kinematics within each individual field differ. 

As expected from the analysis of the two MagES disk fields, the azimuthal and vertical velocity dispersions (panels \textit{d} and \textit{f} of Fig.~\ref{fig:nbody}) in all fields are substantially lower than the observations, with the in-plane radial velocity (panel \textit{e}) dispersion also lower in all but two fields mid-way along the arm. This is likely a reflection of the underestimated velocity dispersions within the model more generally. The model azimuthal velocity (panel \textit{a}) is also significantly higher than the observations in the innermost feature field. This is likely due to the too-low velocity dispersion of the model, as it is this field where the all three components of velocity dispersion are most significantly underestimated. 

The model in-plane radial velocities (panel \textit{b}) do show the same general shape as the observations, with a drop to approximately $-35$km~s$^{-1}$ in field 13, and an increasing velocity moving further along the feature. However, the model radial velocity increases much too sharply compared to the observations, with the outermost field having a predicted radial velocity close to 40~km~s$^{-1}$. This is clearly inconsistent with the strong negative in-plane radial velocity measured along the entire length of the arm. The vertical velocity (panel \textit{c}) follows the same trend as the observations, but offset in magnitude: while increasing along the length of the feature, with the exception of the innermost field it is consistently lower than the observations by \textasciitilde10-15~km~s$^{-1}$. The overall qualitative agreement between model velocity trends and observations suggest it is plausible that the northern arm could be formed solely as a consequence of the tidal force of the Milky Way; however, the quantitative disagreements indicate there must be differences between this specific model realisation -- and associated perturbation -- compared to the actual LMC. 

\subsection{Simpler model suites}\label{sec:simple}
Whilst the $N$-body model has some qualitative agreement with the observed kinematic trends, it is nonetheless a single model that does not include the SMC. To learn more about the origin of the northern arm, a suite of models is required to (i) probe the allowed range of physical parameters, such as varying galaxy masses and the effect of SMC interactions, and (ii) account for the effects of uncertainties on the LMC and SMC central positions and systemic velocities on the orbits of the Clouds. As it is prohibitively computationally expensive to run such large model suites as full $N$-body models, we instead generate a suite of simpler models. We note that while there are limitations associated with these simpler models, which we discuss further below, they are valuable as an initial exploration of the allowable parameter space.

Our models are inspired by those presented in \cite{belokurovCloudsArms2019} who modelled the LMC disk as a collection of test-particles initially on circular orbits in a single plane. In order to account for the velocity dispersion of the LMC disk, as well as the disk thickness, we instead initialise the LMC disk as an exponential disk made of test-particles. We model the LMC potential as a rigid exponential disk and a Hernquist \citep{hernquistAnalyticalModelSpherical1990} dark matter halo. For the exponential disk, we use a disk mass of $2\times10^9$M$_\odot$, a scale radius of 1.5~kpc, and a scale height of 0.4~kpc. For the Hernquist profile, we assume a mass of $1.5\times10^{11}$M$_\odot$ \citep[motivated by the results of][]{erkalTotalMassLarge2019} and a scale radius of 20~kpc in our fiducial model. We also consider a lighter LMC model where the Hernquist profile mass is $1.5\times10^{10}$M$_\odot$ and the scale radius is 1~kpc. In both cases, the scale radius is chosen so that the circular velocity is approximately 90~km~s$^{-1}$ at 10~kpc. The disks are initialised using \textsc{agama} \citep{vasilievAGAMAActionbasedGalaxy2019}. We note that since all of the features examined in this work are focused on the outskirts of the LMC ($>10$~kpc), we only include particles with apocenters larger than 7~kpc when initialising the disk for computational efficiency.

The Milky Way is modelled as a 3 component system with a bulge, disk, and dark matter halo similar to the \texttt{MWPotential2014} from \citet{bovyGalpyPythonLIBRARY2015}. We use an NFW halo \citep{navarroUniversalDensityProfile1997} with a mass of $8\times10^{11}$M$_\odot$, a scale radius of $16$~kpc, and a concentration of $15.3$. For the disk, we use a Miyamoto-Nagai potential \citep{miyamotoThreedimensionalModelsDistribution1975} with a mass of $6.8\times10^{10}$M$_\odot$, a scale radius of $3$~kpc, and a scale height of $0.28$~kpc. For the bulge, we use a Hernquist profile with a mass of $5\times10^9$M$_\odot$ and a scale length of $0.5$~kpc. We also consider a more massive Milky Way case where the NFW mass is raised to $1.2\times10^{12}$~M$_\odot$ with all other parameters kept the same. 

The SMC is modelled as a rigid Hernquist profile with a mass of $2.5\times10^9$M$_\odot$ and a scale radius of $0.043$~kpc. We also consider a more massive SMC with a mass of $5\times10^9$M$_\odot$ and a scale radius of 1.26~kpc. In both cases, the scale radius is chosen so that the SMC has a circular velocity of $60$~km~s$^{-1}$ at $2.9$~kpc \citep[motivated by the results of][]{stanimirovicNewLookKinematics2004}. As the entire SMC mass is enclosed within this radius in our models, this results in much smaller scale radii than in e.g. \citet{beslaRoleDwarfGalaxy2012a}, who model an initially more massive SMC which experiences mass loss through repeated interactions with the LMC.   

As in \cite{erkalTotalMassLarge2019}, we treat each system (i.e. MW, LMC, SMC), as a particle sourcing a potential. This allows us to account for the motion of the Milky Way in response to the LMC. We account for the dynamical friction of the Milky Way on the LMC using the results of \citet{jethwaMagellanicOriginDwarfs2016a}. The LMC and SMC are initialised at their present day locations, then rewound for 1~Gyr in the presence of each other and the Milky Way. At this time, the LMC disk is initialised with \textasciitilde$2.5\times10^6$ tracer particles, and the system is evolved to the present. During initialisation, the LMC disk is aligned such that its geometry matches that from \citetalias{choiSMASHingLMCTidally2018} -- equal to that assumed for our observations in Section~\ref{sec:kinematics}. Due to the rigid nature of the disk potential, the orientation of the disk does not evolve during the simulation. No tracer particles are placed within the SMC potential.

We verify that the present-day kinematics of the inner LMC ($R$<10~kpc; noting that particles with apocentres <7~kpc are not included in our simulation) remain consistent with those observed in the equilibrium LMC disk at these radii (i.e. $V_{\text{circ}}$\textasciitilde90~km~s$^{-1}$, $V_r\sim V_z\sim0$, $z\sim0$, and each of $\sigma_\theta,\sigma_r,\sigma_z$ approximately constant). This indicates our simulations are suitable for comparison with our observations, and that any deviations from equilibrium at larger radii in our simulations are genuinely the result of perturbations from the Milky Way, SMC, or both.

As a summary of our setup, Table~\ref{tab:modpars} shows the properties of each  model set. For the `base-case' model set -- our best estimate of realistic parameters for each of the LMC, SMC, and MW -- we run 100 realisations, sampling from within Gaussian uncertainties on the LMC and SMC distances and systemic velocities as presented in Table~\ref{tab:modinitprops}. Also presented in Table~\ref{tab:modinitprops} are the current-day relative velocities and positions of the SMC compared to the LMC, in the frame of the LMC disk, that result from our sampling of the 12-dimensional LMC/SMC parameter space. 

\begin{table}
	\caption{Simulation parameters for each simpler model ensemble.}
	\label{tab:modpars}
	\begin{adjustbox}{max width=\columnwidth}
		\begin{threeparttable}
			\begin{tabular}{lllll}
				\hline
				Ensemble & Realisations & LMC mass & MW mass & SMC mass \\ \hline
				base-case  & 100 & $1.5\times10^{11}$M$_\odot$\tnote{a} & $8\times10^{11}$M$_\odot$\tnote{b}     & $2.5\times10^{9}$M$_\odot$\tnote{c} \\
				No SMC     & 12 & $1.5\times10^{11}$M$_\odot$  & $8\times10^{11}$M$_\odot$      & -              \\
				Light LMC  & 12 & $1.5\times10^{10}$M$_\odot$\tnote{d}    & $8\times10^{11}$M$_\odot$      & -              \\
				Heavy MW   & 12 & $1.5\times10^{11}$M$_\odot$  & $1.2\times10^{12}$M$_\odot$\tnote{e} & -              \\
				Heavy SMC  & 12 & $1.5\times10^{11}$M$_\odot$  & $8\times10^{11}$M$_\odot$      & $5\times10^{9}$M$_\odot$\tnote{c}		\\ \hline
			\end{tabular}
			\begin{tablenotes}[para]\footnotesize
				\item[a]\protect\citealt{erkalTotalMassLarge2019}; \item[b]\protect\citealt{bovyGalpyPythonLIBRARY2015}; \item[c]\protect\citealt{harrisSpectroscopicSurveyRed2006a}; 	\item[d]\protect\citetalias{vandermarelThirdEpochMagellanicCloud2014}; \item[e]\protect\citealt{bland-hawthornGalaxyContextStructural2016}. 
			\end{tablenotes}
		\end{threeparttable}
	\end{adjustbox}
\end{table}

\begin{table*}
	\caption{Model parameters for the present-day systemic properties of the LMC and SMC. Parameters are sampled from a Gaussian distribution centred on the peak value, with a $1\sigma$ width equal to the literature uncertainty on that parameter. The bottom half of the table presents the present-day distribution of the 3D position and velocity of the SMC relative to the LMC, which results from sampling the systemic properties of the Clouds reported in the upper section of the table. 
	We report the median value, with the uncertainty values corresponding to the 1$\sigma$width of the distribution.} 
	\label{tab:modinitprops}
	\begin{adjustbox}{max width=\textwidth}
	\begin{tabular}{llllp{0.6\linewidth}}
		\hline
		Variable & Value & Unit & Reference & Comment \\ \hline
		LMC $\alpha_0$ & $79.88$ & degrees & \protect\citetalias{vandermarelThirdEpochMagellanicCloud2014} & RA of the LMC COM. Taken from their `PMs+Old $V_{\text{LOS}}$ Sample' result. Held fixed.  \\
		LMC $\delta_0$ & $-69.59$ & degrees & \protect\citetalias{vandermarelThirdEpochMagellanicCloud2014} & DEC of the LMC COM. Taken from their `PMs+Old $V_{\text{LOS}}$ Sample' result. Held fixed.  \\
		LMC $V_{\text{LOS},0}$ & $261.1\pm2.2$ & km~s$^{-1}$ & \protect\citetalias{vandermarelThirdEpochMagellanicCloud2014} & LOS velocity of the LMC COM. Taken from their `PMs+Old $V_{\text{LOS}}$ Sample' result.   \\
		LMC $\mu_{\alpha,0}$ & $-1.895\pm0.024$ & mas~yr$^{-1}$ & \protect\citetalias{vandermarelThirdEpochMagellanicCloud2014} & Proper motion in the $\alpha\cos(\delta)$ direction of the LMC COM. Taken from their `PMs+Old $V_{\text{LOS}}$ Sample' result.   \\
		LMC $\mu_{\delta,0}$ & $0.287\pm0.054$ & mas~yr$^{-1}$ & \protect\citetalias{vandermarelThirdEpochMagellanicCloud2014} & Proper motion in the $\delta$ direction of the LMC COM. Taken from their `PMs+Old $V_{\text{LOS}}$ Sample' result.   \\
		LMC $D_0$ & $50.1\pm2.5$ &~kpc & \protect\citet{freedmanFinalResultsHubble2001} & Distance to the LMC COM. Used in the   \protect\citetalias{vandermarelThirdEpochMagellanicCloud2014} analysis. Whilst more recent (and precise) distance estimates are available \protect\citep[e.g.][]{pietrzynskiDistanceLargeMagellanic2019}, we permit $D_0$ to vary over this range in order to investigate a larger range of allowed LMC orbits.   \\
		SMC $\alpha_0$ & $13.38$ & degrees & \protect\citet{subramanianThreeDimensionalStructureSmall2012a} & RA of the SMC COM. Held fixed. \\
		SMC $\delta_0$ & $-73.0$ & degrees & \protect\citet{subramanianThreeDimensionalStructureSmall2012a} & DEC of the SMC COM. Held fixed.\\
		SMC $V_{\text{LOS},0}$ & $145.6\pm0.6$ & km~s$^{-1}$ & \protect\cite{harrisSpectroscopicSurveyRed2006a} & LOS velocity of the SMC COM.  \\
		SMC $\mu_{\alpha,0}$ & $0.772\pm0.063$ & mas~yr$^{-1}$ & \protect\citet{kallivayalilThirdEpochMagellanicCloud2013} & Proper motion in the $\alpha\cos(\delta)$ direction of the SMC COM.   \\
		SMC $\mu_{\delta,0}$ & $-1.117\pm0.061$ & mas~yr$^{-1}$ & \protect\citet{kallivayalilThirdEpochMagellanicCloud2013} & Proper motion in the $\delta$ direction of the SMC COM. \\
		SMC $D_0$ & $62.1\pm1.9$ & km~s $^{-1}$ & \protect\citet{graczykARAUCARIAProjectDistance2013} & Distance to the SMC COM. \\ \hline
		$r_{\text{SMC}}$ & $23.5\pm1.5$ & kpc & - & Total distance between the LMC and SMC centres of mass. 
		\\
		$V_{\text{tot}}$ & $122.6\pm32.0$ & km~s$^{-1}$ & - & Total velocity of the SMC relative to the LMC. 
		\\ \hline
	\end{tabular}
	\end{adjustbox}
\end{table*}

\begin{figure}
	\includegraphics[width=\columnwidth]{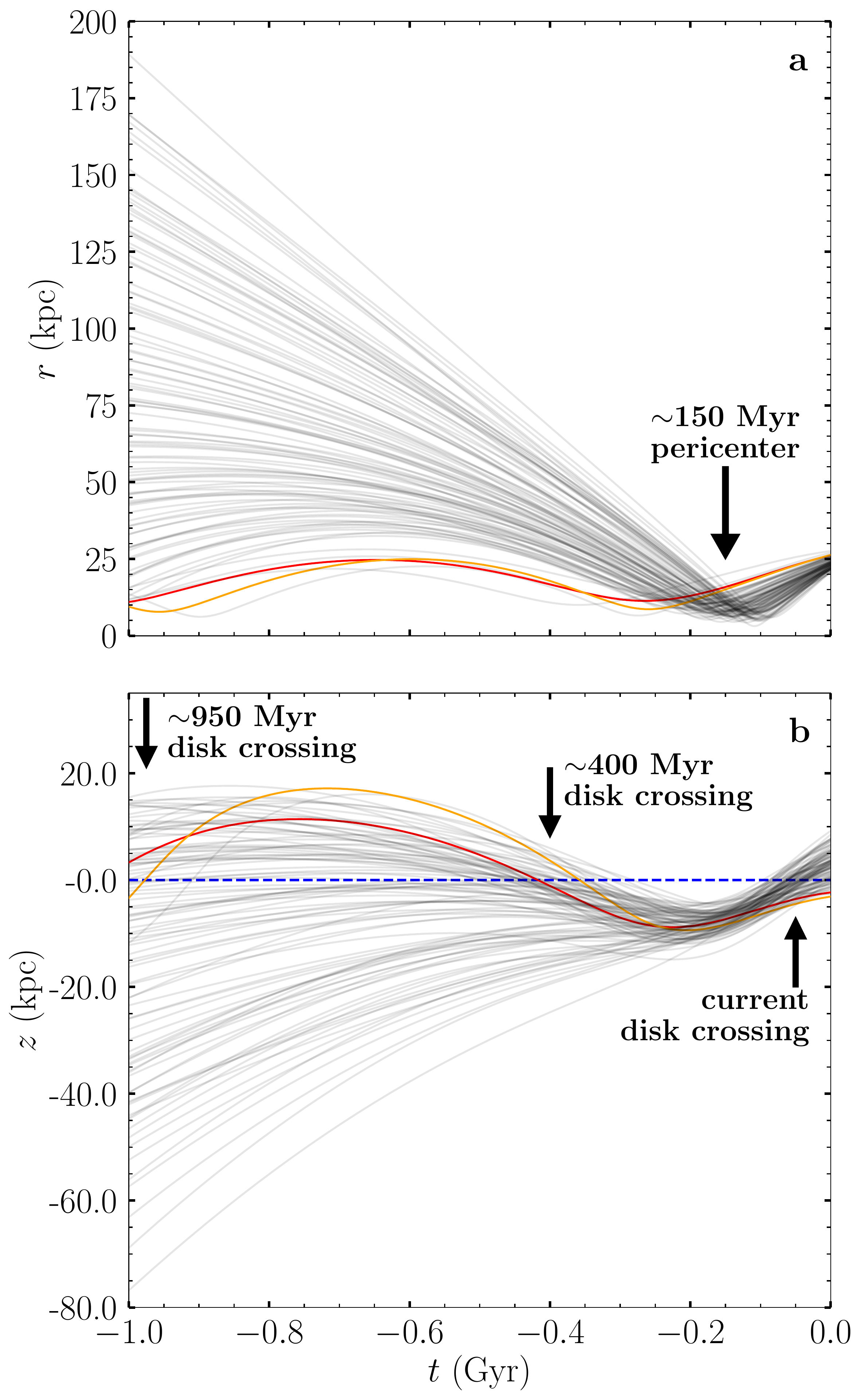}
	\caption{Total distance $r$ (top) and out-of-plane distance $z$ (bottom) of the SMC from the LMC as a function of time from the present day, up to the 1~Gyr cutoff of our models. Each grey line represents a single realisation of the base-case model suite, associated with an allowable orbit sampling from the uncertainties in the present-day systemic motions of the Clouds. Where these lines cross the dashed blue line at $z=0$ in the bottom panel indicates an SMC crossing of the LMC disk plane. Minima in the top panel indicate SMC pericentric passages around the LMC. Red and orange lines represent individual realisations which experience, respectively, one and two SMC crossings of the LMC disk plane prior to today, which are discussed further in Section \ref{sec:origin}.}
	\label{fig:orbits}
\end{figure}

Sampling these parameters results in a range of allowable orbits for the SMC around the LMC, and both Clouds around the Milky Way. In general, the orbit of the Clouds around the Milky Way does not vary too significantly between realisations, with the Clouds always just past their first pericentric passage around the Milky Way \citep[c.f.][]{kallivayalilThirdEpochMagellanicCloud2013}. However, the orbit of the SMC around the LMC can vary significantly. Fig.~\ref{fig:orbits} shows both the total distance $r$ (top panel)\footnote{$r$, the total distance, is distinct from $R$, which is the in-plane cylindrical radius.} and the height above the disk plane $z$ (bottom panel) of the SMC from the LMC centre of mass as a function of time during each of the 100 realisations of the base-case model setup. In general, whilst the orbit of the SMC within the past \textasciitilde250~Myr from today is broadly consistent across all realisations, the orbital history beyond this can diverge quite significantly depending on the specific location in parameter space of each realisation. In the following, we report statistics after running an additional 900 realisations of the base-case model setup (noting these are not full realisations initialised with test particles, but instead simply trace the orbits of the LMC and SMC COM), sampling from the present-day positions and systemic motions of the Clouds as in Table~\ref{tab:modinitprops}, to give 1000 total past orbits.

We find the SMC is currently in the process of crossing the LMC disk plane, with \textasciitilde81\% of orbits having already had a crossing at a distance of $17\pm5$~kpc, and the remaining \textasciitilde19\% set to cross the plane in the very near future. This crossing is likely to affect the dynamics of the LMC disk in the future, but is sufficiently recent (with the median crossing time only $45_{-26}^{+25}$~Myr ago) that it will not have a significant effect on the present-day kinematics of the disk as a whole. We additionally see that in all realisations, the SMC has had a recent pericentric passage around the LMC $147^{+42}_{-31}$~Myr ago \citep[in agreement with][]{zivickProperMotionField2018b}. However, as seen in Fig.~\ref{fig:orbits}, while these are relatively close pericenters, with $r_{\text{peri}} = 8.0^{+2.4}_{-2.0}$~kpc, we also find that they occur significantly below the plane of the LMC disk, with $z_{\text{peri}} = -6.8_{-2.6}^{+2.5}$~kpc.
	
Beyond this, the orbit of the SMC varies significantly. Approximately 51\% of our realisations have a second SMC crossing of the LMC disk $398^{+84}_{-68}$~Myr ago, which occurs across a broad range ($28.8^{+11.4}_{-9.2}$~kpc) of distances\footnote{as these are disk crossings, $z=0$ and thus the in-plane radial distance $R$ is equal to the total distance $r$}. The remaining \textasciitilde49\% of orbits either i) do not quite cross the disk plane but do closely approach it during this time period, or ii) remain significantly behind the LMC’s disk plane for the entirety of the 1~Gyr over which our models are run. A handful of models (\textasciitilde9\%) additionally have a third disk crossing $906_{-163}^{+61}$~Myr ago, though we note a much larger fraction would experience this crossing if our models were rewound for a greater length of time than the 1~Gyr for which they are currently run. Due to the increasing uncertainty in the SMC’s orbit at earlier times, the particulars of this third crossing are much less robustly constrained than the \textasciitilde400~Myr crossing, with a crossing distance of $53.8_{-46.3}^{+13.1}$~kpc. In fact, \textasciitilde20\% of these third crossings pass within 10~kpc of the LMC center (i.e. this occurs in \textasciitilde2\% of the total model set). At around this time, we additionally find a small fraction (\textasciitilde4\%) of our models show a second SMC pericentric passage, again noting this fraction would be substantially increased were our models rewound further than 1~Gyr. These passages have similar pericentric distances to the most recent pericentric passage ($r_{\text{peri}} = 6.2^{+3.8}_{-2.3}$~kpc), but occur at smaller out-of-plane distances ($z_{\text{peri}} = 2.2^{-1.1}_{+2.5}$~kpc) due to the similarly-timed disk crossing in these realisations. 
	  
Whilst these statistics are for the base-case model setup, we find broadly similar results for the heavy-SMC model setup -- that is, all realisations experience a SMC pericentre \textasciitilde150~Myr ago at a reasonably large out-of-plane distance, a moderate fraction of realisations experience a second SMC crossing of the LMC disk plane \textasciitilde400~Myr ago, and a modest number of realisations have additional disk crossings and pericentric passages \textasciitilde1~Gyr ago (noting again the number of such realisations would increase were our models rewound for a greater length of time).

It is important to note that the relative simplicity, and in particular the lack of self-gravity (see \S\ref{sec:caveats} for a detailed discussion), of our models mean the interactions described above are only estimates; more realistic models of the Magellanic/Milky Way system will be required to confirm the precise orbit of the SMC relative to the LMC. Nevertheless, our models provide a useful first look at understanding the likely relative importance of different interactions on the northern arm. We defer detailed discussion of the overall effects of these interactions on the LMC disk, as well as which regions of parameter space correspond to different orbits of the SMC, to a forthcoming paper which incorporating MagES data across a larger region of the LMC disk (Cullinane et al. in prep). 

\subsubsection{Model Caveats}\label{sec:caveats}
Whilst the ability to explore the large allowable parameter space is a significant advantage of our simple model suites, this approach does have limitations. Particularly significant is the lack of self-gravity, which has two significant effects on the system. The first of these is that the gravitational potentials used to model the dark matter halo of each galaxy are unable to deform in response to one another \citep[e.g.][]{garavito-camargoQuantifyingImpactLarge2021}. This can potentially influence both the global orbits of the Clouds, and the response of stars within them. The second is that model particles describing stars in the LMC disk cannot directly affect one another -- i.e. the LMC disk potential is also fixed in shape and orientation -- which can affect the response of the stellar disk to interactions (particularly those which might introduce overdensities to the disk). We discuss in turn these effects on each pair of galaxies in the MW/LMC/SMC system below.

We first discuss the effect of self-gravity on the MW/LMC pair, as we can to some extent quantify these effects through comparison of the simpler models to the $N$-body model. In the $N$-body model the gravitational potentials of both galaxies, which in the simpler models are rigid profiles, are allowed to move in response to one another \citep[i.e. the reflex motion of the Milky Way in response to the LMC is captured][]{gomezItMovesDangers2015}; but this is a global shift in position as opposed to a change in shape. However, models capturing this deformation process \citep[see e.g. Fig.~10 of][]{erkalTotalMassLarge2019} demonstrate the shape of the MW potential is not significantly affected even during the infall of a massive ($1.5\times10^{11}$M$_\odot$) LMC; and at the distance of our outermost field ($R_{\text{LMC}}$\textasciitilde23~kpc), the deformation of the LMC potential is also minimal. 

In terms of the disk, as the stellar density is highest near the base of the feature, in the $N$-body model the higher concentration of particles here better maintains the disk kinematics. This contributes to the $N$-body having a strong negative in-plane radial velocity in field 13, but one much closer to zero in field 11: the stronger LMC gravitational potential at smaller galactocentric radii, in combination with the stronger self-gravity of the disk, helps maintain the disk kinematics near equilibrium levels. In contrast, the lower stellar density in the outskirts of the feature mean the self-gravity of the disk contributes negligibly to the overall gravitational potential, with these regions therefore more easily perturbed. However, we find in \S\ref{sec:genmod} below that $V_r$ is less significantly perturbed in the simpler model suites all along the arm compared to the $N$-body model: somewhat unexpected given the lack of self-gravity in these models should allow for larger perturbations in their kinematics. We therefore conclude the effect of self-gravity in the MW/LMC pair is not responsible for the largely negative in-plane radial velocity along the arm.

We next discuss the effect of self-gravity on the LMC/SMC pair. Several studies have investigated the effects of close interactions with a smaller satellite (like the SMC) on a larger host (like the LMC) in fully self-gravitating systems \citep[e.g.][]{berentzenNumericalSimulationsInteracting2003,bekkiFormationOffcenterBar2009,beslaRoleDwarfGalaxy2012a,yozinTidalinducedLopsidednessMagellanictype2014,pardyTidallyInducedOffset2016}. These studies typically assess the effects of a near-direct collision between the two galaxies: that is, a crossing of the host’s disk plane which occurs at relatively small host galactocentric radii. A common finding of each of these studies is that such interactions can introduce asymmetries in the disk of the host, and offsets of up to \textasciitilde2.5~kpc between the dynamical centres of the disk and a central bar. This results in an off-centre and potentially tilted bar, as is observed in the LMC today. Of greater interest to this paper is that these crossings can also produce density waves and features similar to spiral arms out to large radii \citep[$\gtrsim10$~kpc:][]{berentzenNumericalSimulationsInteracting2003,beslaLowSurfaceBrightness2016} relatively shortly after the crossing time \citep[100-200~Myr:][]{berentzenNumericalSimulationsInteracting2003,pardyTidallyInducedOffset2016}, with these features persisting for \textasciitilde Gyr after the crossing \citep{berentzenNumericalSimulationsInteracting2003,yozinTidalinducedLopsidednessMagellanictype2014}. Our simple models would not fully capture these effects.

However, we do note this specific type of interaction -- that is, a disk plane crossing at small galactocentric radii -- is not typical of the interactions observed in our models, with this only occurring in \textasciitilde9\% of our models and \textasciitilde900~Myr ago, where uncertainties in the orbit of the SMC are very large. The more recent disk crossing observed in our models, which occurs \textasciitilde400~Myr ago and in \textasciitilde51\% of our models, occurs at a much larger radius ($28.8^{+11.4}_{-9.2}$~kpc) than is typically modelled in these studies. In fact, \citet{bekkiFormationOffcenterBar2009} finds that interactions at larger galactocentric radii distances ($R\sim5$-$10$kpc) are unable to produce an off-centre stellar bar in the LMC; and \citet{poggioMeasuringVerticalResponse2021}, while studying the impact of the Sagittarius dwarf on the Milky Way, find that disk crossings at large radii (i.e. which do not align with a simultaneous pericentric passage) affect the MW disk significantly less than crossings at smaller radii. We therefore expect the \textasciitilde400~Myr disk plane crossing will have a comparatively small effect on the LMC disk as a whole.

In addition, as discussed above, our models suggest the SMC’s recent pericentric passage \textasciitilde150~Myr ago, occurs at a relatively large out-of-plane distance ($-6.8_{-2.6}^{+2.5}$kpc), with the SMC only approximately now crossing the LMC disk plane. Thus, whilst the radius of the pericentric passage is similar to the interactions modelled in the above studies, we expect its effect on the LMC disk to be commensurately reduced, though \cite{laporteResponseMilkyWay2018a}, in studying the MW/LMC system, find out-of-plane pericentres may introduce mild ($z\sim1$~kpc) warping of the host galaxy disk which would not be captured in our simpler models.

Likely a more significant effect of the SMC’s recent pericentre is an indirect one: studies of the MW/LMC system \citep[e.g.][]{garavito-camargoHuntingDarkMatter2019} suggest pericentres produce both local and global dark matter wakes in the halo of the host galaxy. These wakes can induce torques on the satellite galaxy \citep{tamfalRevisitingDynamicalFriction2021}, thus affecting its orbit. Along similar lines, \cite{kallivayalilThirdEpochMagellanicCloud2013} note that such dynamical friction effects would result in a more eccentric orbit of the SMC around the LMC, though the magnitude of this effect is not explicitly calculated. As our simple models do not capture these effects, this is a source of increased uncertainty in the orbit of the SMC, and thus its interactions with the LMC, beyond the recent pericentric passage. 

Finally, we briefly discuss the effect of self-gravity on the MW/SMC pair. In contrast to the MW/LMC pair, we do not capture the effect of dynamical friction from the Milky Way on the SMC. This may affect the recent orbit of the SMC, which in turn would affect specifics of interactions between the LMC and SMC. However, we expect the direct effect of the LMC -- being much closer and having likely experienced repeated interactions with the SMC prior to the current infall to the Milky Way potential -- is more significant in this case.

We additionally note that the gravitational potential used to represent the SMC in our models is relatively simple, particularly given recent findings that indicate it is currently being tidally disrupted by the LMC \citep[e.g.][]{zivickProperMotionField2018b,deleoRevealingTidalScars2020a}. More detailed modelling which captures this disruption, as well as mass loss from the SMC over time (due to likely repeated interactions with the LMC) would be necessary to fully describe these effects and assess how such a varying potential affects the SMC’s orbit. 

The above simplifications mean our models do not capture all the subtleties of interactions between the Clouds, and thus cannot definitively establish the origin of substructures such as the northern arm. However, we stress our aim is qualitative, not quantitative, agreement with observations; and our simpler models do permit an exploration of the allowable parameter space which can indicate the plausibility of various interactions in forming substructures. This ability to isolate which interactions are more or less likely to contribute to the origin of substructures is valuable, as these can be investigated using more detailed models in the future. 

\subsection{Simple model kinematics along the northern arm}\label{sec:modkin_int} 
\begin{figure*}
	\includegraphics[width=\textwidth]{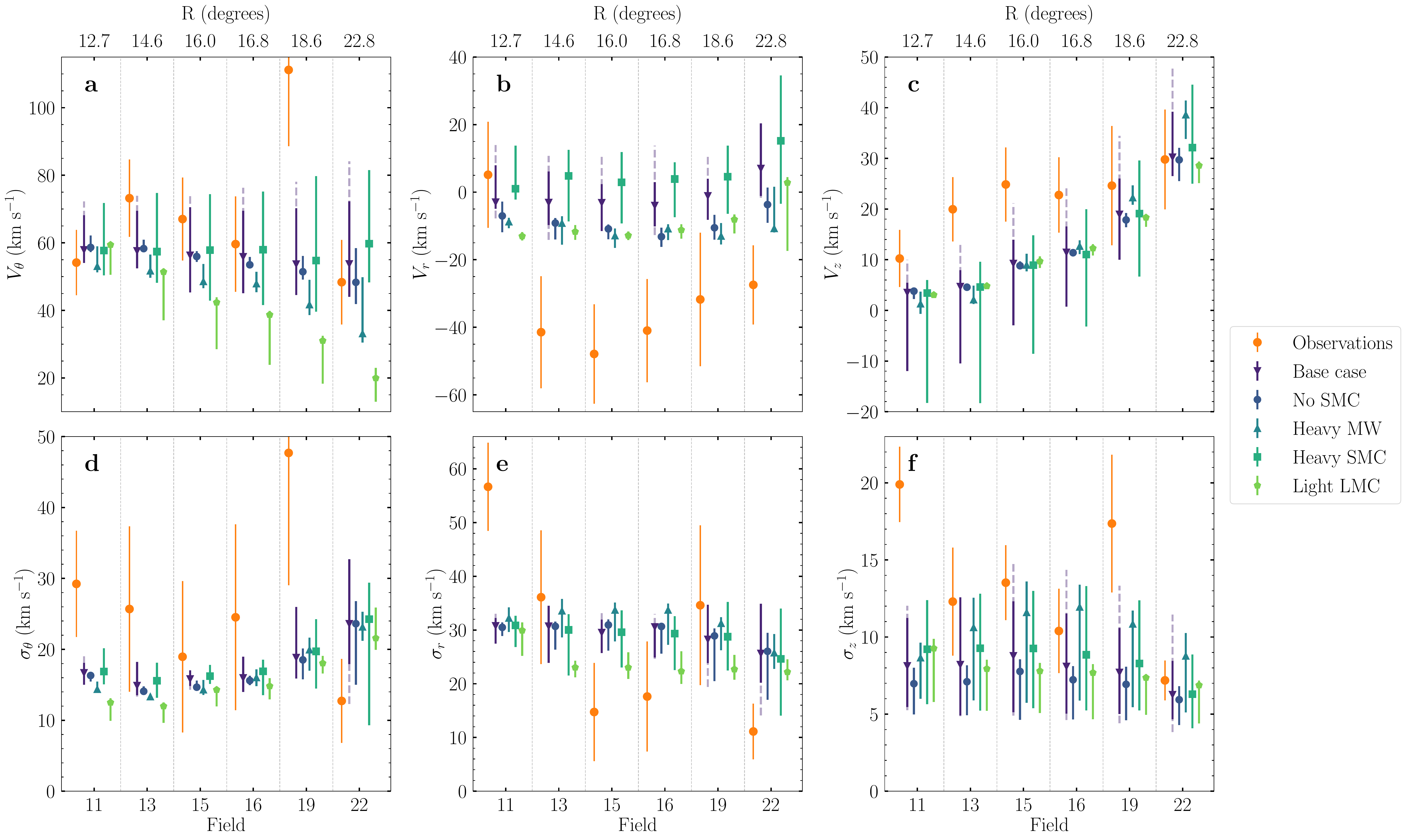}
	\caption{Model velocities and dispersions for MagES fields along the arm-like feature for the simpler model suites, calculated assumming a \protect\citetalias{choiSMASHingLMCTidally2018} disk geometry. Top panels show, in order, the azimuthal, radial, and vertical velocity component, with bottom panels showing the corresponding velocity dispersion in each component. Orange points show the observations and associated $1\sigma$ uncertainties. Coloured model points show ensemble medians, and error bars show ensemble ranges, with each suite represented by a different colour and symbol. Points without error bars have sufficiently small ranges that these are not observable. For clarity, fields are artificially spaced equally along the x-axis, and each suite of model points slightly offset. The top axis lists the LMC galactocentric radius of the fields. Dashed extensions of the error bars for the base-case ensemble (purple) show the full range of data from the 100 realisations relative to the range associated with the subsample of 12 realisations used for the other model suites.}
	\label{fig:modsuite}
\end{figure*}

\begin{figure}
	\includegraphics[width=\columnwidth]{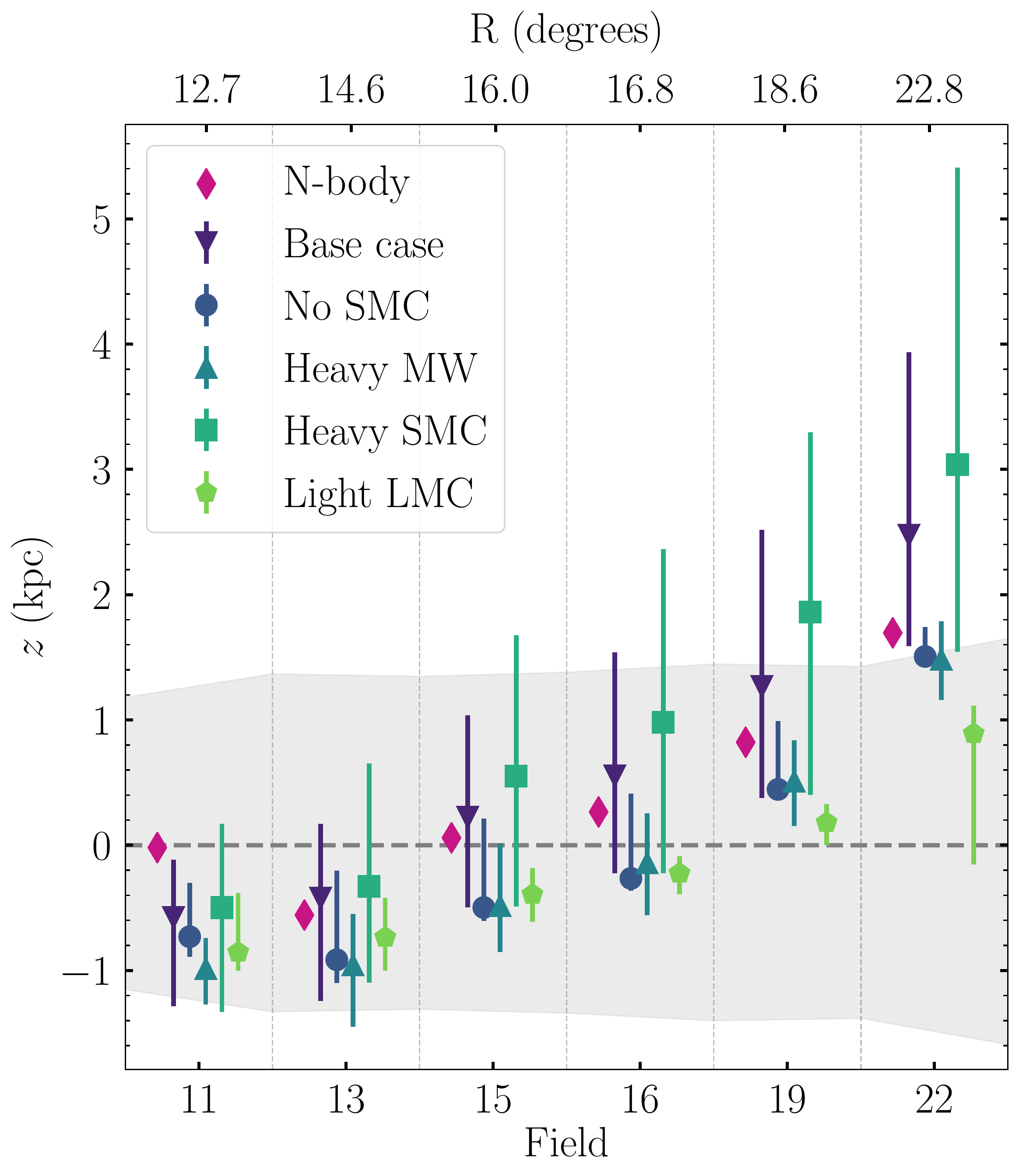}
	\caption{Model out-of-plane distance ($z$) for fields along the northern arm-like feature, calculated relative to a \protect\citetalias{choiSMASHingLMCTidally2018} disk geometry. Magenta diamonds show results from the $N$-body model results. Coloured model points show ensemble medians, and error bars show ensemble ranges, with each suite represented by a different colour and symbol. Points without error bars have sufficiently small ranges that these are not observable. For clarity, fields are spaced equally along the x-axis, and each suite of model points slightly offset. The top axis lists the LMC galactocentric radius of the fields. The dashed grey line indicates the $z$=0 assumption utilised for the observed data, with the shaded grey region indicating the distance range associated with the uncertainty in the median $G_0$ magnitude along the northern arm as in \S\ref{sec:phot}.}
	\label{fig:moddist}
\end{figure}

We first discuss the base-case suite of 100 models (represented by purple points in Figs.~\ref{fig:nbody}-\ref{fig:moddist}). Kinematics for this suite are presented in Fig.~\ref{fig:nbody}. The distribution of results across the ensemble are represented by box-and-whisker plots, displaying the 5\textsuperscript{th}, 25\textsuperscript{th}, 50\textsuperscript{th}, 75\textsuperscript{th}, and 95\textsuperscript{th} percentiles within each field. For a given field, the spread in ensemble kinematics --  \textasciitilde10~km~s$^{-1}$ in $V_r$ and $V_z$, 10-20~km~s$^{-1}$ in $V_\theta$, and 5-10~km~s$^{-1}$ in each velocity dispersion component -- is due entirely to differences in the orbits of the Clouds parameterised by sampling from within the uncertainties in their central positions and motions. These variations can be a similar order of magnitude to the observational uncertainties within the fields, and demonstrate the importance of sampling these parameters as compared to running a single model realisation. 

We find a number of key differences between the ensemble kinematics and the $N$-body model. The azimuthal and in-plane radial velocity dispersions (panels \textit{d} and \textit{e} of Fig.~\ref{fig:nbody}) are up to 10-15~km~s$^{-1}$ higher than the $N$-body model, and the vertical velocity dispersion (panel \textit{f}) up to 5~km~s$^{-1}$ higher. This is by design, as the model suites are initialised with higher velocity dispersions to more closely match recent MagES measurements in the outer LMC disk \citepalias[see][]{C20}. However, the velocity dispersion in each component remains underestimated in field 11 (closest to the LMC disk) relative to observations.

The ensemble azimuthal velocity (panel \textit{a}) is lower than the $N$-body model, and remains flat at approximately the value measured in the outer LMC disk in \citetalias{C20}. This is more consistent with the observations, particularly in the inner- and outermost fields along the feature where the $N$-body model is most discrepant. The vertical velocity (panel \textit{c}) is also more consistent with observations in the three outermost feature fields. However, in the three fields closest to the LMC disk, the vertical velocity is not significantly different from the $N$-body model and remains lower than the observations. 

The greatest difference in model kinematics is in the in-plane radial velocity (panel \textit{b}). In the base-case ensemble, this remains almost flat along the length of the feature, with only a very mild (\textasciitilde5~km~s$^{-1}$) drop and subsequent increase in the mean radial velocity along the feature. This is significantly different from the very negative values seen in the observations and the inner fields of the $N$-body model. However, the lack of a steep increase in radial velocity along the feature in the model suite is a trend that is less discrepant with observations than the $N$-body model in the outermost feature fields.

In order to understand the drivers of these kinematic differences, we now compare the different simple model ensembles to assess the impact of varying galaxy masses on the kinematics of the northern arm. Fig.~\ref{fig:modsuite} shows the three disk velocity components ($V_\theta$, $V_r$, and $V_z$), as well as dispersions in each of these components, for both the model ensembles and the observations within each field. Each point represents the median of the 12 realisations, with error bars showing the full range across each suite. We include the base-case ensemble in this comparison, but for consistency sample only the same 12 realisations as included for the other suites. Pale dashed extensions to the base-case ensemble results show the full range from all 100 realisations of the ensemble, for comparison. Fig.~\ref{fig:moddist} shows, in the same format, the out-of-plane distance ($z$) for each of the model ensembles.

\subsubsection{General comments}\label{sec:genmod} 
In general, most of the model suites follow the same overall kinematic trends, and do not provide substantially better matches to the observations than the base-case model. The largest differences between model suites typically occur in the outermost field, with the median response typically very similar for most of the length of the feature. This as expected: at smaller radii, the LMC potential dominates, and any external perturbing potential, regardless of origin, does not significantly affect the kinematics. In contrast, at large galactocentric radii the perturbing potential can dominate, leading to different kinematics depending on the origin of the perturbation.

We find the out-of-plane distances (Fig.~\ref{fig:moddist}) are typically small (<3~kpc) for each of the model suites, consistent with our findings in \S\ref{sec:phot} that the northern arm roughly follows the plane of the LMC disk. We have tested utilising the median out-of-plane distance for each field in each model suite as the assumed distance for the MagES fields when deriving the observed kinematics in the plane of the disk. However, we find the resulting differences in kinematics are typically less than \textasciitilde5~km~s$^{-1}$: well within the uncertainties of the measurements calculated assuming the fields exactly follow the LMC disk plane. We also find the median distance dispersion within each field is negligibly different (\textasciitilde0.15~kpc) between model suites, with a mild (\textasciitilde0.3~kpc) decrease in the median distance dispersion along the length of the feature. This is, within uncertainty, consistent with our measurements of constant $G_0$ dispersion along the first half of the feature. It is further suggestive that the increase in the $G_0$ dispersion measured in the outer regions of the feature is due to contamination as speculated in \S\ref{sec:phot}, and not a genuine thickening of the feature -- if so, we would expect the distance dispersion of the models to increase along the length of the northern arm. 

Notably, regardless of model suite, each component of velocity dispersion is underestimated in the innermost feature field. In the case of the vertical velocity dispersion, this is due to the initial conditions of the model, which even within the outer LMC disk is lower than observations at \textasciitilde10~km~s$^{-1}$. This is due to the modest scale height of the disk used in our model and the relatively small contribution of the disk to the gravitational field in the outer parts of the disk. It is possible that previous encounters with the SMC may be needed to inflate this dispersion. Alternatively, the outer disk of the LMC may be thicker than assumed in our model.

We also note that whilst selection of the 12 realisations attempts to sample the full possible kinematic distribution, comparison of the set of 12 realisations to the full suite of 100 base-case models as in Fig.~\ref{fig:modsuite} reveals the full range is somewhat underestimated, with the upper limits of $V_\theta$, $V_r$, $V_z$, and $\sigma_z$ underestimated by up to 30\%. In general, this does not improve the consistency between the model kinematics and observations, with the possible exception of $V_z$ in the inner fields. Median values are not significantly different in the restricted set, and lower limits are typically well-sampled. The largest differences in median values are only on the order of \textasciitilde10\%, with overestimated $\sigma_z$ medians and underestimated $V_\theta$ medians along the length of the arm. 

\subsubsection{Effect of the MW mass}\label{sec:MWmass}
The effect of the heavy MW (indicated by dark green points in Figs.~\ref{fig:modsuite} and~\ref{fig:moddist}) is most apparent in the azimuthal velocity (panel \textit{a} of Fig.~\ref{fig:modsuite}), with a mild decrease in the median $V_\theta$ (by \textasciitilde15~km~s$^{-1}$) along the length of the feature. However, we note this remains consistent with the observations within uncertainty. Further, the $V_\theta$ medians are typically underestimated as compared to a full suite of 100 realisations. Consequently, this does not preclude a heavy MW from matching the observations. 

Comparing the in-plane radial and vertical velocities (panels \textit{b} and \textit{c} respectively), in the innermost fields the heavy MW suite is not significantly different from the base-case suite (indicated by purple points). A larger difference occurs in the outermost fields along the feature, with the heavy MW generating slightly higher vertical velocities, and slightly more negative in-plane radial velocities. It is not surprising the strongest effects are felt in the outermost fields: it is here where the MW gravitational potential is strongest compared to the LMC gravitational potential, and can induce the strongest perturbations.

The heavy MW has a negligible effect on the azimuthal velocity dispersion (panel \textit{d}), and minimally (\textasciitilde2~km~s$^{-1}$) increases the in-plane radial velocity dispersion (panel \textit{e}) -- not sufficient to meet the large measured velocity dispersions in the innermost observed field. It does have an increased (\textasciitilde5~km~s$^{-1}$) vertical velocity dispersion(panel \textit{f}) in all but the innermost field, which provides a closer match to observations than any of the other model suites. The out-of-plane distance (Fig.~\ref{fig:moddist}) is also negligibly different from the no-SMC and $N$-body models, remaining within 1.5~kpc of the assumed \citetalias{choiSMASHingLMCTidally2018} inclined disk geometry. Out-of-plane distances of this magnitude can be accommodated within our photometric uncertainties. As discussed in \S\ref{sec:phot}, the typical uncertainty in our median $G_0$ magnitudes along the feature is \textasciitilde0.05~mag. At the distance of the feature, this corresponds to an \textasciitilde1.5~kpc uncertainty in the derived distance. 

We can understand the increased effects of the heavy MW in the $z$-direction when considering the LMC’s orbit and inclination relative to the MW during its infall. The orientation of the LMC is such that the northern half of the disk is inclined closer to the MW plane. As such, there is an increasingly strong gravitational force from the MW along the length of the northern arm. Further, the orbit of the LMC is such that it is approaching the MW from underneath the MW disk plane. The force of the MW thus pulls forward in the positive $z$-direction on the LMC disk, increasing $V_z$ particularly in the outermost fields where this pull is strongest relative to the LMC potential. This may also explain the increased vertical velocity dispersion $\sigma_z$ as compared to the lighter MW models, and the increasingly positive out-of-plane distance along the feature. 

As the primary effect of the MW on the LMC is in the positive $z$-direction, we hypothesise the MW is also responsible for the asymmetric LOS velocity distributions observed in the northern LMC disk in \citetalias{C20}. The distribution of $V_{\text{LOS}}$ for likely-Magellanic stars in MagES fields 18 and 12 were found to have tails to low LOS velocities; the low inclination of the LMC disk implies stars in these tails have positive vertical velocities of up to \textasciitilde40~km~s$^{-1}$. This is similar to the positive $V_z$ velocities found along the northern arm. We therefore suggest stars in the northern LMC disk showing this perturbation signature are, like the northern arm, disturbed during the LMC's infall to the MW.  

\subsubsection{Effect of the SMC}\label{sec:smcmass}
We now consider the effect of the SMC, comparing the no-SMC (indicated by dark blue points in Figs.~\ref{fig:modsuite} and~\ref{fig:moddist}) and heavy SMC (indicated by turquoise points) model suites to the regular SMC base-case ensemble (indicated by purple points). The median azimuthal and vertical velocities (panels \textit{a} and \textit{c} of Fig.~\ref{fig:modsuite} respectively) are negligibly affected by the presence of the SMC, although some individual realisations have quite large differences from the median. In the case of the heavy SMC, certain individual realisations have \textasciitilde20~km~s$^{-1}$ higher azimuthal velocities in the outermost feature fields, and \textasciitilde20~km~s$^{-1}$ lower vertical velocities in the innermost feature fields. However, those realisations are very inconsistent with observations -- the associated negative vertical velocities being \textasciitilde$5\sigma$ inconsistent with the positive vertical velocities measured. We can conclude the strong perturbations associated with these individual realisations are not realistic, despite being within the allowed uncertainties for the central positions and systemic velocities of the SMC in particular, given these large discrepancies are only observed in model suites including the SMC. The SMC also negligibly affects the median velocity dispersion in any of the three components (panels \textit{d-f}), with the only difference being an increased allowable range of azimuthal and radial velocity dispersions in the outermost fields compared to the no-SMC models.

Instead, the SMC has the strongest effect on the in-plane radial velocities (panel \textit{b}), with model suites including the SMC having higher median radial velocities than the no-SMC suite, and the heavy SMC suite generating the largest increase of up to \textasciitilde20~km~s$^{-1}$. Notably, this perturbation is in the wrong direction: these median kinematics are further from the negative observed radial velocities than the no-SMC suite (although we note some individual realisations of the base-case and heavy SMC suites do overlap the range of the no-SMC suite). This suggests recent interactions with the SMC, as captured by these models, are not the source of the perturbation generating the northern arm-like feature. 

In the innermost two fields along the northern arm, the base-case and heavy SMC suites have slightly smaller (\textasciitilde0.4~kpc) median out-of-plane distances than the no-SMC suite (Fig.~\ref{fig:moddist}). However, in fields further along the feature, these distances significantly increase, with the median out-of-plane distance \textasciitilde3~kpc in front of the LMC disk in the outermost field, and some individual realisations >5~kpc from the assumed disk plane. Notably, we find the realisations which produce the largest out-of-plane distances are the same realisations that produce very negative vertical velocities in the innermost fields, strongly inconsistent with observations. Even so, the median out-of-plane distances are moderately larger than the \textasciitilde1.5~kpc uncertainties in distance accommodated by our photometric uncertainties in \S\ref{sec:phot}. Whilst some individual realisations of these models do have out-of-plane distances within this range, a majority of these model realisations are ruled out as these geometries would result in brighter $G_0$ magnitudes along the arm, inconsistent with those measured. This provides further evidence that recent SMC interactions are not responsible for formation of the northern arm. 

\subsubsection{Effect of the LMC mass}\label{sec:lmcmass}
Whilst a number of recent studies have indicated the total LMC mass is large \citep[$\geq$10$^{11}$M$_\odot$: e.g. ][]{erkalTotalMassLarge2019,penarrubiaTimingConstraintTotal2016}, we additionally explore the formation of the northern arm assuming a factor-of-ten lighter LMC (indicated by light green points in Figs.~\ref{fig:modsuite} and~\ref{fig:moddist}) as used in traditional models of the LMC assuming tidal truncation \citep[see e.g. section 2 of][for a review]{garavito-camargoHuntingDarkMatter2019}. The most significant kinematic difference this induces is in the azimuthal velocity (panel \textit{a} of Fig.~\ref{fig:modsuite}), which displays a strong drop from \textasciitilde60~km~s$^{-1}$ in the innermost field to only \textasciitilde20~km~s$^{-1}$ in the outermost field. This is a result of the model setup, rather than a physical perturbative effect from the MW. As the model suites are initialised to match the rotation curve of the LMC ($V_{\text{circ}}$\textasciitilde90~km~s$^{-1}$) at 10~kpc, this necessitates an enclosed mass of nearly $1.5\times10^{10}$M$_\odot$ at this radius. In order to facilitate this, and maintain the total mass of $1.5\times10^{10}$M$_\odot$ for the model, there is negligible dark (or baryonic) matter beyond this radius. As a result, the azimuthal velocity drops off with approximately 1/$R$ dependence as expected given the lack of matter beyond this radius. Given this is strongly inconsistent with the approximately flat azimuthal velocities measured, we can conclude the LMC is not this light, and must be at least $1.5\times10^{11}$M$_\odot$ in order to maintain the flat rotation curve observed across these large galactocentric radii. All other kinematic component medians do not differ significantly under the light-LMC case as compared to the other model suites. 

\subsection{Origin of the northern arm}\label{sec:origin}
As discussed in \S\ref{sec:MWmass}, the increasingly positive vertical velocity observed along the northern arm is consistent with a MW origin, with qualitatively similar trends observed along the arm in all models, including those omitting the SMC. The heavy-MW model suite also produces the closest $\sigma_z$ to that observed, indicating the strength of the Milky Way’s gravitational force in this direction. 
More difficult to understand is the strongly negative radial velocity observed along the arm, which none of our models replicate. The model suite producing the closest match to these kinematics is the heavy MW suite (indicated by dark green points in Figs.~\ref{fig:modsuite} and~\ref{fig:moddist}); individual model realisations in this suite provide the most negative radial velocities along the length of the feature, albeit significantly weaker than those observed (approximately $-10$~km~s$^{-1}$, compared to approximately $-40$~km~s$^{-1}$). Also notable is the $N$-body model (magenta points in Figs.~\ref{fig:nbody} and~\ref{fig:moddist}), which does have a significant negative radial velocity in field 13 (approximately $-35$~km~s$^{-1}$), but which increases rapidly resulting in a strongly positive radial velocity in the outermost feature field. 

\begin{figure*}
	\includegraphics[width=0.33\textwidth]{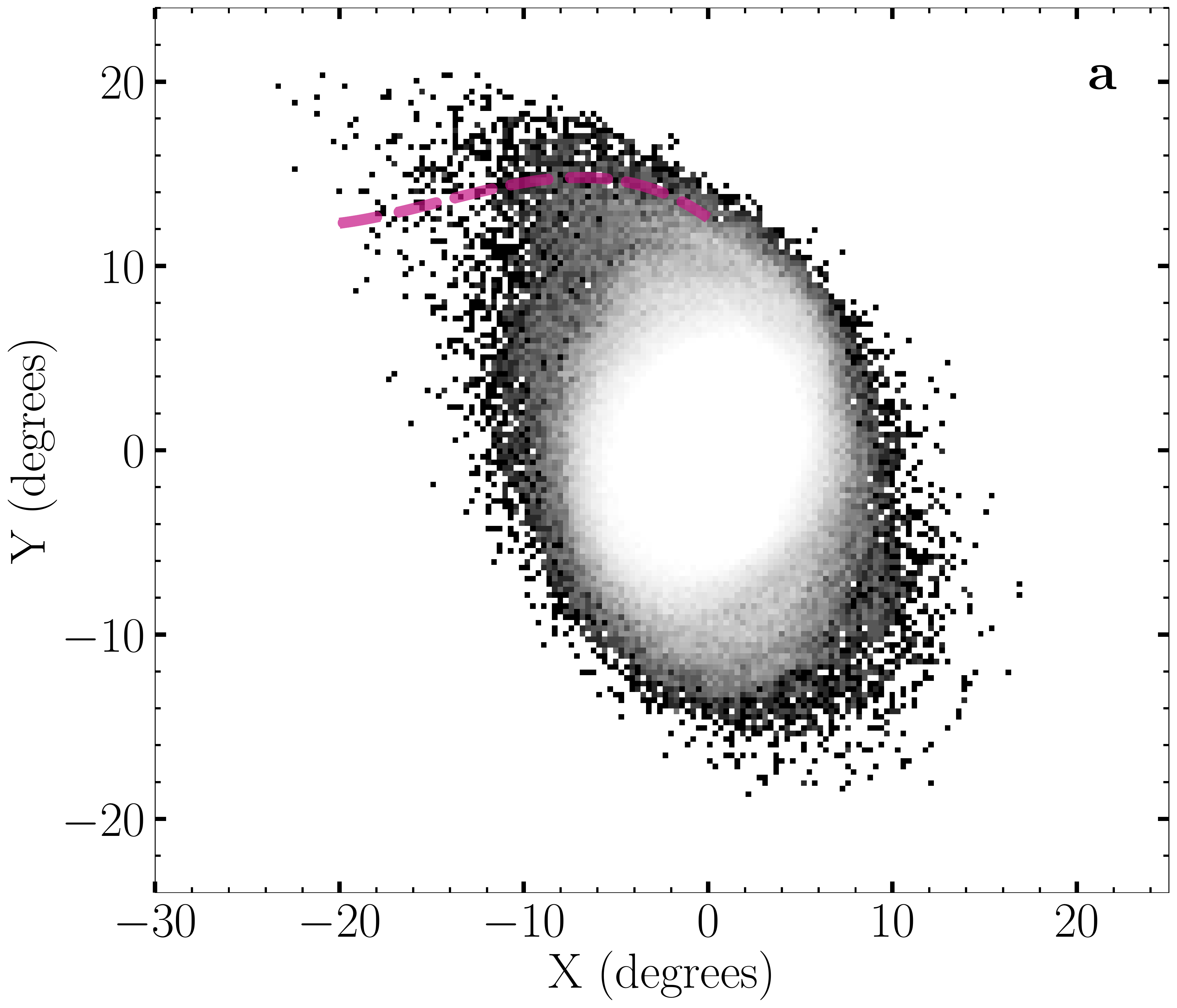}	\includegraphics[width=0.30\textwidth]{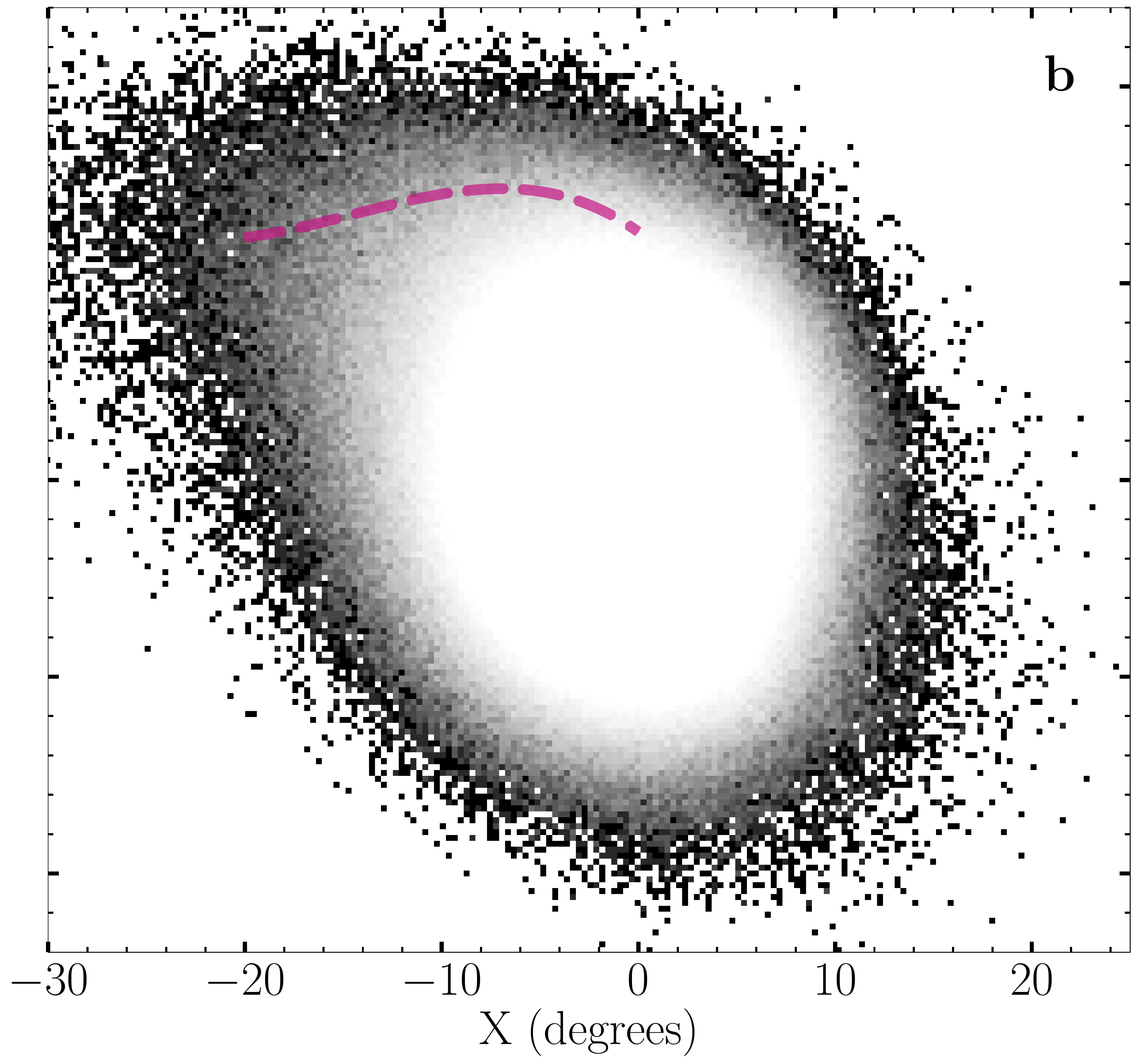}
	\includegraphics[width=0.35\textwidth]{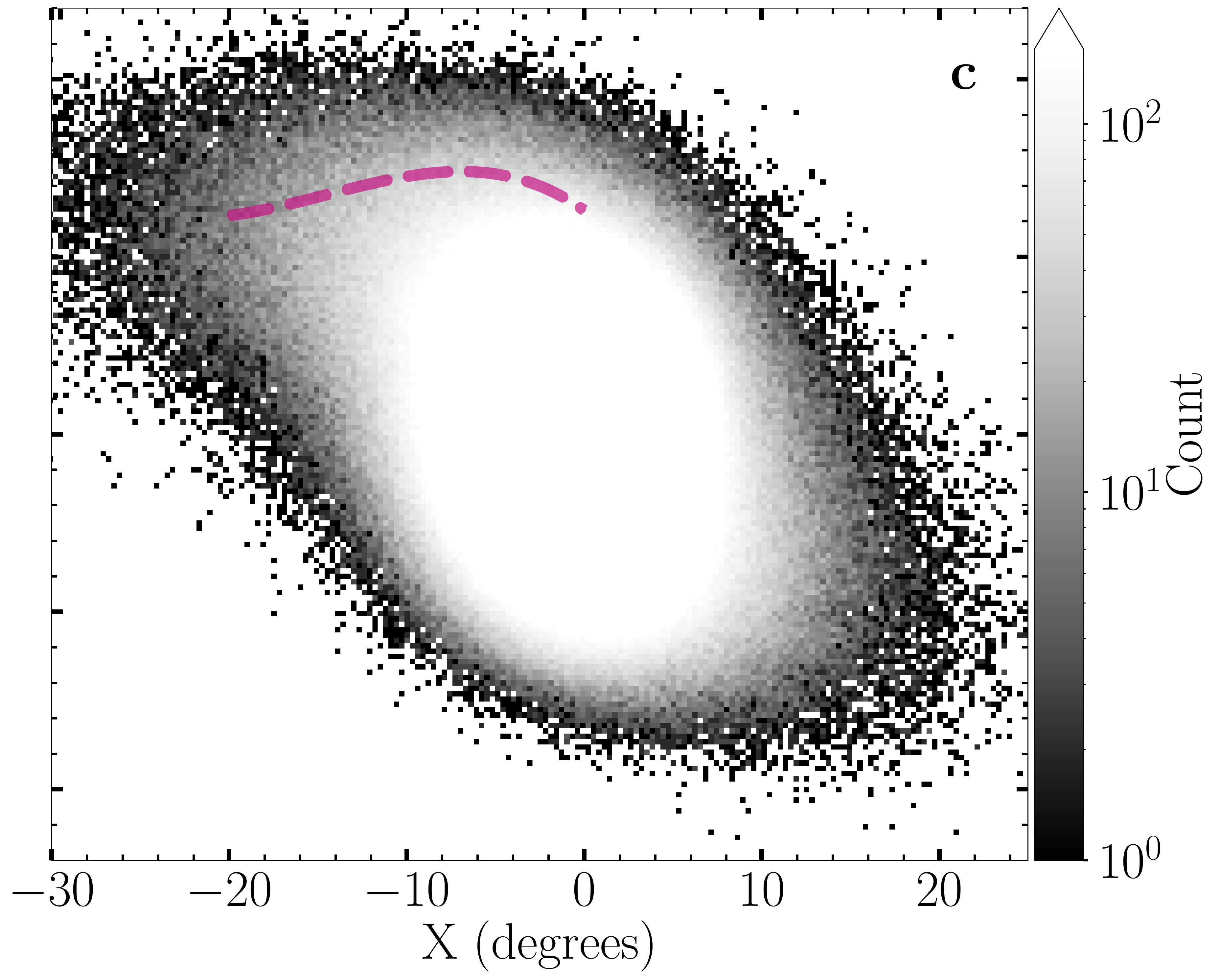}
	\caption{Density plots of model particles for different model suites. Panels show the $N$-body model (\textit{a}), an individual realisation of the base-case suite (\textit{b}), and the same realisation in the heavy MW suite (\textit{c}). Dashed magenta lines in each panel show the observed feature track of the northern arm. The two realisations of the simpler model suites produce relatively flat northern overdensities in the Y direction, similar to that observed, while the debris track in the $N$-body model increases strongly in Y along the length of the feature, to larger values than those observed.}
	\label{fig:heavymw}
\end{figure*}

As discussed in \S\ref{sec:caveats}, we do not expect the lack of self-gravity between the LMC and MW in the simpler model suites to significantly affect the kinematics of the northern arm. However, we do note that the geometry of the feature differs significantly in the simpler models as compared to the $N$-body model. Fig.~\ref{fig:heavymw} shows the density of model particles for the same individual realisation of the base-case and heavy MW model suites, in addition to the $N$-body model. Notably, the debris forming the arm-like feature in the $N$-body model (panel \textit{a}) has a significantly different geometry to the simpler model realisations (panels \textit{b} and \textit{c}), with the structure increasing steeply in Y-position along the length of the feature. This means that when comparing measurements at the same X/Y positions as the observed MagES fields, the regions compared in the $N$-body model are not along the stream of debris actually forming the arm-like structure, and as a result there are fewer model particles within each field. This is in contrast to the simpler model realisations, where the northern overdensities are not significantly different from the observed feature track, albeit without the observed gap between the northern arm and the disk. As a result, it is perhaps not surprising that the kinematics of the $N$-body model are somewhat different to those in the simpler model ensembles, and the observations: different areas of the feature, under the influence of different gravitational forces, are being compared. Nonetheless, as discussed above, when a feature track is fitted to the $N$-body model and equivalent $\phi_1/\phi_2$ locations along the northern arm are compared, the resultant kinematics are not substantially different from those derived when equivalent X/Y positions are compared. 

Given individual realisations of the heavy MW ensemble produce the closest kinematics to those observed, it might be inferred that an even heavier MW is necessary in order to produce the strong observed perturbations in $V_r$. However, we note there is an upper limit on the MW mass beyond which the LMC and SMC become bound, and have experienced multiple previous pericentric passages around the MW. That scenario is inconsistent with results that the Clouds are only now on their first infall into the MW potential \citep{kallivayalilThirdEpochMagellanicCloud2013}. Given the 50\% increase in MW halo mass in our models only has a relatively small effect on $V_r$ (reducing these by \textasciitilde5~km~s$^{-1}$ compared to the no-SMC models with a regular-mass MW), the MW mass required to reproduce the observed kinematics would likely exceed that binding threshold. This implies a heavy MW likely contributes to, but is not the only required condition for, reproduction of the feature kinematics. 

\begin{figure}
	\includegraphics[width=\columnwidth]{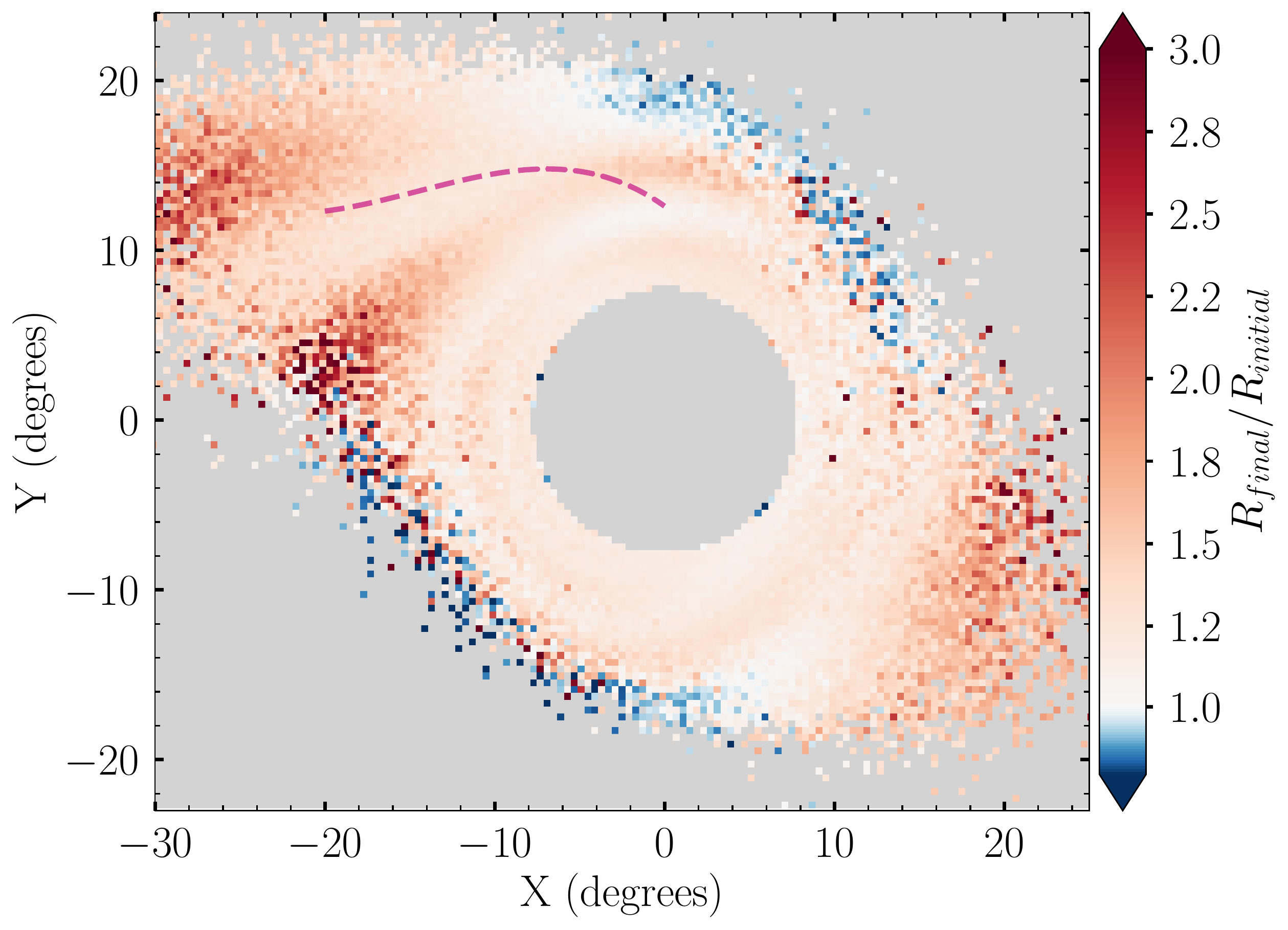}
	\caption{Binned map of particles within a single realisation of the heavy MW model, colour-coded by the mean ratio within each bin of the current LMC galactocentric radius ($R_{\text{final}}$) to the initial particle galactocentric radius 1~Gyr ago during model initialisation ($R_{\text{initial}}$). The dashed magenta line shows the observed feature track of the northern arm. The central 8$^\circ$ of the LMC disk is masked to emphasise the variation in $R_{\text{final}}/R_{\text{initial}}$ in the outskirts of the LMC.}
	\label{fig:radius}
\end{figure}

Further, Fig.~\ref{fig:radius} shows the distribution of model particles for an individual heavy MW realisation, colour-coded by the ratio of each particle's current LMC galactocentric radius $R_{\text{final}}$, to its origin radius $R_{\text{initial}}$. Particles in the region of the northern arm generally move outwards over the course of the simulation, with $R_{\text{final}}/R_{\text{initial}}$\textasciitilde1.2 along most of the arm. This implies particles located at large distances along the arm originate at marginally larger galactocentric radii than particles at the base of the arm: consistent with the mild negative metallicity gradient observed along the arm. However, the fact that $R_{\text{final}}/R_{\text{initial}}$>1 along the length of the arm indicates the MW acts to push stars that form the northern arm outwards from the LMC disk. Notably, immediately below the observed feature track and crossing its base, model particles move strongly outwards: $R_{\text{final}}/R_{\text{initial}}$ reaches up to \textasciitilde3. This may contribute to forming an overdensity along the feature track, with particles immediately below the feature track pushed strongly outwards to form the feature and generate a gap between the feature and the observed LMC disk. This scenario, however, does not explain the strongly negative radial velocities observed along the arm -- indicating models including only the Milky Way do not capture the full perturbation to the LMC.

We next consider the potential effects of recent interactions with the SMC on the northern arm, discussing first the recent pericentric passage of the SMC around the LMC \textasciitilde150~Myr ago. As discussed in \S\ref{sec:simple}, while this is a close pericentre (with the SMC passing within $8.0^{+2.4}_{-2.0}$~kpc of the LMC centre), it is not coincident with a disk plane crossing: the SMC remains \textasciitilde7~kpc below the LMC disk plane during the encounter. As such, we find the SMC does not substantially affect the LMC disk during this interaction and, as discussed in \S\ref{sec:caveats}, the inclusion of self-gravity in the models is unlikely to significantly change this conclusion. In addition, we point out that for every model, the projected location of the pericentric passage is towards the southwest of the LMC: almost directly opposite to the northern arm. At this radius, the circular velocity of the LMC (which as seen in \S\ref{sec:kinematics}, remains constant even along the arm) implies a timescale of \textasciitilde300~Myr for the stars most strongly perturbed by this interaction to reach the north-eastern disk. This is approximately double the \textasciitilde150~Myr that has passed since the pericentric passage, further indicating this interaction is unlikely to be the origin of the northern arm. 

Interactions with a greater possibility of contributing to the formation of the northern arm are SMC crossings of the LMC disk plane, as these directly affect the nearby stars as the SMC passes through the disk. In the \textasciitilde50\% of our base-case and heavy SMC model realisations which experience disk crossings in the past 1~Gyr (beyond that which is currently occurring), we find the LMC disk is most strongly affected by the disk crossing \textasciitilde400~Myr ago. This crossing can occur across a broad range of distances ($28.8^{+11.4}_{-9.2}$~kpc: see \S\ref{sec:simple}) from the LMC centre, but those which occur at the smallest radii have the largest effect on the LMC disk -- and a much more significant effect than the recent SMC pericentric passage in the regions of interest. A handful of models (\textasciitilde9\%) have yet another SMC disk crossing \textasciitilde900~Myr ago, which can occur across a very wide distance range (1$\sigma$ limits of 7.5 and 66.9~kpc). While crossings which occur at distances toward the upper end of this range are unlikely to significantly affect the LMC, the \textasciitilde20\% which pass within 10~kpc of the LMC center could potentially affect the northern arm, and we consider these in addition to the \textasciitilde400~Myr crossing in the discussion that follows.

\begin{figure*}
	\setlength\tabcolsep{0.5pt}
	\begin{tabular}{cccc}
		\includegraphics[height=0.3\textwidth]{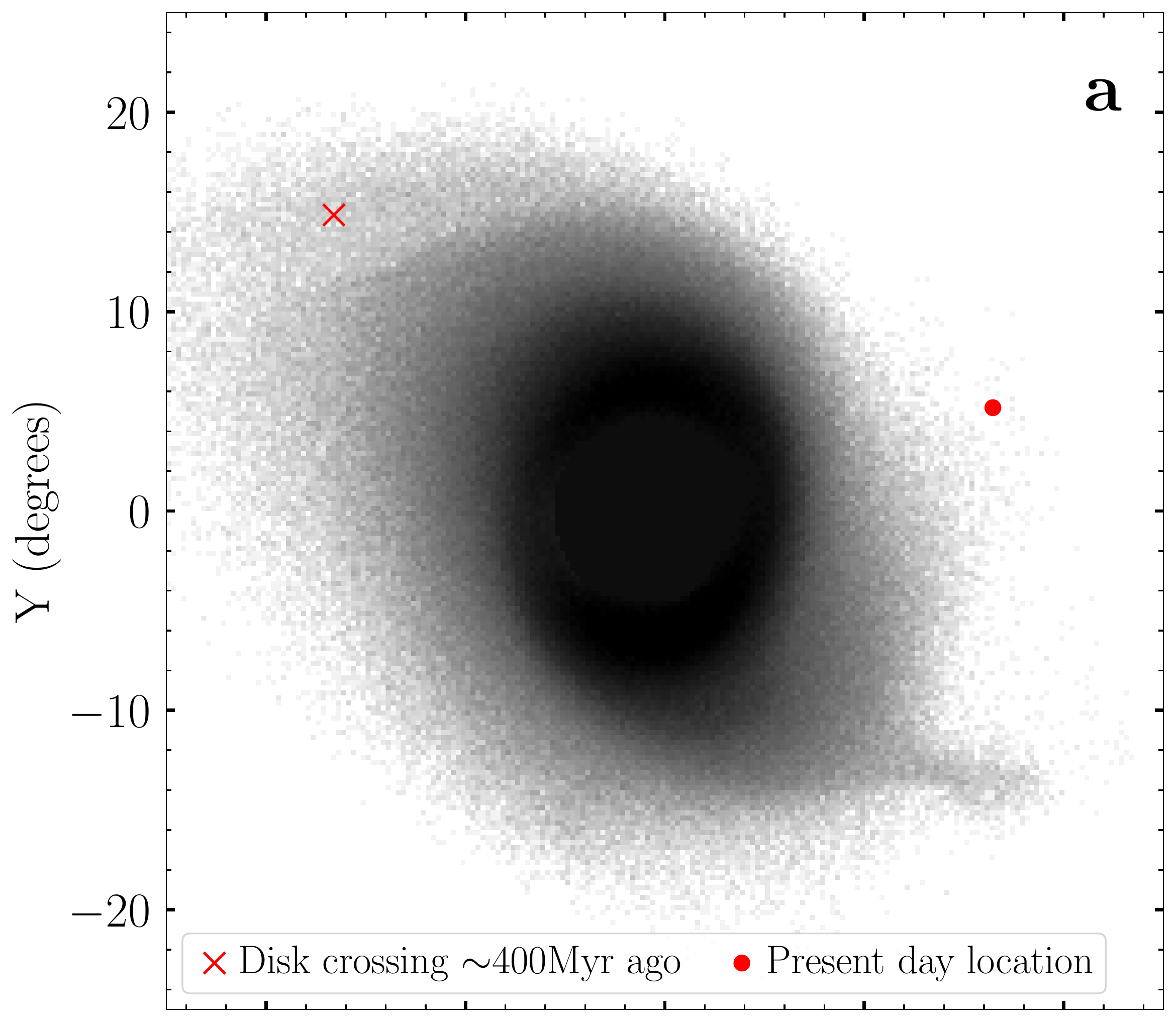} & \includegraphics[height=0.3\textwidth]{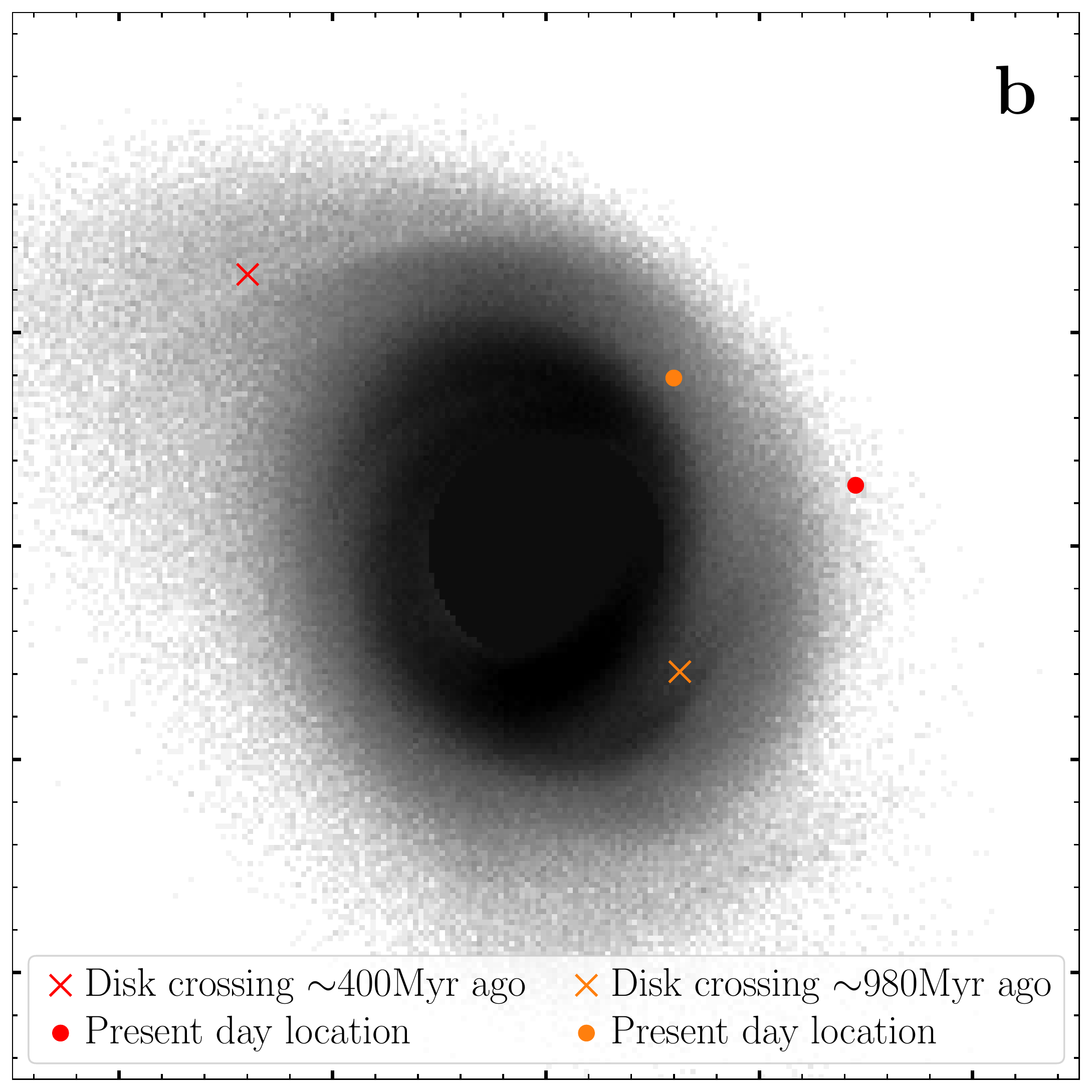} & \includegraphics[height=0.3\textwidth]{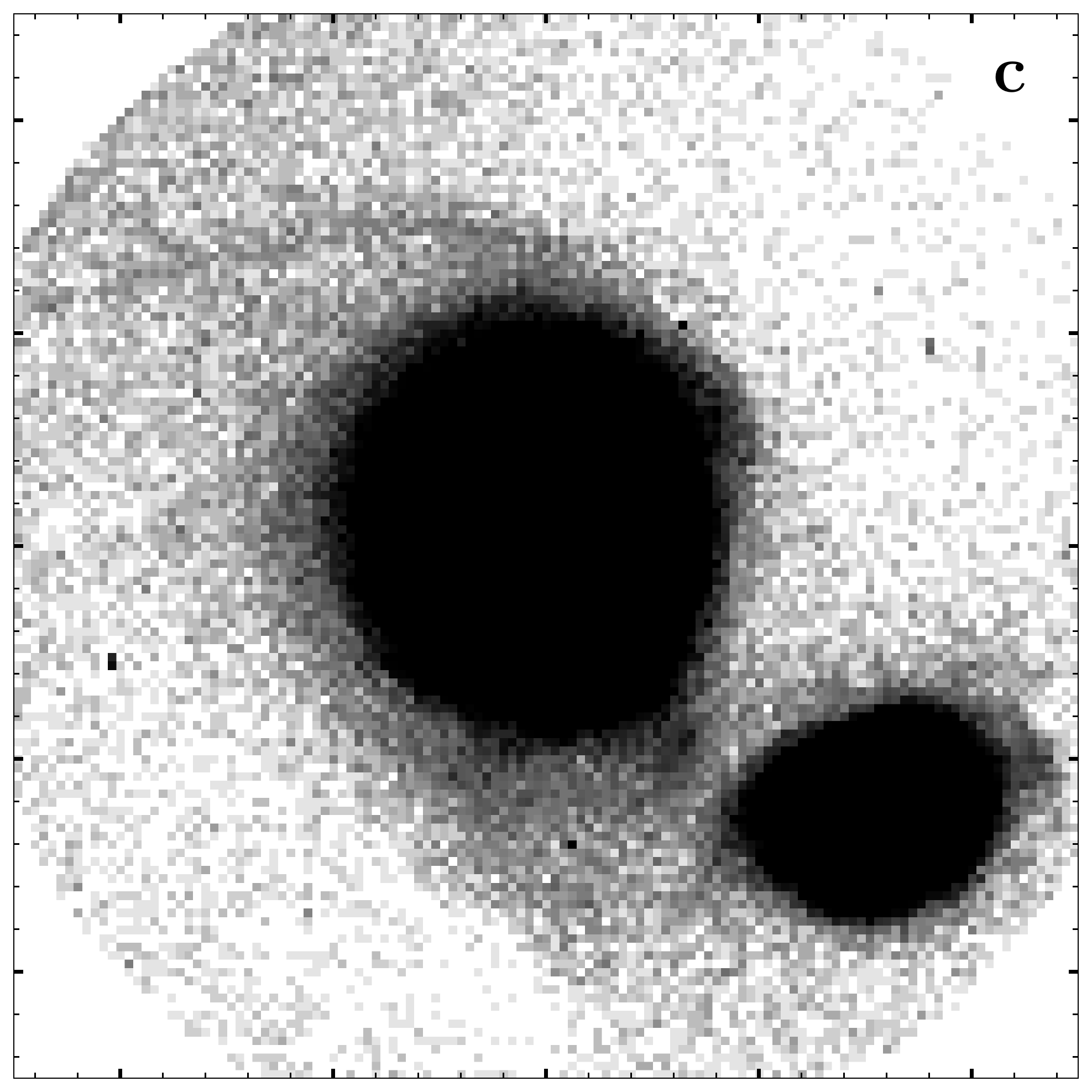} &
		\includegraphics[height=0.3\textwidth]{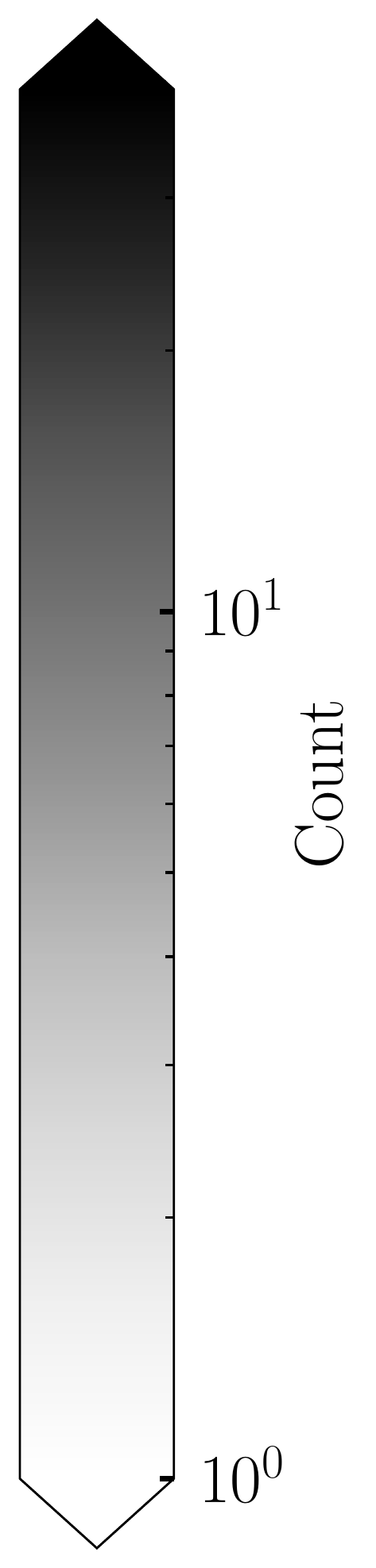}\\ 
		\includegraphics[height=0.332\textwidth]{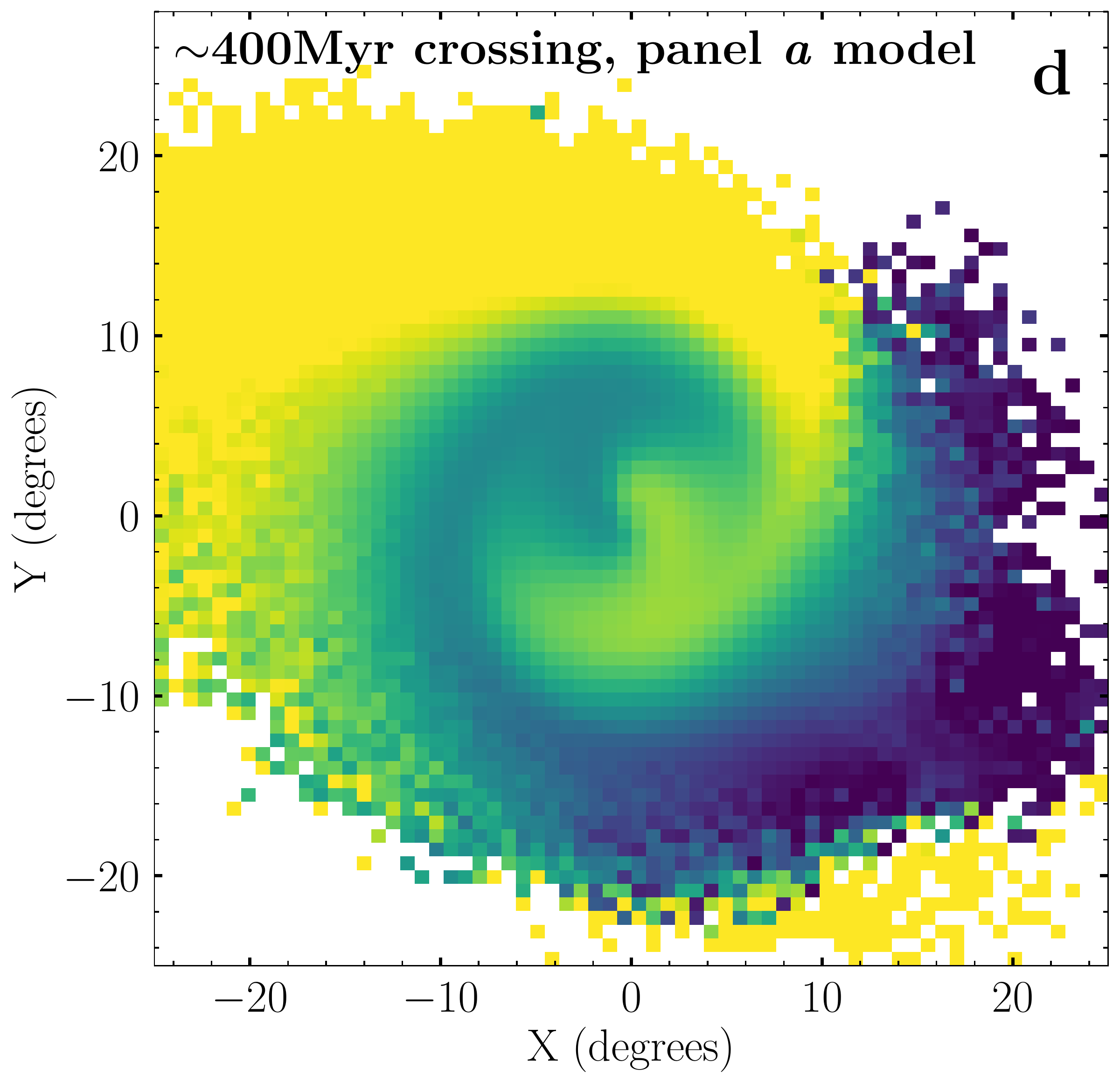} & \includegraphics[height=0.332\textwidth]{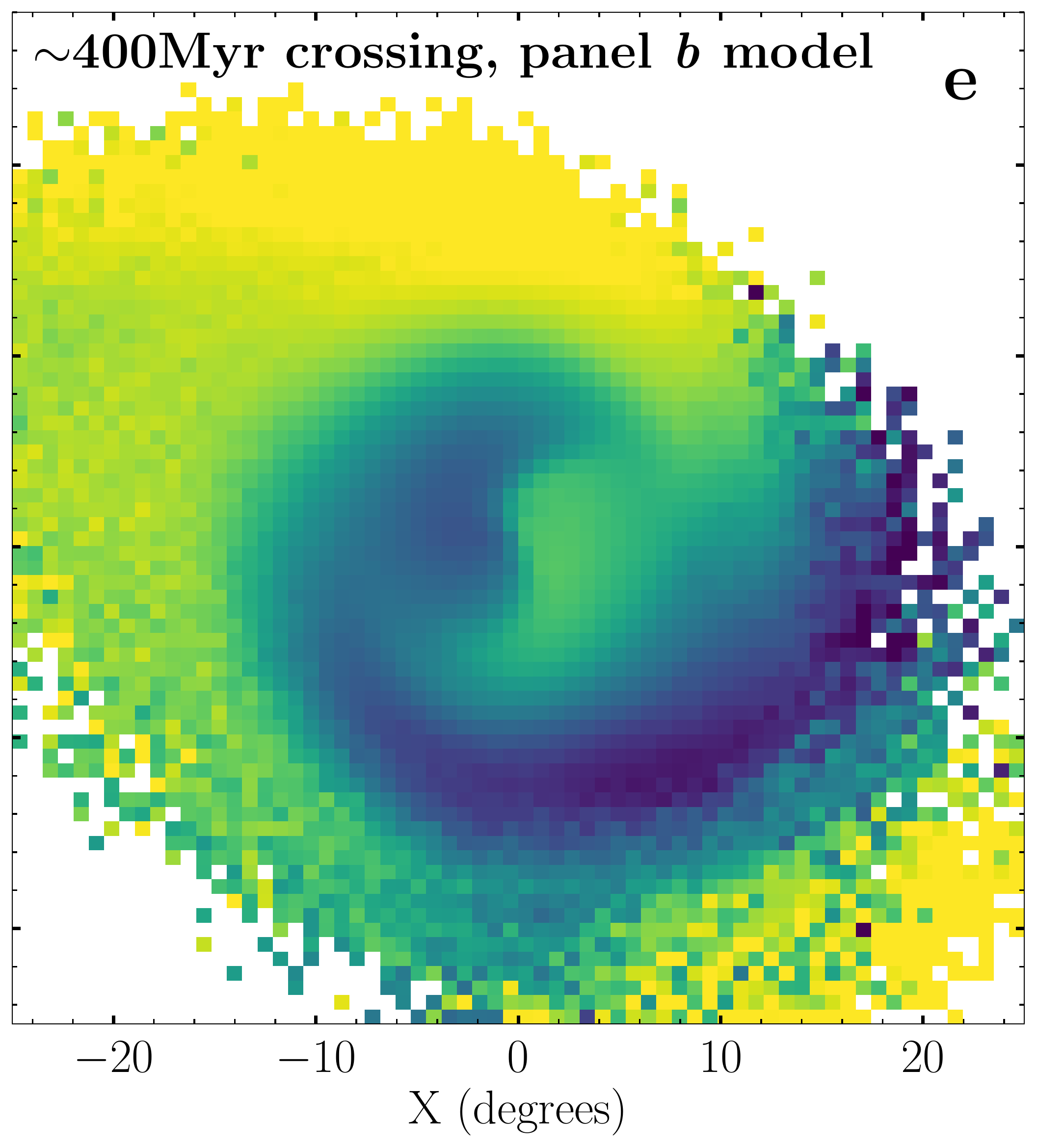} &
		\includegraphics[height=0.332\textwidth]{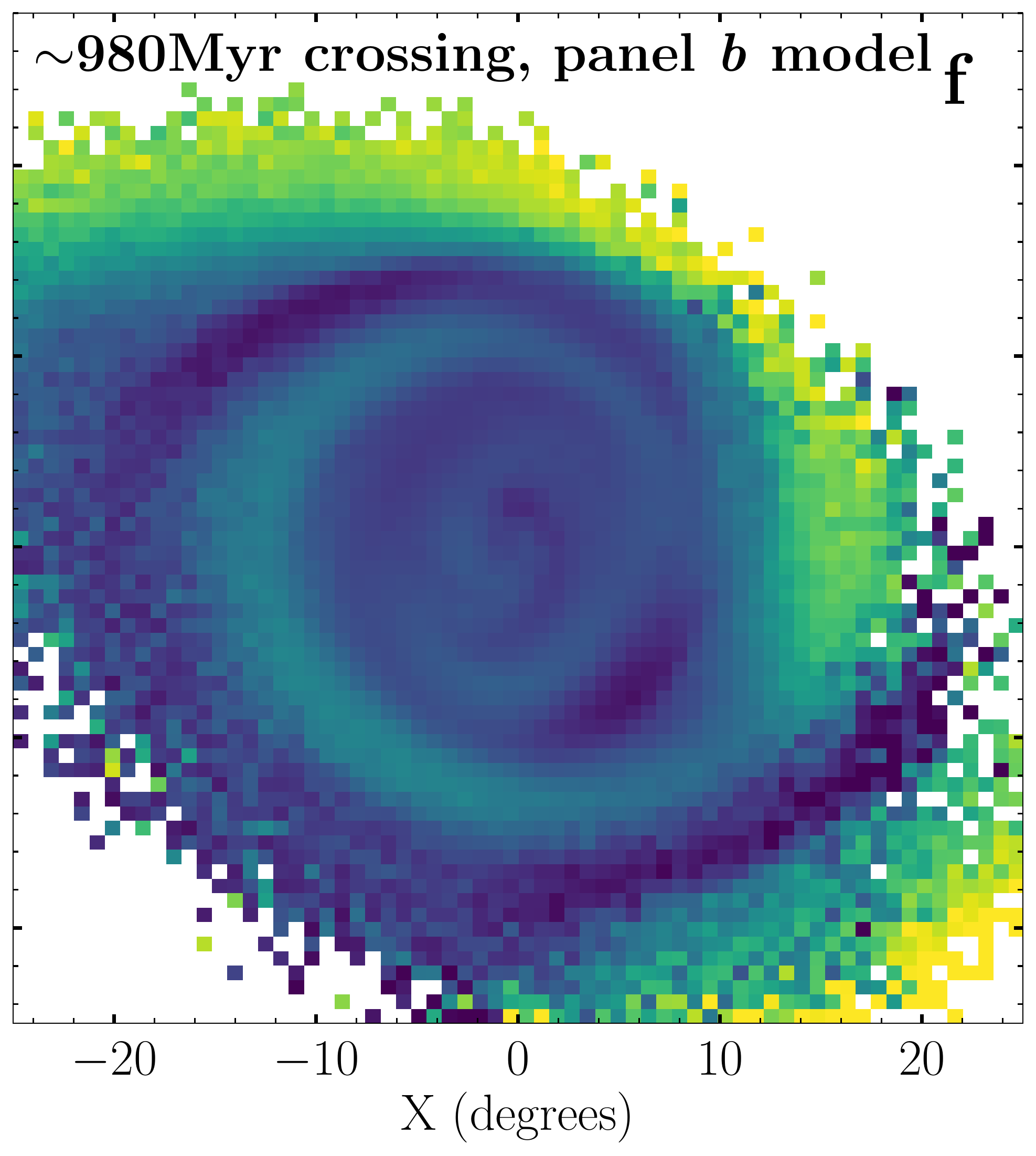} &
		\raisebox{15pt}{\includegraphics[height=0.3\textwidth]{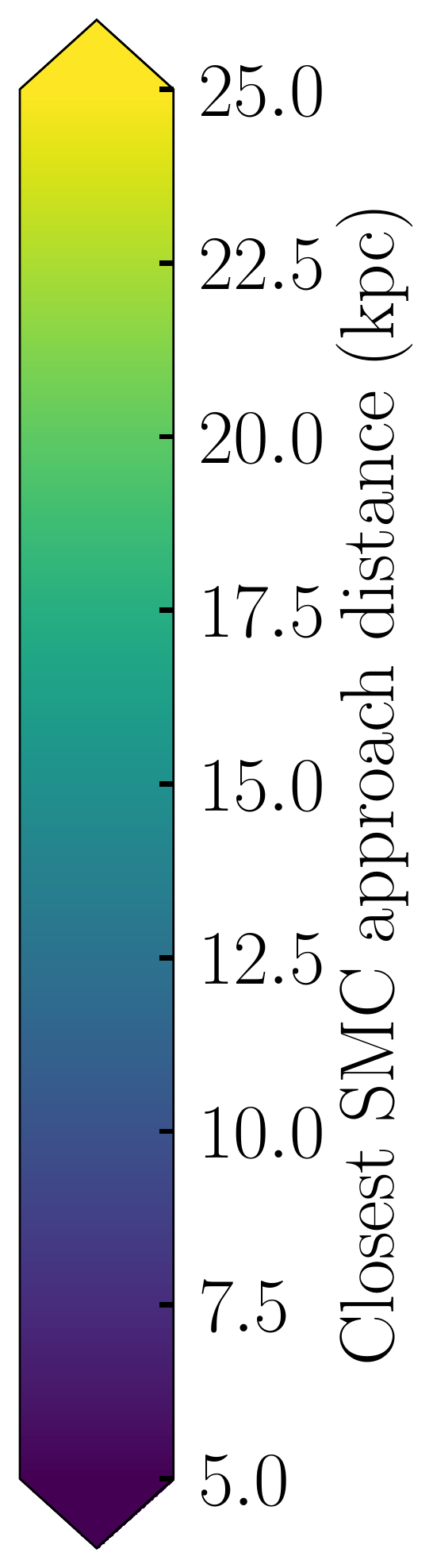}} 	
	\end{tabular}
	
	\caption{Upper panels: Density of particles for base-case model realisations having experienced one (left) or two (center) SMC crossings of the LMC disk plane, compared to the density of observed LMC stars selected using very similar criteria to \protect\cite{gaiacollaborationGaiaEarlyData2021a} (right). Locations of crossings are marked by coloured x-signs, with the present-day location of the crossing (different to the present-day location of the SMC itself) marked with circles of the corresponding colour. Lower panels: Current model particle distribution, colour-coded by the particle distance from the SMC at the time of each SMC crossing of the LMC disk plane. In order, panels show the realisation with a single crossing \textasciitilde400~Myr ago (left) corresponding to the density map in panel \textit{a}, the \textasciitilde400~Myr crossing (centre) in the two-crossing model corresponding to the density map in panel \textit{b}, and the \textasciitilde980~Myr crossing in the two-crossing model (right) corresponding to the density map in panel \textit{b}. Stars closely perturbed during the most recent SMC disk crossing \textasciitilde400~Myr ago, in both realisations, now comprise the western LMC disk (which appears truncated in Gaia maps of the periphery, as in panel \textit{c}.}
	\label{fig:smctrunc}
\end{figure*}

Fig.~\ref{fig:smctrunc} presents results from two realisations of the base case model, as highlighted in Fig.~\ref{fig:orbits}, demonstrating the effect of these disk crossings on the LMC disk. The left panels present a realisation which has only experienced the most recent \textasciitilde400~Myr disk crossing -- which in this model occurs at a distance of \textasciitilde18~kpc from the LMC center -- and the centre panels present one of the few realisations which has experienced two SMC disk crossings (excluding that which is currently occurring) within the past 1~Gyr. These occur \textasciitilde360 and \textasciitilde980~Myr ago, at LMC distances of \textasciitilde14.5~kpc and \textasciitilde8~kpc respectively. The upper panels show the original locations of these disk plane crossings, and the present-day location of the crossing, computed by rotating the location of the original disk-crossing within the LMC's disk plane assuming a circular velocity of 90~km~s$^{-1}$. Note these are different to the present-day location of the SMC itself. Lower panels show the binned current LMC particle distribution, colour-coded by the distance of each particle from the SMC at the time of each disk crossing.

We first discuss the SMC crossing of the LMC disk plane \textasciitilde400~Myr ago. Considering first the realisation which has experienced only this crossing, shown in the leftmost panels of Fig.~\ref{fig:smctrunc}, we find the geometry of the northern arm in this realisation is very similar to that in both the base-case and heavy MW model suite realisations in Fig.~\ref{fig:heavymw}. This indicates the crossing does not significantly impact the geometry of the northern arm. Further, as seen in panels \textit{d} and \textit{e} of Fig.~\ref{fig:smctrunc}, particles which today form the northern arm are not closely perturbed by the SMC during its crossing of the LMC disk plane. The fact that this crossing typically occurs at large LMC galactocentric radii (typically double that of the recent pericentric passage) further indicates that, as discussed in \S\ref{sec:caveats}, the inclusion of self-gravity in the models is unlikely to change this result. This fact, in conjunction with the fact that model realisations which experience this crossing (and indeed most realisations in both the base-case and heavy SMC model suites) do not produce negative in-plane radial velocities as observed along the northern arm, lead us to conclude the disk crossing \textasciitilde400~Myr ago is likely not the origin of the northern arm.

Instead, we do note particles most closely perturbed during this disk plane crossing have in fact moved clockwise with the LMC’s rotation (from red cross to red circle in panels \textit{a} and \textit{b} of Fig.~\ref{fig:smctrunc}), and are now located in the western outskirts of the LMC disk: the same region as the observed apparent truncation in the western LMC disk at a radius of \textasciitilde10$^\circ$ in panel \textit{c} of Fig.~\ref{fig:smctrunc}. The MagES collaboration is currently investigating this truncation feature, and the potential role of the SMC in its formation, in more detail (Cullinane et al. in prep).

We next consider the model realisation which experiences disk crossings both \textasciitilde400 and \textasciitilde900~Myr ago, focussing on the older disk crossing which occurs in this model \textasciitilde980~Myr ago. Panel \textit{f} reveals some particles closely perturbed in this crossing are now located in the vicinity the northern arm. This is evidence that historical interactions with the SMC can potentially influence stars which now form the northern arm. We do find that the few realisations in both the base-case and heavy-SMC models which have experienced this older disk crossing still do not produce the negative in-plane radial velocities as observed along the northern arm. However, it is possible for these early disk crossings to occur at small LMC radii (see \ref{sec:simple}); and in such a case, as discussed in \S\ref{sec:caveats}, our relatively simple models would not capture the full effect of the interaction due to the lack of self-gravity incorporated in the models. Notably, \citet{beslaLowSurfaceBrightness2016} find that multiple LMC/SMC close passages over the course of 6~Gyr can produce significant overdensities and apparent spiral arms in the outer LMC disk, particularly in its northern outskirts at similar distances to the location of the arm today, though they do not report kinematics for these features. It is thus plausible that early interactions with the SMC may have perturbed stars which today form the northern arm, producing both the characteristic gap between the arm and the nearby northern LMC disk, and the strongly negative in-plane radial velocities observed, neither of which are replicated in our simpler models. More realistic models are thus required to confirm this possibility, and better constrain these early interactions between the Clouds.

In summary, we posit the following scenario for the formation of the northern arm. Prior to the Clouds’ infall into the MW potential and up to \textasciitilde1~Gyr ago, historical interactions between the LMC and SMC, potentially including disk crossings at small LMC galactocentric radii, perturb stars that, at the present day, comprise the northern outskirts of the LMC, inparting a strongly negative radial velocity to the stars which will eventually form the arm. Over the last \textasciitilde Gyr, the Clouds have fallen into a relatively massive MW potential, which acts to further perturb these stars -- particularly in the $z$-direction -- whilst they rotate around the LMC, producing the arm-like feature seen today. Recent interactions between the LMC and SMC during the past Gyr, particularly the SMC's recent pericentric passage \textasciitilde150~Myr ago and an SMC crossing of the LMC disk plane \textasciitilde400~Myr ago, likely do not strongly affect the stars that form the northern arm, but do closely impact stars which today form a truncation in the western LMC disk.

\section{Summary}\label{sec:concs}
We have performed a detailed investigation of the arm-like feature in the extreme northern outskirts of the LMC first discovered by \citet{mackey10KpcStellar2016}. Our analysis utilises spectroscopic data for red clump and red giant branch stars from seven MagES fields located along the full length of the feature to obtain [Fe/H] abundances, and in conjunction with Gaia EDR3 data, the first 3D kinematics for individual stars within the arm. We also use Gaia photometry of the red clump to probe the structure of the arm. 

We find the northern arm generally follows the inclination of the LMC disk plane, and has a similar thickness to the outer LMC disk. The median metallicity near the base of the arm is consistent with that in the nearby outer LMC disk, and we find weak evidence for a mild negative gradient in [Fe/H], decreasing from approximately $-0.9$ at 11~kpc from the LMC centre, to approximately $-1.2$ at an LMC galactocentric radius of \textasciitilde22~kpc in the outermost MagES feature field. We therefore conclude the arm is comprised of LMC disk material.

The kinematics of the northern arm also indicate it is comprised of perturbed LMC material. The azimuthal velocity remains reasonably constant along the feature, at approximately \textasciitilde60~km~s$^{-1}$: similar to that measured in the outer LMC disk. In contrast, the in-plane radial velocity and out-of-plane vertical velocities are strongly perturbed. Both of these velocity components are near zero at the base of the arm, consistent with the equilibrium values in the outer LMC disk. However, the in-plane radial velocity drops to approximately $-40$~km~s$^{-1}$ just two degrees from the base of the arm, remaining near this value along its length, and the vertical velocity steadily increases to \textasciitilde30~km~s$^{-1}$ along the length of the arm. The velocity dispersion in each component decreases along the length of the arm, from values comparable to those in the outer LMC disk near the base of the arm, to roughly half this in the outermost MagES field. 

In order to understand the formation of the northern arm, we develop a new suite of dynamical models, sampling from uncertainties in the LMC and SMC central locations and systemic motions, and investigating the effect of different LMC/SMC/MW masses on the structure and kinematics of the feature. Our models describe the LMC as a collection of \textasciitilde$2.5\times10^6$ tracer particles within a rigid two-component potential, and the SMC as a rigid Hernquist potential. The geometry of the LMC disk plane is aligned with that from \citetalias{choiSMASHingLMCTidally2018}.  Both Clouds are initialised at their present day locations, then rewound for 1~Gyr in the presence of each other and the Milky Way. The tracer particle distribution of the LMC disk is then generated, and the system allowed to evolve to the present.

In order to explore the large and complex parameter space of the Magellanic system, these models are necessarily somewhat simplified -- they lack self-gravity, as well as dynamical friction between the LMC and SMC, and the Hernquist potential used to describe the SMC does not capture its tidal disruption. Each of these simplifications can affect the orbits of, and thus interactions between, the two Clouds; we therefore perform only qualitative comparisons with observations. We do note, however, that potential LMC-SMC interactions of interest -- particularly the recent SMC pericentre \textasciitilde150~Myr ago and a possible crossing of the LMC disk plane by the SMC \textasciitilde400~Myr ago -- are relatively recent, and are suggested to occur at reasonable distances from the LMC COM. The resultant effects (or lack thereof) on the northern arm are thus unlikely to differ significantly from those predicted by the simple model suites, although full N-body simulations are required to confirm this. 

We find models with a heavy MW ($1.2\times10^{12}$M$_\odot$) and without an SMC have the closest match to the observed kinematics, reproducing the same qualitative velocity trends as those observed. In these models, the LMC's infall to the Milky Way's gravitational potential produces the increasingly positive out-of-plane velocity along the arm. However, even this model is insufficient to fully reproduce the feature kinematics: most significantly, the observed in-plane radial velocity is \textasciitilde20-30~km~s$^{-1}$ more negative than in the model. 

Our models also suggest that, under the conditions explored, recent (i.e. within the past Gyr) interactions with the SMC do not strongly contribute to the formation of the northern arm. Model LMC particles most significantly disturbed in these interactions, including the recent SMC pericentre \textasciitilde150~Myr ago and a potential recent crossing of the LMC disk plane by the SMC \textasciitilde400~Myr ago, are today located predominantly in the southern and western LMC disk, and are far from the northern arm. Further, model realisations in which the SMC plays a more important dynamical role (particularly those including a heavy SMC) become increasingly inconsistent with observations, with positive in-plane radial velocities and negative vertical velocities (in contrast to the negative in-plane radial velocities and positive vertical velocities observed). However, as it is likely the LMC and SMC are a long-lived binary pair, it is possible that historical interactions with the SMC prior to the past \textasciitilde900~Myr have perturbed LMC stars which now form the northern arm. Indeed, such interactions could be responsible for the strongly negative observed in-plane radial velocity, which is not replicated in any of our models. 

In summary, we suggest the following origin for the northern arm. Prior to the Clouds’ infall into the MW potential \textasciitilde1~Gyr ago, interactions between the LMC and SMC perturbed the kinematics of stars in what is now the northern outskirts of the LMC, generating negative in-plane radial velocities. Over the last \textasciitilde Gyr, as the LMC has fallen into the Milky Way potential (where a higher Milky Way mass is preferred), these stars have been further perturbed, producing the characteristic shape of the northern arm, and the positive out-of-plane velocities observed along its length. Self-gravitating models that are able to more accurately trace the dynamical influence and evolution of the SMC over longer timescales will be required to quantitatively test this scenario.

\section*{Acknowledgements}
We thank Eugene Vasiliev for helpful advice on how to use \textsc{agama}. This work has made use of data from the European Space Agency (ESA) mission {\it \textit{Gaia}} (\url{https://www.cosmos.esa.int/gaia}), processed by the {\it \textit{Gaia}} Data Processing and Analysis Consortium (DPAC, \url{https://www.cosmos.esa.int/web/gaia/dpac/consortium}). Funding for the DPAC has been provided by national institutions, in particular the institutions participating in the {\textit{Gaia}} Multilateral Agreement. Based on data acquired at the Anglo-Australian Observatory. We acknowledge the traditional owners of the land on which the AAT stands, the Gamilaraay people, and pay our respects to elders past, present and emerging. This research has been supported in part by the Australian Research Council (ARC) Discovery Projects grant DP150103294. ADM is supported by an ARC Future Fellowship (FT160100206).

\section*{Data availability}
The data underlying this article will be shared on reasonable request to the corresponding author.


\bibliographystyle{mnras}
\bibliography{p2refs} 




\bsp	
\label{lastpage}
\end{document}